\documentclass[a4paper,12pt]{article}
\usepackage{epsfig,graphicx,xcolor,soul,amsbsy,amssymb,latexsym,amsfonts,amsmath,tcolorbox,setspace}
\usepackage{pstricks}
\usepackage{color}

\usepackage{eurosym}
\usepackage[a4paper]{geometry}
\usepackage{graphicx}
\usepackage{subfigure}
\usepackage{xcolor}
\usepackage{amsmath,amssymb,amsfonts,tcolorbox,pstricks,mathtools, setspace}
\numberwithin{equation}{section}

\newcommand{\bea}{\begin{eqnarray}\displaystyle}
\newcommand{\eea}{\end{eqnarray}}
\newcommand{\nn}{\nonumber \\}
\newcommand{\nonestar}{${\cal N}=1^*$ }
\newcommand{\rmd}{ {\rm d} }
\newcommand\F{\widetilde{F}}
\newcommand\Z{{\cal Z}}

\newcommand{\figref}[1]{Fig.~\protect\ref{#1}}

\title{\begin{flushright}{\vspace{-2.5cm}\small SNUST 15-02}\end{flushright}\vspace{0.8cm}
\bf{M Strings, Monopole Strings, Modular Forms}\\[15pt]}
\author{\Large  Stefan~Hohenegger\footnote{\tt s.hohenegger@ipnl.in2p3.fr}~,~Amer Iqbal\footnote{\tt  amer@cmsa.fas.harvard.edu}~,~ Soo-Jong Rey\,\footnote{\tt sjrey@snu.ac.kr}}
\date{}

\begin{document}

\maketitle

\begin{center}
\renewcommand{\thefootnote}{\fnsymbol{footnote}}\vspace{-0.5cm}
${}^{\footnotemark[1]}$ Universit\'e Claude Bernard (Lyon 1)\\UMR 5822, CNRS/IN2P3, Institut de Physique Nucl\'eaire, Bat. P. Dirac\\ 4 rue Enrico Fermi, F-69622-Villeurbanne, \rm FRANCE\\[0.4cm]
${}^{\footnotemark[2]}$ Department of Physics \& Department of Mathematics\\
LUMS School of Science \& Engineering, U-Block, D.H.A, Lahore, \rm PAKISTAN\\[0.4cm]
${}^{\footnotemark[2]}$ Center of Mathematical Sciences and Applications \\ Harvard University, Cambridge, MA
02138 USA\\[0.4cm]
${}^{\footnotemark[3]}$ School of Physics and Astronomy \& Center for Theoretical Physics\\
          Seoul National University, Seoul 151-747 \rm KOREA\\[0.4cm]
${}^{\footnotemark[3]}$ Fields, Gravity \& Strings, Center for Theoretical Physics of the Universe \\
Institute for Basic Sciences, Daejeon 305-811 \rm KOREA\\[1cm]
\end{center}

\begin{abstract}
We study relations between M-strings (one-dimensional intersections of M2-branes and M5-branes) in six dimensions and m-strings (magnetically charged monopole strings) in five dimensions. For specific configurations, we propose that the counting functions of BPS bound-states of M-strings capture the elliptic genus of the moduli space of m-strings. We check this proposal for the known cases, the Taub-NUT and Atiyah-Hitchin spaces, for which we find complete agreement. We further analyze the modular properties of the M-string free energies and find that they do not transform covariantly under $SL(2,\mathbb{Z})$. However, for a given number of M-strings, we show that there exists a unique combination of unrefined genus-zero free energies that transforms as a Jacobi form under a congruence subgroup of $SL(2,\mathbb{Z})$. These combinations correspond to summing over different numbers of M5-branes and make sense only if the distances between them are all equal. We explain that this is a necessary condition for the m-string moduli space to be factorizable into relative and center-of-mass parts.

\end{abstract}

${}$\\[500pt]
%${}$}

\tableofcontents

\onehalfspacing

%--------------------------------------------------------
\section{Introduction and Summary}

The dynamics of six-dimensional quantum field theories has a very rich structure since they contain not only particles but also string degrees of freedom. Yet they give rise to consistent superconformal field theories (SCFTs) at the conformal fixed points, with well-defined local energy-momentum tensors. Using F-theory \cite{Vafa:1996xn} on elliptically fibered Calabi-Yau three-folds (CY3folds), such SCFTs have recently been classified \cite{Heckman:2013pva}-\cite{Heckman:2015bfa}. In this framework, the strings \cite{strings} arise from D3-branes wrapping a $\mathbb{P}^1$ inside the base of the elliptically fibered CY3fold, while in the corresponding M-theory description ({\it i.e.} once compactified to a five-dimensional space-time), they correspond to M5-branes wrapping a divisor \cite{Witten:1996qb}.

In this paper, we study these string degrees of freedom more carefully, focusing on two different incarnations that are related by U-duality: The first one was pioneered in \cite{Haghighat:2013gba}, where the one-dimensional intersection of an M2-brane ending on an M5-brane was  dubbed {\it M-string}. In the higher-dimensional F-theory description, this corresponds to a D3-brane wrapping a $\mathbb{P}^1$ with normal bundle ${\cal O}(-2)$ inside the base of the elliptically fibered CY3fold. Replacing the $\mathbb{P}^1$ by a chain of $\mathbb{P}^{1}$'s corresponds to configurations of multiple parallel M5-branes with M2-branes suspended between them. The corresponding CY3fold is an elliptic fibration over a resolved $A_{N-1}$ surface blown up at $N$ points, which can also be realized as an $A_{N-1}$ fibration over $\mathbb{T}^2$. In the latter case, the M-theory compactification gives rise to five-dimensional ${\cal N}=1^{*}$ $SU(N)$ gauge theory. Upon further compactification on a circle, we obtain the four-dimensional ${\cal N}=2^{*}$ gauge theory, whose (complexified) gauge coupling corresponds to the area of the base $\mathbb{T}^2$. The partition function of the Bogomol'nyi-Prasad-Sommerfield (BPS) excitations of the M-strings was worked out in an infinite class of configurations in \cite{Haghighat:2013gba,Haghighat:2013tka,Hohenegger:2013ala} and it was then mapped to the gauge theory partition function.

Another incarnation of string degrees of freedom in the five-dimensional theories can be obtained in a dual formulation. The five-dimensional S-duality maps (electrically charged) particle states to (magnetically charged) {\it monopole string} (m-string) states. The details of this map and in particular the BPS spectra are rather involved~\cite{Douglas:2010iu}-\cite{Tachikawa:2011ch}. The S-duality then implies that  degeneracies of the BPS m-string states can be extracted from the five-dimensional ${\cal N}=1^{*}$ partition function \cite{Kim:2011mv}. In \cite{Bak:2014xwa, Harvey:2014nha}, the elliptic genus (see \cite{EG} for the definition) for the m-strings was directly studied by the string worldsheet path integral approach. For example, the elliptic genus of the Taub-NUT space as the moduli space of charge $(1,1)$ monopoles in $SU(3)$ gauge theory \cite{taub-nut} was computed in \cite{Harvey:2014nha, Bak:2014xwa} and found to agree with the index computed in \cite{Kim:2011mv} for all instances where they are comparable.

In this paper, we show that there exists a natural and direct correspondence between the M-strings and the m-strings and propose that the BPS degeneracies of bound-state of M-strings provide the elliptic genus of the moduli space of corresponding m-strings. More concretely, if we denote the relative moduli space of m-strings of charge $(k_1,\cdots,k_{N-1})$ as $\widehat{\cal M}_{k_1,\cdots, k_{N-1}}$ and the corresponding (equivariantly regularized) elliptic genus $\phi_{\widehat{\cal M}_{k_1,\cdots,k_{N-1}}}(\tau,m,\epsilon_1)$, we propose
\begin{align}
&\lim_{\epsilon_2\mapsto 0}
{\widetilde{F}^{(k_1,\cdots,k_{N-1})}(\tau,m,\epsilon_1,\epsilon_2) \over \widetilde{F}^{(1)}(\tau, m, \epsilon_1, \epsilon_2)}
=\phi_{\widehat{{\cal M}}_{k_1,\cdots,k_{N-1}}}(\tau,m,\epsilon_1) && \mbox{for}&&\mbox{gcd}(k_1,k_2,\cdots,k_{N-1})=1\,.\label{RelPropIntro}
\end{align}
Here, $\widetilde{F}^{(k_1,\cdots,k_{N-1})}$ is the counting-function of M-string bound states of configurations with $k_{i}$ $(i=1, \cdots, N-1)$ M2-branes connecting the $i$-th and $(i+1)$-th M5-brane. The parameters $\epsilon_{1,2}$  are  equivariant deformation parameters. From the point of view of $\widehat{\cal M}_{k_1,\cdots,k_{N-1}}$, $\epsilon_1$ corresponds to the action of a $U(1)$ isometry, which is used to  equivariantly regularize the elliptic genus.  We first confirm (\ref{RelPropIntro}) for the case of the charge $(1,1)$ m-string for $SU(3)$ gauge group whose relative moduli space is known to be the Taub-NUT space. The elliptic genus of the latter was recently calculated in \cite{Harvey:2014nha, Bak:2014xwa} and we will see that the universal part of the Taub-NUT elliptic genus which does not depend on the size of the asymptotic circle is precisely given by $\widetilde{F}^{(1,1)}/\widetilde{F}^{(1)}$. We also consider the case of Atiyah-Hitchin space whose elliptic genus was calculated in \cite{Bak:2014xwa}. We show that part of its elliptic genus, which counts states in the neutral sector, is precisely captured by $\widetilde{F}^{(2)}/\widetilde{F}^{(1)}$.

The functions $\widetilde{F}^{(k_1,\cdots,k_{N-1})}$ can be determined from the M-string partition functions for $N$ parallel M5-branes  ${\cal Z}_{N}(\tau,m,t_{f_{a}},\epsilon_1,\epsilon_2)$ (see \cite{Haghighat:2013gba,Hohenegger:2013ala,Haghighat:2013tka}) for given K\"ahler parameters $t_{f_a}  (a=1, \cdots, N-1)$, which is interpretable as the grand-canonical counting function. Here, $\tau$ corresponds to the complex structure of a torus $\mathbb{T}^2$ on which the M5-branes are compactified and $\epsilon_{1,2}$ are  equivariant deformation parameters. Specifically, we have
\begin{align}
\widetilde{F}^{(k_1,\cdots,k_{N-1})}(\tau,m,\epsilon_1,\epsilon_2)
=\mbox{coefficient of $Q_{f_{1}}^{k_{1}}\cdots Q_{f_{N-1}}^{k_{N-1}}$  in} \ \sum_{\ell=1}^{\infty}\frac{\mu(\ell)}{\ell}\,
\log {\cal Z}_{N}(\ell\tau,\ell m,\ell t_{f_{a}},\ell \epsilon_1,\ell \epsilon_2)\,.\nonumber
\end{align}
In this expansion, ${\bf Q}:= (Q_{f_1}, \cdots, Q_{f_{N-1}})$ denote the fugacities $(e^{2 \pi i t_{f_1}}, \cdots, e^{2 \pi i t_{f_{N-1}}})$ where the K\"ahler parameters  $(t_{f_1}, \cdots, t_{f_{N-1}})$ act as the respective chemical potentials.

From the viewpoint of the M-string partition function ${\cal Z}_{N}(\tau,m,t_{f_{a}},\epsilon_1,\epsilon_2)$, the limit $\epsilon_2\to 0$ in (\ref{RelPropIntro}) corresponds to the Nekrasov-Shatashvili (NS) limit \cite{Nekrasov:2009rc, Mironov:2009uv}, which is required for the five-dimensional S-duality correspondence to the m-strings to work. Put differently, the aforementioned five-dimensional S-duality transformation is possible only for certain values of the $\Omega$-deformation parameters $(\epsilon_1, \epsilon_2)$. Indeed, the m-string in five dimensions is an extended object and hence its ground-state should possess $ISO(2)$ boost isometry. From the viewpoint of the M-string configuration, this isometry is in general broken by the equivariant deformations. To restore it, the NS-limit needs to be taken.

Another hint for the necessity of the NS-limit comes from the modular properties of the free energies $\widetilde{F}^{(k_1,\cdots,k_{N-1})}(\tau,m,\epsilon_1,\epsilon_2)$. For general $(k_1, \cdots, k_{N-1})$, the latter do not have any particular modular properties, not even under some congruence subgroup $\Gamma$ of $SL(2,\mathbb{Z})$. This means they do not transform in a nice way under the transformations
\begin{align}
(\tau, m, \epsilon_1,\epsilon_2)\mapsto \Big(\frac{a \tau + b}{c\tau + d}, \frac{m}{c\tau+d}\,,\frac{\epsilon_1}{c\tau+d}\,,\frac{\epsilon_2}{c\tau+d}\Big)\,\,\,\,\,\,\,\,\,\text{where}\,\,\,\,\,\,\, \left(\begin{array}{cc} a & b \\ c & d\end{array}\right)\in \Gamma\subset SL(2,\mathbb{Z})\, .  \nonumber
\end{align}
Note that we should require covariance under this transformations were we to identify them with the elliptic genus of a hyperk\"ahler manifold of complex dimension $2K$ (see {\it e.g.} \cite{Kawai:1993jk})~\footnote{In fact, in order to identify them with elliptic genera, we must require $\Gamma=SL(2,\mathbb{Z})$, which is the reason for the restriction to $\text{gcd}(k_1,\ldots,k_{N-1})=1$ in (\ref{RelPropIntro}), as we shall discover.}. However, in the NS limit $\epsilon_2 \rightarrow 0$, the function $\epsilon_2 \widetilde{F}^{(k_1,\cdots,k_{N-1})}(\tau, m, \epsilon_1)$ behaves almost like a Jacobi form of weight $-1$ and index $K=\sum_{a=1}^{N-1}k_{a}$ with respect to the variables $(\tau, m)$. Indeed, if in addition we also send $\epsilon_1\to 0$, they become quasi-modular functions: while not being fully covariant as they stand, modular covariance can be restored at the expense of making them non-holomorphic functions. Furthermore, for a given integer $K$, there exists a function $T^{(K)}(\tau,m)$, which is a unique linear combination of all $\widetilde{F}^{(k_1,\cdots,k_{N-1})}(\tau,m)$ with $\sum_{a}k_a=K$ in the limit $\epsilon_1, \epsilon_2 \rightarrow 0$. This function turns out to be a {\sl holomorphic} modular form. We find a pattern concerning general construction of all $T^{(K)}(\tau, m)$. They are weak Jacobi forms of weight $-2$ and index $K$ under some particular congruence subgroup of $SL(2,\mathbb{Z})$.

Physically, an important aspect of the $T^{(K)}(\tau,m)$ is that they combine free energies of all possible connected configurations for a fixed number $K$ of constituent M-strings. In order to render a physical meaning to such combinations, we have to work at a point where all K\"ahler moduli $t_{f_a}$ in the M-string setup are all equal ({\it i.e.} the M5-branes are separated to an equal distance). It is unclear what such combinations refer to in the M-string framework. We suggest that such a prescription is more naturally interpretable in the U-dual configuration of the m-strings. Indeed, recalling that $K$ m-string moduli space is given by
\begin{align}
{\cal M}(K) = \mathbb{R}^3 \times (\mathbb{S}^1_{\rm com} \times \widehat{\cal M}_{\rm rel}(K) )/\mathbb{Z}_K,
\end{align}
we see that the $\mathbb{Z}_K$ acts on $\mathbb{S}^1$ (corresponding to the center of mass moduli) as well as the moduli space of the relative motion $\widehat{\mathcal{M}}_{\text{rel}}(K)$. It only becomes factorized in the limit that the m-string tensions are all set  equal. Moreover, only in this limit, the level-matching condition for the m-string elliptic genus is obeyed.

The rest of this paper is organized as follows. In section 2, we recapitulate the brane configuration relevant for the description of the M-strings. We generalize the discussion of \cite{Bak:2014xwa} and explain various deformations while interpolating between the M-string and the monopole string (m-string). In section 3, we review the construction of the M-string partition function. In section 4, we analyze the modular properties of the M-string free energy and give some explicit examples for the simplest configurations with the lowest number of stretched M2-branes. In section 5, we study the properties of the M-string free energy in the NS limit. We study the charge $(1,1)$ and charge $(2)$ configurations in detail and relate the corresponding M-string partition functions to the elliptic genus of Taub-NUT and Atiyah-Hitchin space, respectively. In section 6, we study combinations of free energies corresponding to different M-string configurations and study the modular properties of the genus-zero part. In appendix A and B, we recapitulate relevant aspects of the magnetic monopoles and of the noncompact hyperk\"ahler geometries that we use  in this paper.
In appendix C, we collect the general expression for the free energy. In appendix D, we collect explicit expressions of  free energies for some of the lower charge configurations. In appendix E, we review modular objects. In appendix F, we collect lengthy expressions of the free energies for higher charge  configurations. \\[10pt]

\noindent
{\bf Note added:} While this paper was being completed, the paper \cite{Haghighat:2015coa} appeared on the ArXiv, which has partial overlap with the ideas in sections 2 and 3.
%------------------------------------------------------
\section{Brane Configurations}
The problem of counting BPS excitationa in \nonestar theories can be formulated using configurations in M-theory and their Type IIA reductions, the first  equivariantly deformed versions of which were first given in \cite{Haghighat:2013gba} for M-strings. Here, we consider another configuration that allows an interpretation in terms of m-strings. Indeed, depending on U-duality frames chosen for the Type IIA reduction, the BPS states can be interpreted as arising either from M-strings or m-strings. In this section, we elaborate on this point and explain different hyperk\"ahler geometries of the moduli space of the BPS states that result from different U-duality frames.

%--------------------------------------------------------
\subsection{Supersymmetry}
We study brane configurations in M-theory, consisting of $N$ parallel M5-branes with a number of $K$ different M2-branes stretched between them, in addition to a number $M$ of M-waves in $\mathbb{R}^{1,10}$. The world-volumes of multiple M5-branes are oriented along $(0,1,2,3,4,5)$ directions. When the branes coincide, the spacetime (Poincar\'e) symmetry $ISO(1,10)$ is broken to $ISO(1,5) \times Spin_R(5)$, which is further broken to $Spin_R(4)$ when the branes are split linearly along the $(6)$ direction.
We consider a split by a finite distance and place the $N$ parallel M5-branes at $ - \infty < a_1 \le a_2 \le \cdots \le a_N < + \infty$. The moduli space and R-symmetry then become
\begin{align}
&(\mathbb{R}^5)^N / S_N  &&\longrightarrow&&  (\mathbb{R}^4)^N/S_N  \nonumber \\
&Sp_R(4)  &&\longrightarrow&&  Spin_R(4).
\end{align}
The M5-brane preserves the supersymmetry generated by the 32-component spinor $\epsilon$ satisfying the projection condition
\bea
\Gamma^0 \Gamma^1 \Gamma^2 \Gamma^3 \Gamma^4 \Gamma^5 \epsilon = \epsilon,
\eea
where $\Gamma^I,  I=0, 1, \cdots, 10$ are $32\times 32$ Dirac matrices. In the signature convention $(\Gamma^0)^2 = - \mathbb{I}, (\Gamma^1)^2 = \cdots = (\Gamma^{10})^2 = + \mathbb{I}$, they obey $\Gamma^0 \Gamma^1 \cdots \Gamma^{10} = \mathbb{I}$.
The BPS excitations on the M5-brane worldvolume are provided by other M-branes.

The world-volumes of the M2-branes are oriented along $(0,1,6)$ directions. They are distributed among $N-1$ intervals formed by separated M5-branes along the (6) direction with multiplicity ${\bf k} = \{k_i \vert i = 1, \cdots, N-1 \}$. They break the worldvolume Poincare symmetry $ISO(1,5)$ to $ISO(1,1) \times Spin(4)$. The R-symmetry $Spin_R(4)$ of the M5-brane worldvolume theory remains intact. The M2-branes break supersymmetry further to those components satisfying the projection condition
\bea
\Gamma^0 \Gamma^1 \Gamma^6 \epsilon = \epsilon.
\eea
The worldvolume of the multiple M-waves are oriented along $(0,1)$ directions. They are distributed among $N$ M5-branes, with multiplicity ${\bf m}= \{ m_i \vert i = 1, \cdots, N-2 \}$. They preserve the $ISO(1,1) \times Spin(4)$ worldvolume symmetry as well as the $Spin_R(4)$ R-symmetry. The M-waves break supersymmetry further to those components satisfying the projection condition
\bea
\Gamma^0 \Gamma^1 \epsilon = \epsilon.
\eea
The brane complex $(N, K, M)$ is a $1/8$-BPS configuration. It then follows that these residual supercharges form a $(4,0)$ supermultiplet of $ISO(1,1)$. To see this, we combine the projection conditions and the relation $\Gamma^0 \cdots \Gamma^{10} = \mathbb{I}$ and
\bea
\Gamma^2 \Gamma^3 \Gamma^4 \Gamma^5 \epsilon = \epsilon, \qquad
\Gamma^7 \Gamma^8 \Gamma^9 \Gamma^{10} \epsilon = \epsilon, \qquad \Gamma^0 \Gamma^1 \epsilon = \epsilon.
\eea
The space transverse to the M2 branes and M-waves is spanned by (2,3,4,5,7,8,9,10) directions, exhibiting $Spin(8)$ rotational symmetry. Introducing M5-branes breaks this further to $Spin_{\parallel}(4) \times Spin_{\perp}(4)$. Decomposing each $Spin(4)$ to chiral $SU(2)$ and anti-chiral $SU(2)$, respectively, the $1/8$-BPS supercharges form the representation:
\bea
\Big[\text{Spin}_{\parallel}(4)\times \text{Spin}_{\perp}(4)\Big]_{\text{Spin}(1,1)}:\,\,\,({\bf 2}, {\bf 1}, {\bf 2}, {\bf 1})_{+1/2}.
\eea
%

%%%%%%%%%%%%%%%%%%%%%%%%%%%%%%%%%%%%%%%%%%%%%%%%%%%%%%%%
\begin{figure} [tbp]
\centering
\vskip-3cm
%\subfigure[]{
\makebox[\textwidth][c]{\includegraphics[width=0.8\textwidth]{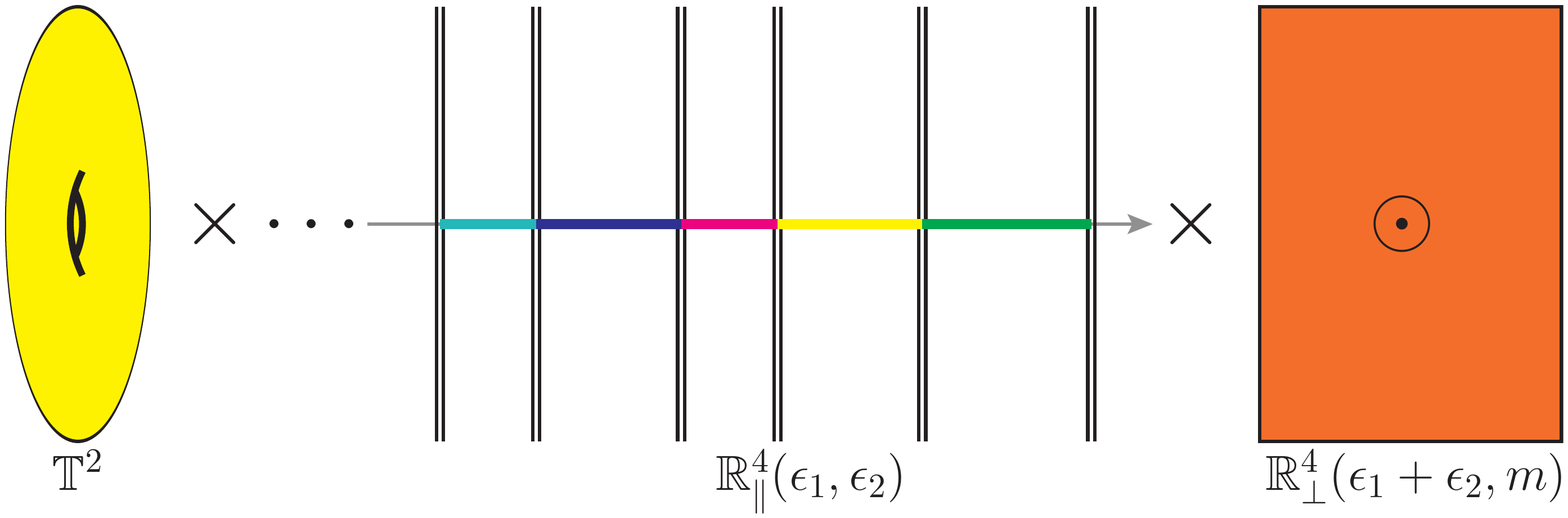}}%
%	\includegraphics[angle=0,width=14cm]{Fig-BraneConf.pdf}
%}
\vskip-2.3cm
\caption{\sl Brane configuration:
The M5-branes are all located at the origin in $\mathbb{R}^4_\perp$,
wrapped around $\mathbb{T}^2$ and stretched along the $(6)$-direction. }
\end{figure}
%%%%%%%%%%%%%%%%%%%%%%%%%%%%%%%%%%%%%%%%%%%%%%%%%%%%%%%%

\noindent
We shall compactify the
$(1)$ direction to a circle of radius $R_1$
so that both M2-branes and M-waves have finite energies. To unambiguously count these energies, we also compactify the $(0)$ direction to a circle of radius $R_0$. Transverse to the M2-branes and M-waves, the $(2,3,4,5)$ directions and the $(7,8,9,10)$ directions are $\mathbb{R}^4_\parallel$ and $\mathbb{R}^4_\perp$, respectively. See figure 1 for illustration of the brane configuration.

%--------------------------------------------------------
\subsection{Omega Background}
To count the BPS states in the M-strings frame, it is necessary to remove contributions due to the noncompact flat directions. This is achieved by formulating the theory on the generalized $\Omega$-background \cite{Nekrasov:2002qd} together with an addition $U(1)_{m}$ corresponding to the mass deformation in the ${\cal N}=2^{*}$ gauge theory, which rotates $\mathbb{R}_{\parallel}^4$ and $\mathbb{R}^4_\perp$ simultaneously by a $U(1)_{\epsilon_1} \times U(1)_{\epsilon_2}\times U(1)_{m}$ action with respect to the $(0)$-direction \cite{Haghighat:2013gba}: If we denote the complex coordinates on $\mathbb{R}_{\parallel}^{4}$ by $(z_{1},z_{2})=(x_2 + i x_3, x_4 + i x_5)$ and on $\mathbb{R}_{\perp}^{4}$ by $(w_1,w_2) = (x_7 + i x_8, x_9 + i x_{10})$, then
\bea
U(1)_{\epsilon_1} \times U(1)_{\epsilon_2} \times U(1)_{m}: \quad
&& (z_1, z_2) \ \quad \rightarrow \quad (e^{2 \pi i \epsilon_1}\,z_1, e^{2 \pi i \epsilon_2}\,z_2) \nonumber \\
&& (w_1, w_2) \quad \rightarrow
\quad \ (e^{2\pi i m- \pi i (\epsilon_1 + \epsilon_2)}\,w_1, e^{-2\pi i m- \pi i (\epsilon_1 + \epsilon_2)}\,w_2).
\nonumber
\eea
The corresponding brane configuration in the M-theory frame is given by
\begin{equation}
\begin{array}{c|cc|cccc|c|cccc}
& {\color{blue} (0)} & {\color{blue} (1)}& 2 & 3 & 4 & 5 & {\color{red} 6 } & \ 7 \ & 8 \ & 9 \ & 10 \\
\hline
M5 \ & = & =  & = &=&=&{\color{blue} =} &&&&& \\
M2 \ & = & =& & & & & {\color{red} =} &&&& \\
M\sim & = & =& && & &&&& \\
\epsilon_1 & & & \circ & \circ &  & & & \circ & \circ & \circ & \circ \\
\epsilon_2 & & & & &\circ & \circ & & \circ &\circ & \circ & \circ\\
m & & &  &  & &  & & \circ &\circ & \circ & \circ
\end{array}
\end{equation}
We put parentheses on the $(0)$ and $(1)$ directions to emphasize that these directions are compactified on circles of radii $R_0, R_1$ respectively, which together form a torus $\mathbb{T}^2$. The circles denote the planes that are twisted by the $\Omega$-deformation when we go around the $(0)$-direction.

We remark that at the outset the mass deformation $m$ was associated with the twist around the $(1)$-direction while the $\Omega$ deformation parameters $(\epsilon_1, \epsilon_2)$ were associated with the twist around the $(0)$-direction. Here, we implicitly included an appropriate action of the mapping class group $SL(2, \mathbb{Z})$ of the torus $\mathbb{T}^2$ so that both twists act in the $(0)$-direction. This is always possible and in fact corresponds to the Type IIA frame.

Wrapped around the $(0)$ direction,  all M5-branes are at the fixed point in $\mathbb{R}^4_\perp$, and the M2-branes and M-waves are at the fixed point in $\mathbb{R}^4_\parallel$. They can be interpreted as multi-instantons on $\mathbb{R}^4_\parallel$ and, roughly speaking, their configurations are described by the Hilbert scheme of points. With these deformations, it follows that the \nonestar partition function becomes  equal to the elliptic genus of the $(4,0)$ supersymmetric nonlinear sigma model whose target space is the noncompact hyperk\"ahler manifold of the multi-instanton moduli space with a suitable choice of vector bundle.

%---------------------------------------------------------
\subsection{Nekrasov-Shatashvili Limit}

Our goal is to map the counting of BPS states of M-strings in six dimensions to the counting of BPS states of m-strings in five dimensions. We will achieve this by first taking the NS limit and then taking an appropriate S-duality action $S \in SL(3, \mathbb{Z})$ that maps the compactified M-string to the compactified m-string and vice versa \footnote{This was also independently observed in \cite{Haghighat:2015coa}.}.

The $(0)$-circle is twisted by the $\Omega$-rotation as well as the mass deformation. On the other hand, the $(1)$-direction is an untwisted Kaluza-Klein circle, which the M-string wraps. By the S-duality action, we would like to map this M-string configuration, which is a particle state on $\mathbb{R}^4_\parallel$, to an m-string configuration, which is a string state on $\mathbb{R}^4_\parallel$.

With the two-parameter $\Omega$-background, however, there is an obstruction to perform the S-duality rotation.
The S-duality action requires a transitive $\mathbb{S}^1$ action, which means that the deformed background has to have the isometry $ISO(2) \times U_{\epsilon_2}(1) \subset ISO(4)_\parallel$. This isometry is regained precisely by the NS limit in which $\epsilon_2$ is set to zero while $\epsilon_1$ is finite.  With the transitive isometry restored, we can now provisionally compactify the $(5)$-direction to a circle of radius $R_5$ and wrap the M-strings and M-waves around it
\footnote{When computing an index or the elliptic genus, we also take the time $(0)$ to be compactified on a circle with radius $\beta$. }.
This is depicted by the following brane configuration in the M-theory frame:
\begin{equation}
\begin{array}{c|cc|cccc|c|cccc}
& {\color{blue} (0)} & {\color{blue} (1)}& 2 & 3 & 4 & {\color{blue} (5) } & {\color{red} 6 } & \ 7 \ & 8 \ & 9 \ & 10 \\
\hline
M5 \ & = & =  & = &=&=&{\color{blue} =} &&&&& \\
M2 \ & = & (=)& & & &(=) & {\color{red} =} &&&& \\
M\sim & = & (=)& && & (=) &&&& \\
%\epsilon_1 & & & & & \circ & \circ & & \circ & \circ & \circ & \circ \\
\epsilon_1 & & & \circ & \circ & &  & & \circ &\circ & \circ & \circ \\
m & & &  &  & &  & & \circ &\circ & \circ & \circ
\end{array}
\end{equation}
The 3-torus $\mathbb{T}^3$ formed by the Euclidean $(0), (1), (5)$ circles is invariant under the action of the mapping class group $SL(3, \mathbb{Z})$ if all directions were untwisted. In the present case, the $(0)$-circle is twisted by the $\Omega$-background rotation, thus breaking the full $SL(3, \mathbb{Z})$ to $SL(2,\mathbb{Z})$ corresponding to the automorphism group of $\mathbb{T}^2$ formed by the $(1,5)$ directions.

%%%%%%%%%%%%%%%%%%%%%%%%%%%%%%%%%%%%%%%%%%%%%%%%%%%%%%%%
\begin{figure}[tbp]
\centering
\vskip-0.3cm
%\subfigure[]{
	\includegraphics[angle=0,width=5cm]{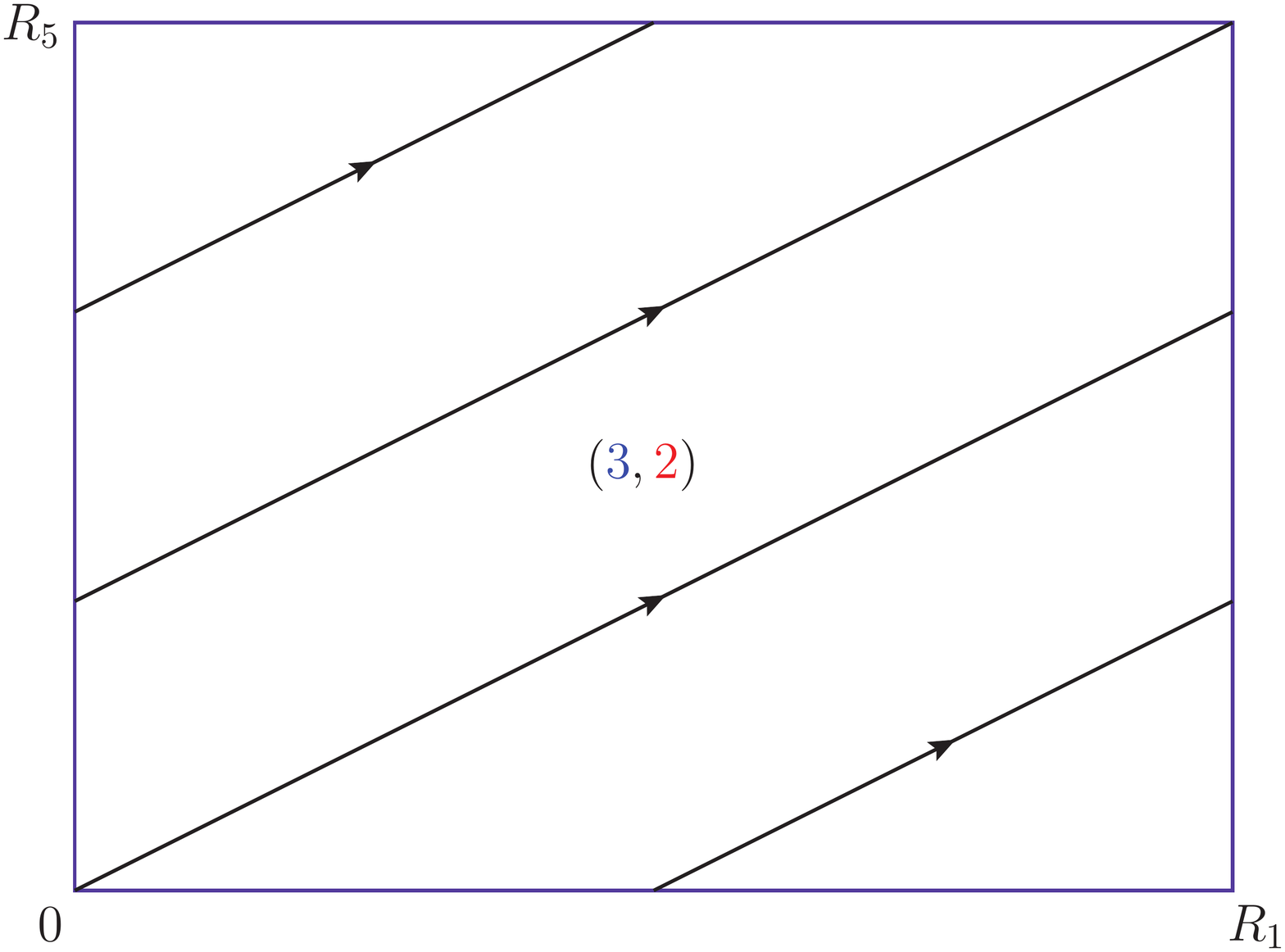}
%}
\vskip-0cm
\caption{\sl One cycle of the M2-brane wraps around the two-torus formed by $(x_1, x_5)$ by $(w_1, w_5) = (3, 2)$ times. The resulting M-string tension is given by
$T = T_{M_2} (R_2 w_1 + R_5 w_5)$. Likewise, the M-wave propagates along the same cycle of the M2-brane. Note that the cycle lies within the M5-worldvolume.}
\end{figure}
%%%%%%%%%%%%%%%%%%%%%%%%%%%%%%%%%%%%%%%%%%%%%%%%%%%%%%%%

Since both the $(1)-$ and $(5)-$directions are compactified, the orientation of the M2-branes and M-waves within this two-dimensional subspace must be specified.
Here we consider wrapping/propagation of the M2-brane and M-wave along the $(1)$-circle direction. However, since the $(5)$ direction is also compactified, the M2-branes and M-waves can also wrap/propagate along the $(5)$-circle direction. Consequently, the M2-brane and M-wave wraps/propagates on a commensurate cycle $(w_1, w_5)$ of the $(1,5)$ torus. This is illustrated in figure 2. Under the S-duality, the two relatively coprime quantum numbers $w_1$ and $w_5$ are interchanged each other. With the $(0)$ direction is taken time direction, M2-branes wrapping on $(1)$ or $(5)$ directions are the M-strings and the m-strings, respectively. We see that the S-duality indeed exchanges the six-dimensional M-strings and five-dimensional m-strings. For a finite $R_5$, if $R_1$ is  much smaller than $R_5$, the low-lying BPS excitations are M-strings; if $R_1$ is much larger than $R_5$, the low-lying excitations are m-strings. 

%------------------------------------------------------------------------------
\subsection{Refined Topological Strings in the Nekrasov-Shatashvili Limit}

In section 6, we shall be taking the NS limit ($\epsilon_{2}\mapsto 0$) of the free energy which computes the degeneracies of M-string BPS configurations suspended between the M5-branes. This free energy is obtained from the topological string partition function of a CY3fold. Here we briefly study the effect of this limit on a topological string partition function of a generic toric CY3fold.

Denote by $\mathcal{Z}_{X}(\omega,\epsilon_1,\epsilon_2)$ be the refined topological string partition function of a CY3fold $X$ and let $F_{X}(\omega,\epsilon_1,\epsilon_2)=\mbox{ln}\,\mathcal{Z}_{X}$ be the free energy. For any toric CY3fold  $\mathcal{Z}_{X}$ can be written in terms of degeneracies of BPS states coming from M2-branes wrapping the holomorphic cycles in $X$ \cite{Gopakumar:1998ii,Gopakumar:1998jq,Hollowood:2003cv}.  These degeneracies $N_{\beta}^{j_{L},j_{R}}$ are labeled by the charge $\beta\in H_{2}(X,\mathbb{Z})$ of the curve on which the M2-brane is wrapped and the $SU(2)_{L}\times SU(2)_{R}$ (the little group) spins. The free energy and the partition function in terms of $N_{\beta}^{j_{L},j_{R}}$ are given by
\begin{align}
F_{X}(\omega,\epsilon_1,\epsilon_2)&=\sum_{\beta\in H_{2}(X,\mathbb{Z})}
\sum_{n=1}^{\infty}{1 \over n}
\sum_{j_{L},j_{R}}
\frac{
e^{-n\,\int_{\beta}\omega}
N^{j_{L},j_{R}}_{\beta}(-1)^{2j_{L}+2j_{R}}
\mbox{Tr}_{j_{L}}(\sqrt{q t})^{n\,j_{L,3}}
\mbox{Tr}_{j_{R}}(\sqrt{q / t})^{n\,j_{R,3}}}
{(q^{\frac{n}{2}}-q^{-\frac{n}{2}})(t^{\frac{n}{2}}-t^{-\frac{n}{2}})}
\end{align}
and
\begin{align}
{\cal Z}_{X}(\omega,\epsilon_{1},\epsilon_{2})&=\prod_{\beta\in H_{2}(X,\mathbb{Z})}\prod_{j_{L,R},j_{3,L,R}}\prod_{m_{1},m_{2}=1}^{\infty}\Big(1-e^{-\int_{\beta}\omega}q^{j_{3,L}+j_{3,R}+m_{1}-\frac{1}{2}}t^{j_{3,L}-j_{3,R}+m_{2}-\frac{1}{2}}\Big)^{K^{j_L,j_R}_{\beta}}\,,
\end{align}
respectively, where $K^{j_L,j_R}_{\beta}=(-1)^{2j_{L}+2j_{R}}N^{j_L,j_R}_{\beta}$, while $q=e^{2\pi i\epsilon_1}$ and $t=e^{-2\pi i\epsilon_2}$.

The free energy is a sum over both single-particle and multi-particle states from the spacetime viewpoint and can be written as
\bea
F_{X}(\omega,\epsilon_1,\epsilon_2)&=&\sum_{n=1}^{\infty}\frac{\Omega(n\omega,n\epsilon_1,n\epsilon_2)}{n}\,.
\eea
The function $\Omega(\omega,\epsilon_1,\epsilon_2)$ computes the multiplicities of single particle bound states and can be obtained from the partition function using the plethystic logarithm:
\bea
\Omega(\omega,\epsilon_1,\epsilon_2)&=&\sum_{\beta\in H_{2}(X,\mathbb{Z})}
\sum_{j_{L},j_{R}}
\frac{
e^{-\int_{\beta}\omega}
N^{j_{L},j_{R}}_{\beta}(-1)^{2j_{L}+2j_{R}}
\mbox{Tr}_{j_{L}}(\sqrt{q t})^{j_{L,3}}
\mbox{Tr}_{j_{R}}(\sqrt{q / t})^{j_{R,3}}}
%{(q^{\frac{1}{2}}-q^{-\frac{1}{2}})(t^{\frac{1}{2}}-t^{-\frac{1}{2}})}
{(\sqrt{q}-\sqrt{q}^{-1})(\sqrt{t}-\sqrt{t}^{-1})}\\\nonumber
&=&\mbox{PLog}{\cal Z}_{X}(\omega,\epsilon_1,\epsilon_2)=\sum_{k=1}^{\infty}\frac{\mu(k)}{k}\mbox{ln}{\cal Z}_{X}(k\omega,k\epsilon_1,k\epsilon_2)\,,
\eea
where $\mu(k)$ is the M\"obius function and $\Omega(\omega,\epsilon_1,\epsilon_2)$ computes the multiplicities of single particle bound states. This is the function we will study in the next sections for the case of M-strings and m-strings.

The NS limit of the free energy is given by
\bea\label{NS1}
\lim_{\epsilon_{2}\mapsto 0}\frac{\partial}{\partial t_{a}}\epsilon_{2}F_{X}(\omega,\epsilon_1,\epsilon_2)=-\sum_{\beta\in H_{2}(X,\mathbb{Z})}
\sum_{n=1}^{\infty}{1 \over n}
\sum_{j}
\frac{
e^{-n\,\int_{\beta}\omega}
H^{a}_{\beta}\,n^{j}_{\beta}(-1)^{2j}
\mbox{Tr}_{j}\sqrt{q}^{\ n\,j_{3}}}
%{(q^{\frac{n}{2}}-q^{-\frac{n}{2}})}
{(\sqrt{q}^{n}-\sqrt{q}^{-n})}\eea
where
\bea
\sum_{j}(-1)^{2j}n^{j}_{\beta}\mbox{Tr}_{j}q^{j_{3}}&=&\sum_{j_{L},j_{R}}
N^{j_{L},j_{R}}_{\beta}(-1)^{2j_{L}+2j_{R}}
\mbox{Tr}_{j_{L}}q^{j_{L,3}}
\mbox{Tr}_{j_{R}}q^{j_{R,3}}\\\nonumber
H^{a}_{\beta}&=&\frac{\partial}{\partial t_{a}} \left( \int_{\beta}\,\omega \right) \ \ \in \mathbb{Z}_{\geq  0}\,.
\eea
Recall that $n^{j}_{\beta}$ is the number of particles with spin $j$ with respect to the diagonal $SU(2)\subset SU(2)_{L}\times SU(2)_{R}$ and charge $\beta$. Hence, they count the physical states.

In (\ref{NS1}), we differentiated with respect to the K\"ahler parameter $t_a$ in order to get the usual multi-covering expansion. This allows the exponential of (\ref{NS1}) to be expressed as a product form, which might have interesting modular properties. Consequently, the 
topological string partition function in the NS limit becomes
\bea
{\cal Z}^{a}_{X}(\omega,\epsilon_1)=\mbox{exp}\Big(\lim_{\epsilon_{2}\mapsto 0}\frac{\partial}{\partial t_{a}}\epsilon_{2}F_{X}\Big)=\prod_{\beta\in H_{2}(X,\mathbb{Z})}\prod_{j,j_3}\prod_{m=1}^{\infty}\Big(1-e^{-\int_{\beta}\omega}q^{j_{3}+m-\frac{1}{2}}\Big)^{(-1)^{2j}H^{a}_{\beta}n^{j}_{\beta}}\,.
\eea
Thus, for each K\"ahler parameter $t_a$, we have an NS limit partition function ${\cal Z}^{a}_{X}$.

In section 6, we study the NS limit of the BPS counting function of configurations of M2-branes suspended between M5-branes,  $\widetilde{F}^{(k_1,\cdots k_{N-1})}$ . Since we will not be looking at the total partition functions but only a fixed subsector of it, in the rest of this paper, we will regard the NS limit to be simply $\epsilon_2\mapsto 0$ without any accompanying derivative.

%%%%%%%%%%%%%%%%%%%%%%%%%%%%%%%%%%%%%%%%%%%%%%%%%%%%%%%%%%%%%%%
\section{M-strings and ${\cal N}=1^{*}$ Theory}
The partition function of five-dimensional ${\cal N}=1^{*}$ gauge theory on $\mathbb{S}^{1}\times \mathbb{R}^{4}$ corresponds to an an index that counts the degeneracies of BPS bound-states of W-bosons with instanton particles. In \cite{Kim:2011mv}, this index was computed. After the five-dimensional S-duality, the partition function can also be interpreted as counting the degeneracies of BPS bound-states of m-strings with winding modes. This S-dual description was further studied in \cite{Bak:2014xwa}, order by order in the $Q_\tau = e^{2\pi i\tau}$  expansion, and it was shown that this index can be related to the elliptic genus of Atiyah-Hitchin and Taub-NUT spaces. We will recapitulate this in detail in section 5 and will see that in precise manner the M-strings free energy, in the NS limit, captures the elliptic genus of the Atiyah-Hitchin and Taub-NUT spaces to all orders in $Q_\tau=e^{2\pi i\tau}$.

%-------------------------------------------------
\subsection{Refined Topological String Partition Function}
\begin{figure}[h]
  \centering
  \includegraphics[width=3.5in]{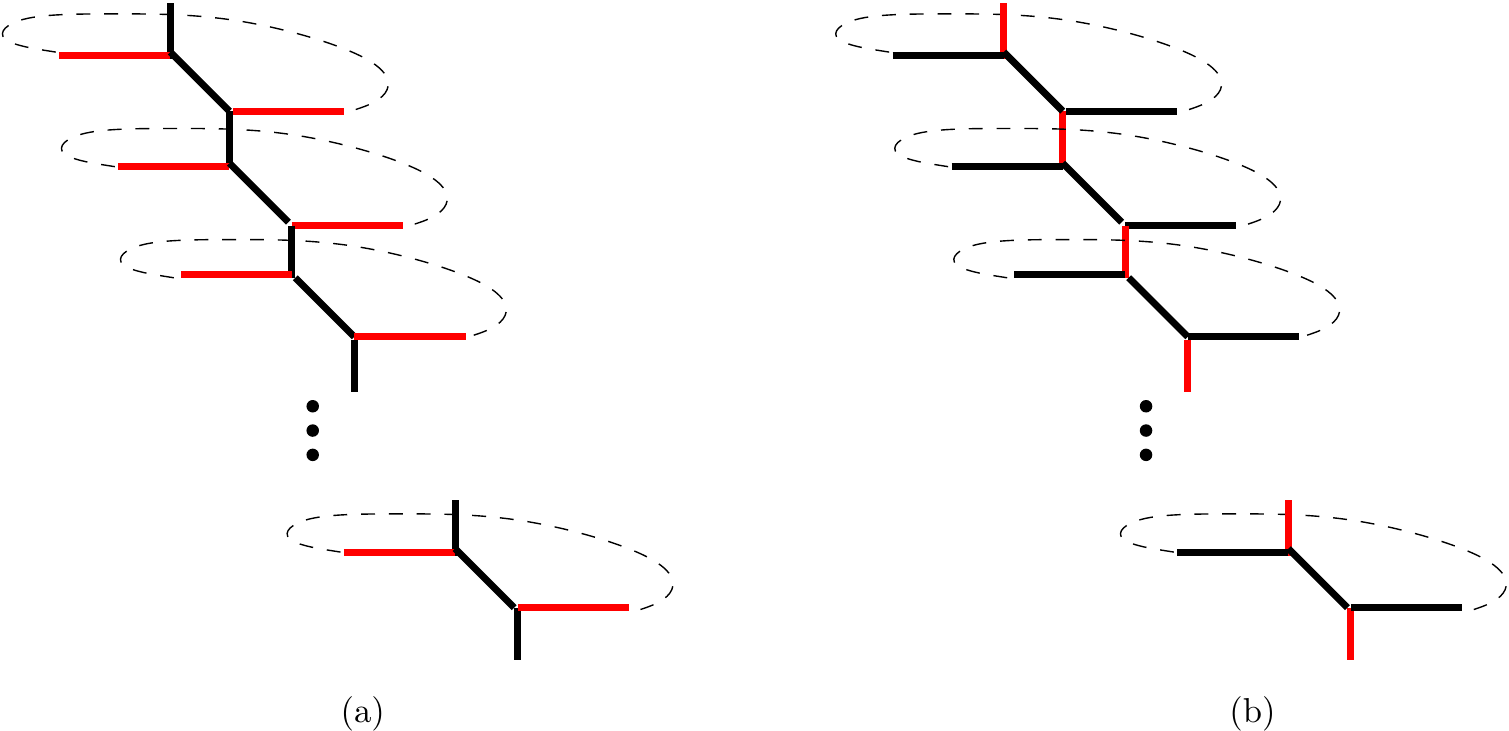}\\
 \caption{Topic diagram of the CY3fold which gives five-dimensional ${\cal N}=1^*$ theory with two choices for preferred direction. Here, $m$ is the K\"ahler parameter of the $\mathbb{P}^1$ which corresponds to the $(1,-1)$ line and $(\tau-m)$ is the K\"ahler parameter of the horizontal (red) line in (a). }\label{web1}
\end{figure}

Certain five-dimensional gauge theories can be geometrically engineered by M-theory compactified on elliptic CY3fold. The latter, called $X_{N}$ in the following, is given by a resolved $A_{N-1}$ singularity fibered over a genus-one curve of complex structure $\tau$. The toric diagram of $X_{N}$ is shown in \figref{web1}.
%\vskip0.5cm
%\vskip0.5cm

The duality between toric CY3folds and $(p,q)$ 5-brane webs in type IIB string theory \cite{Leung:1997tw} therefore maps the CY3fold $X_{N}$ to a $(p,q)$ 5-brane web which in turn is dual, after compactification on $S^{1}$, to the brane setup discussed in the last section.

The full partition function of the gauge theory, which consists of a perturbative and an instanton part, is given by the refined topological string partition function of $X_{N}$ and can be calculated using the topological vertex \cite{Hollowood:2003cv,Haghighat:2013gba}. In the refined topological vertex formalism, a preferred direction in the toric diagram needs to be chosen such that edges oriented in the preferred direction cover all the vertices of the toric diagram. In the associated gauge theory, this preferred direction corresponds to the curve whose K\"ahler parameter is identified with the gauge coupling. Hence, different choices of the preferred direction correspond to dual gauge theories geometrically engineered by the same CY3fold. In \figref{web1}, we indicated the preferred direction with red color (horizontal in \figref{web1}(a) and vertical in \figref{web1}(b)).

A deformation of the $(p,q)$ 5-brane web in $X_N$ corresponds to a deformation of the five-dimensional theory.  In particular, the mass deformation in the five-dimensional theory corresponds to the choice given by \figref{web1}(a) and the corresponding refined topological string partition function is given by
\bea\label{index}
\Z_{N}&:=& \Z^{\text{classical}}_N \Z^{0}_{N}\sum_{k\geq  0}Q_{\tau}^{k}\sum_{\sum_{\alpha=1}^{N}|\nu_{\alpha}|=k}\prod_{\alpha,\beta=1}^{N}
\left[ \prod_{(i,j)\in \nu_{\alpha}}\frac{1-y\,Q_{\alpha\beta}q^{-\nu_{\beta,j}^{t}+i}\,t^{-\nu_{\alpha,i}+j-1}}
{1-Q_{\alpha\beta}q^{-\nu_{\beta,j}^{t}+i}\,t^{-\nu_{\alpha,i}+j-1}}
\ \ \ \  \right. \nn
&&\left. \hskip6cm \times
\prod_{(i,j)\in \nu_{\beta}}\frac{1-y\,Q_{\alpha\beta}q^{\nu_{\alpha,j}^{t}-i+1}t^{\nu_{\beta,i}-j}}
{1-Q_{\alpha\beta}q^{\nu_{\alpha,j}^{t}-i+1}t^{\nu_{\beta,i}-j}} \right] \, .
\eea
We organized the topological string partition function in a way to make contact with the partition function of the five-dimensional ${\cal N}=1^{*}$ gauge theory. Here, $\Z^{\text{classical}}_N$ is the classical part of the gauge theory,  ${\cal Z}^{0}_{N}$ is the perturbative part
\bea
\Z^{0}_{N}&:=&\{Q_{m}\}^{N}\prod_{1\l  \alpha< \beta\l  N}\frac{\{Q_{\alpha\beta}Q_{m}^{-1}\}\,\{Q_{\alpha\beta}Q_{m}\}}
{\{Q_{\alpha\beta} \sqrt{\tfrac{q}{t}}\}
\{Q_{\alpha\beta} \sqrt{\tfrac{t}{q}}\}}\,,\,\,\,\{x\}=\prod_{i,j=1}^{\infty}(1-x\,q^{i-\frac{1}{2}}\,t^{j-\frac{1}{2}})\, ,
\eea
and rest is the instanton part, in which $\tau$ is interpreted as the four-dimensional gauge coupling constant, $Q_{\tau}=e^{2\pi i \tau}$. The $\Omega$-deformation is to regularize the integral over the instanton moduli space obtained from localization of the gauge theory partition function. Note, however, that the deformation modifies the perturbative part ${\cal Z}^0_N$ as well. The factor $Q_m = e^{2 \pi i m}$ is the mass-deformation parameter of the hypermultiplet. The factors $Q_{\alpha\beta}=e^{2 \pi i t_{\alpha \beta}}$ $(\alpha, \beta = 1, \cdots, N)$ are the moduli parameters of the $(N-1)$ vector multiplets in the Coulomb branch. Recall that, in the $(p,q)$-web description in \figref{web1}(a), $t_{\alpha\beta}= (b_{\alpha}-b_{\beta})$ measures the distance between the $\alpha$-th and $\beta$-th horizontal branes. After the U-duality map to M5-brane gauge theory description, the parameters $b_{\alpha}$, with $\sum_{\alpha=1}^{N}b_{\alpha}=0$, become the Coulomb branch parameters breaking $SU(N)\mapsto U(1)^{N-1}$.

The partition function $\Z_{N}$ is a holomorphic function of the moduli parameters but is in general not modular invariant. It can be made modular invariant at the expense of introducing a holomorphic anomaly \cite{Haghighat:2013gba}, meaning that the partition function cannot be refined while maintaining both the modular symmetry and the holomorphy. In constructing various counting functions, we will be primarily guided by their modular properties and will discuss them in more detail in the following sections.

The dual description of the same partition function can be obtained by choosing the preferred direction (vertical) as shown in \figref{web1}(b). In the topological string description, this corresponds to the exchange of the fiber and the base of the CY3fold $X_N$ through flop transitions. In this case, the refined topological string partition function can be written as:
\bea
\Z_{N}&=&(\Z_{1}(\tau,m,\epsilon_{1},\epsilon_{2}))^{N}\cdot \widetilde{\Z}_{N}(\tau,m,t_{f_{a}},\epsilon_1,\epsilon_2),
\eea
where
\bea\label{fe2}
\Z_{1}(\tau,m,\epsilon_{1},\epsilon_{2})=\frac{1}{\eta(\tau)}\prod_{i,j,k=1}^{\infty}\frac{(1-Q_{\tau}^{k}Q_{m}^{-1}q^{i-\frac{1}{2}}t^{j-\frac{1}{2}})(1-Q_{\tau}^{k-1}Q_{m}q^{i-\frac{1}{2}}t^{j-\frac{1}{2}})}{(1-Q_{\tau}^{k}q^{i-1}t^{j})(1-Q_{\tau}^{k}q^{i}t^{j-1})}\,,
\eea
and
\bea\label{fugacityexpansion}
\widetilde{\Z}_{N}(\tau,m,t_{f_{a}},\epsilon_1,\epsilon_2)&=&\sum_{k_1,\cdots,k_{N-1}\geq  0}Q_{f_{1}}^{k_{1}}\cdots Q_{f_{N-1}}^{k_{N-1}} \ Z_{k_1\cdots k_{N-1}}(\tau,m,\epsilon_+,\epsilon_-)\,,
\eea
Here, we introduced (anti)self-dual combinations of the $\Omega$-deformation parameters:
\begin{align}
&\epsilon_+=\frac{\epsilon_1+\epsilon_2}{2}\,,&&\text{and}&&\epsilon_-=\frac{\epsilon_1-\epsilon_2}{2}\,.
\end{align}

We can express the coefficients in $\widetilde{\Z}_N$ in terms of (products of) Jacobi theta functions. The expansion given in (\ref{fugacityexpansion}) corresponds to an instanton expansion in a dual theory which is engineered by the same CY3fold $X_N$ but in which the base curves are chosen to be the $(-2)$-curves of the resolved $A_{N-1}$ fiber with an elliptic fibration over them. In this dual description, $Q_{f_{a}}=e^{2\pi i t_{f_{a}}}$ $(a=1, \cdots, N-1)$ where $t_{f_{a}}=b_{a}-b_{a+1}$ (with $a=1,\ldots N-1$) are the gauge couplings of the quiver $U(1)^{N-1}$ gauge theories. The partition function with this choice of preferred direction is given by \cite{Haghighat:2013gba}
\begin{align}\label{mm}
\boxed{
\widetilde{\Z}_{N}(\tau,m,t_{f_{a}},\epsilon_1,\epsilon_2)=\sum_{\nu_{1},\mathellipsis, \nu_{N-1}}\left(\prod_{a=1}^{N-1}(-Q_{f_{a}})^{|\nu_{a}|}\right)
\prod_{a=1}^{N-1}\prod_{(i,j)\in \nu_{a}}\frac{\theta_{1}(\tau;z^{a}_{ij})\,\theta_{1}(\tau;v^{a}_{ij})}{\theta_1(\tau;w^{a}_{ij})\theta_1(\tau;u^{a}_{ij})}
}
\end{align}
where
\begin{equation}\nn
\begin{array}{ll}
e^{2\pi i \,z^{a}_{ij}}=Q_{m}^{-1}\,q^{\nu_{a,i}-j+\frac{1}{2}}\,t^{\nu^{t}_{a+1,j}-i+\frac{1}{2}},\qquad &
e^{2\pi i\,v^{a}_{ij}}=Q_{m}^{-1}\,t^{-\nu^{t}_{a-1,j}+i-\frac{1}{2}}\,q^{-\nu_{a,i}+j-\frac{1}{2}},\\
e^{2\pi i \,w^{a}_{ij}}=q^{\nu_{a,i}-j+1}\,t^{\nu_{a,j}^{t}-i},& e^{2\pi i \,u^{a}_{ij}}=q^{\nu_{a,i}-j}\,t^{\nu_{a,j}^{t}-i+1}.
\end{array}
\end{equation}
In this expression, $\theta_{1}(\tau;z)$ is one of the Jacobi theta functions defined in the appendix (\ref{JacobiTheta})

Again the partition function $\widetilde{\Z}_{N}$ is holomorphic in the moduli but not modular invariant in $\tau$, because the instanton expansion coefficients, the ratios of the Jacobi theta function $\theta_{1}(\tau;z)$,  involve the second Eisenstein series $E_2(\tau)$. It can be made modular invariant if $E_{2}(\tau)$ is replaced by the non-holomorphic second Eisenstein series $\widehat{E}_{2}(\tau,\overline{\tau})$ (see (\ref{E2hat}) for the definition).

The nonperturbative partition function (\ref{mm}) was also interpreted as the partition function of a configuration of M2-branes suspended between $N$ M5-branes \cite{Haghighat:2013gba}. These M2-branes would also wrap around $\mathbb{S}^1$, and the winding numbers are dual to the M-waves studied in section 2. The term $Z_{k_{1}\cdots k_{N-1}}$ is the contribution of a configuration in which $k_i$ M-strings are stretched between the $i$-th and the $(i+1)$-th M5-branes. In \cite{Haghighat:2013tka}, it was argued that $Z_{k_{1}\cdots k_{N-1}}$ is the elliptic genus of a two-dimensional quiver gauge theory that captures the M-string worldsheet dynamics.

%-------------------------------------------------------
\subsection{Modular Properties of $Z_{k_{1}\cdots k_{N-1}}$}\label{Sec:ModularZk}

The M-string partition function given by  (\ref{mm}) sets the starting point of our investigation of modular properties of the free energy in the next section. The free energy for a particular configuration of M-strings is a combination of different $Z_{k_{1}\cdots k_{N-1}}$ and hence its modular transformation properties will depend on how $Z_{k_{1}\cdots k_{N-1}}$ transform. So let us first consider how
\bea \label{Zk}
Z_{k_{1}\cdots k_{N-1}}(\tau,m,\epsilon_{+},\epsilon_{-})=(-1)^{k_{1}+\cdots k_{N-1}}\sum_{\nu_{a},|\nu_{a}|=k_{a}}\prod_{a=1}^{N-1}\prod_{(i,j)\in \nu_{a}}\frac{\theta_{1}(\tau;z^{a}_{ij})\,\theta_{1}(\tau;v^{a}_{ij})}{\theta_1(\tau;w^{a}_{ij})\theta_1(\tau;u^{a}_{ij})}
\eea
transforms under an $SL(2,\mathbb{Z})$ action given by
\begin{align}
&(\tau,m, \epsilon_{1},\epsilon_{2})\mapsto \left(\frac{a\tau+b}{c\tau+d},\frac{m}{c\tau+d},\frac{\epsilon_{1}}{c\tau+d},\frac{\epsilon_{2}}{c\tau+d}\right)\,,&&\begin{bmatrix}
    a       & b \\
    c & d
\end{bmatrix} \in SL(2,\mathbb{Z})\,.
\end{align}
Since $Z_{k_{1}\cdots k_{N-1}}$ is a ratio of the products of theta functions, its transformation properties follow from those of $\theta_{1}(\tau,z)$ (with $\begin{bmatrix}
    a       & b \\
    c & d
\end{bmatrix} \in SL(2,\mathbb{Z})$, $m,n\in \mathbb{Z}$):
\bea \label{thetaprop}
\theta_{1}\left(\frac{a\tau+b}{c\tau+d},\frac{z}{c\tau+d}\right)&=&\psi(a,b,c,d)^3\,(c\tau+d)^{\frac{1}{2}}\,e^{\frac{i\pi c z^2}{c\tau+d}}\,\theta_{1}(\tau,z)\\\nonumber
\theta_{1}(\tau,z+n\,\tau+m)&=&(-1)^{m+n}e^{-i\pi n^2\,\tau-2\pi i n\,z}\theta_{1}(\tau,z)\,.
\eea
The multiplier $\psi(a,b,c,d)$ in this  equation is a 24-th root of unity  whose explicit form will not be needed since it cancels in the homogeneous ratio among the Jacobi elliptic function $\theta_{1}(\tau,z)$ of $Z_{k_{1}\cdots k_{N-1}}$. From (\ref{Zk}) and (\ref{thetaprop}), it then follows that for $\ell,r\in \mathbb{Z}$
\bea \label{Zktrans}
Z_{k_{1}\cdots k_{N-1}}(\tau+1,m,\epsilon_{+},\epsilon_{-})&=&Z_{k_{1}\cdots k_{N-1}}(\tau,m,\epsilon_{+},\epsilon_{-})\,,\\\nonumber
Z_{k_{1}\cdots k_{N-1}}(-\tfrac{1}{\tau},\tfrac{m}{\tau},\tfrac{\epsilon_{+}}{\tau},\tfrac{\epsilon_{-}}{\tau})&=&e^{\frac{2\pi i}{\tau}f_{\vec{k}}(m,\epsilon_{+},\epsilon_{-})}Z_{k_{1}\cdots k_{N-1}}(\tau,m,\epsilon_{+},\epsilon_{-})\,,\\\nonumber
Z_{k_{1}\cdots k_{N-1}}(\tau,m+\ell\tau+r,\epsilon_{+},\epsilon_{-})&=&e^{-2\pi i K\,\ell^2\tau+4\pi i m K}Z_{k_{1}\cdots k_{N-1}}(\tau,m,\epsilon_{+},\epsilon_{-})\,,
\eea
where
\begin{align}\label{fk}
f_{\vec{k}}(m,\epsilon_{+},\epsilon_{-})&=K\,m^2+Q_{-}\epsilon_{+}^2+Q_{+}\epsilon_{-}^2\,,
\end{align}
in terms of the shorthand notations:
\begin{align}
&K=\sum_{a=1}^{N-1}k_{a}\,,&&Q_{\pm}:= \pm\Big(\sum_{a=1}^{N-1}k_{a}(k_{a}-\tfrac{1}{2})+\sum_{a=1}^{N-2}k_{a}k_{a+1}\Big)- \frac{K}{2}\,.
\label{DefQpm}
\end{align}

With respect to the variables $(\tau,m)$,  $Z_{k_{1}\cdots k_{N-1}}$ is a Jacobi form with index $K$. With respect to the variables $\epsilon_{\pm}$,  it also has properties very similar to a meromorphic Jacobi form with index matrix in the basis $m,\epsilon_{+},\epsilon_{-}$ given by:
\bea
\begin{bmatrix}
    K       & 0 &0 \\
    0 & Q_{-} & 0\\
    0&0& Q_{+}
\end{bmatrix}\,.
\eea
However, $Z_{k_{1}\cdots k_{N-1}}$ fails to be a multi-variable Jacobi form, since the shift property (third property in (\ref{Zktrans})) that is present for $m$ is not present for $\epsilon_{\pm}$:
\bea \nonumber
Z_{k_{1}\cdots k_{N-1}}(\tau,m,\epsilon_{+}+a\tau+b,\epsilon_{-})&=&(-1)^{k_{1}+\cdots k_{N-1}}\sum_{\nu_{a},|\nu_{a}|=k_{a}}(-1)^{(a+b)\kappa(\vec{\nu})}e^{-2\pi i Q_{-}\,a^2\tau+2\pi i \epsilon_{+} \kappa(\vec{\nu})} \\\nonumber
&&\times
\prod_{a=1}^{N-1}\prod_{(i,j)\in \nu_{a}}\frac{\theta_{1}(\tau;z^{a}_{ij})\,\theta_{1}(\tau;v^{a}_{ij})}{\theta_1(\tau;w^{a}_{ij})\theta_1(\tau;u^{a}_{ij})}\\\nonumber
Z_{k_{1}\cdots k_{N-1}}(\tau,m,\epsilon_{+},\epsilon_{-}+a\tau+b)&=&(-1)^{k_{1}+\cdots k_{N-1}}\sum_{\nu_{a},|\nu_{a}|=k_{a}}(-1)^{(a+b)h(\vec{\nu})}e^{-2\pi i Q_{-}\,a^2\tau+2\pi i \epsilon_{+} h(\vec{\nu})} \\\nonumber
&&\times
\prod_{a=1}^{N-1}\prod_{(i,j)\in \nu_{a}}\frac{\theta_{1}(\tau;z^{a}_{ij})\,\theta_{1}(\tau;v^{a}_{ij})}{\theta_1(\tau;w^{a}_{ij})\theta_1(\tau;u^{a}_{ij})},
\eea
where the short-hand notations are
\begin{align}
&\kappa(\vec{\nu})=\sum_{a=1}^{N-1}(||\nu_{a}||^2-||\nu^{t}_{a}||^2)\,,&&h(\vec{\nu})=\sum_{a=1}^{N-1}(||\nu_{a}||^2+||\nu^{t}_{a}||^2)\,.
\end{align}

If we combine various $Z_{k_{1}\cdots k_{N-1}}$ for different values of $K = (k_1,\cdots,k_{N-1})$, then the index matrices do not simply add up since the $Q_{\pm}$ are quadratic in $k_{i}$ (see (\ref{DefQpm})). However, this situation changes if we take the NS limit $\epsilon_{2}\mapsto 0$, since in this case the index with respect to the remaining parameter $\epsilon_{1}$ is $Q_{-}+Q_{+}=K$ which is linear in $k_{i}$. So, in the NS limit $\epsilon_{2}\mapsto 0$, the index with respect to $(m,\epsilon_{1})$ depends only on the total number of M2-branes $K$ and this remains true for the product of $Z_{k_{1}\cdots k_{N-1}}$ for different $k_{i}$'s.

%%%%%%%%%%%%%%%%%%%%%%%%%%%%%%%%%%%%%%%%%%%%%%%%%%%%%%%%%%%%%
\section{BPS Degeneracies of M-Strings}\label{Sect:DegMonopole}
We shall first analyze in detail the BPS degeneracies of M-strings.

\subsection{M-String Free Energy}
The function $\Omega_{X}(\omega,\epsilon_{1},\epsilon_{2})$, discussed in Section 2, counts the degeneracies of single-particle BPS states in the five-dimensional \nonestar gauge theory, which descends from M-theory compactified on a CY3fold $X$.
For the particular CY3fold $X_{N}$ discussed in section 3, we have
\begin{align}
\Omega_{N}(\tau,m,t_{f_{a}},\epsilon_1,\epsilon_2)=\mbox{PLog}{\cal Z}_{N}(\tau,m,t_{f_a},\epsilon_1,\epsilon_2)=N\underbrace{\mbox{PLog}{\cal Z}_{1}}_{\Omega_{1}}+\underbrace{\mbox{PLog}\widetilde{\cal Z}_{N}}_{\widetilde{\Omega}_{N}}\,.
\end{align}
Here, the second term, $\widetilde{\Omega}_{N}(\omega,\epsilon_{1},\epsilon_{2})$, defines the free energy for counting BPS states of the M-strings and can be written as
\bea\label{bound}
\boxed{
\widetilde{\Omega}_{N}(t_{f_{a}},\tau, m,\epsilon_1,\epsilon_2)=
\sum_{\{k_i\}
=1}^{\infty}Q_{f_{1}}^{k_{1}}\cdots Q_{f_{N-1}}^{k_{N-1}}\, \widetilde{F}^{(k_1,k_2,\cdots,k_{N-1})} (\tau, m,\epsilon_1,\epsilon_2).
}
\eea

In this section, we aim to study the modular and other properties of the function $\widetilde{F}^{(k_1,\cdots,k_{N-1})}$ which counts the degeneracies of the bound-states of multiple M-strings in configurations where $k_i$ $(i=1, \cdots, N-1)$ M2-branes are stretched between the $i$-th and $(i+1)$-th M5-branes
\bea\nonumber
\widetilde{F}^{(k_1,\cdots,k_{N-1})}(\tau, m,\epsilon_1,\epsilon_2)&=&\oint \frac{dQ_{f_{i}}}{2\pi i Q_{f_{i}}^{k_{i}+1}}\cdots \frac{dQ_{f_{N-1}}}{2\pi i Q_{f_{N-1}}^{k_{N-1}+1}}\widetilde{\Omega}_{N}(t_{f_{a}},\tau, m,\epsilon_1,\epsilon_2)\\\nn
&=&\left(\sqrt{q}-\sqrt{q}^{\ -1}\right)^{-1}\left(\sqrt{t}-\sqrt{t}^{\ -1}\right)^{-1} \sum_{n,\ell}Q_{\tau}^{n}Q_{m}^{\ell}C_{n,\ell}(\epsilon_{1},\epsilon_{2})\,.
\eea
As can be seen from \figref{web1}, the fugacities are related by $Q_{\tau}=Q_{m}Q_{1}$. Since the topological string free energy is an expansion in non-negative powers of $Q_{f_{i}},Q_{m}$ and $Q_{1}$, the coefficient $C_{n,\ell}(\epsilon_{1},\epsilon_{2})$ must vanish for $n<|\ell|$.

In the next section, we will consider the NS limit $\epsilon_{2}\mapsto 0$ and then further take the limit $\epsilon_{1}\mapsto 0$. In this limit,  $\widetilde{F}_{N}^{(k_1,\cdots,k_{N-1})}(\tau, m,\epsilon_1,\epsilon_2)$ behaves as
\bea
\widetilde{F}^{(k_1,\cdots,k_{N-1})}(\tau, m,\epsilon_1,\epsilon_2)&=&\frac{1}{\epsilon_{1}\epsilon_{2}}\Big(\sum_{n,\ell}Q_{\tau}^{n}Q_{m}^{\ell}C_{n,\ell}(0,0)\Big)+\cdots\
\eea
where
\bea
C_{n,\ell}(0,0)&=&\sum_{j_{L},j_{R}}N_{n,\ell}^{j_{L},j_{R}}\,(-1)^{2j_{L}+2j_{R}}(2j_{L}+1)(2j_{R}+1).
\eea
We can express $\widetilde{F}^{(k_{1}\cdots k_{N-1})}$ in terms of $Z_{k_{1}\cdots k_{N-1}}$ (given in  .(\ref{Zk})) as follows:
\begin{align}
\widetilde{F}^{(k_{1},\cdots,k_{N-1})}&=&\sum_{d|s}\frac{\mu(d)}{d}G_{\frac{k_{1}}{d}\cdots \frac{k_{N-1}}{d}}(d\tau,d\,m,d\epsilon_{1},d\,\epsilon_{2})\,,\,\,\,\,\, s=\mbox{gcd}(k_{1},k_{2},\cdots,k_{N-1}),\label{Fs}
\end{align}
where we introduced
\begin{align}
G_{r_{1}r_{2}\cdots r_{N-1}}=(-1)^{\sum_{a}r_{a}}\sum_{\ell=1}^{\sum_{a}r_{a}}
{1 \over \ell!} \frac{(-1)^{\ell}}{{\ell^{}}^{{\rm gcd}(r_{a})-1}}\sum_{{k^{i}_{1},\cdots,k^{i}_{N-1}\geq  0}\atop{\sum_{i=1}^{\ell}k^{i}_{a}=r_{a}}}\prod_{i=1}^{
\ell}Z_{k^{i}_{1}\,k^{i}_{2}\,\cdots\,k^{i}_{N-1}}
\end{align}

%---------------------------------------------------------
\subsection{Modular Transformations and Theta Decomposition}

In section~\ref{Sec:ModularZk}, we found that $Z_{k_{1}\cdots k_{N-1}}$ is a Jacobi form of weight zero and index $K$ with respect to the variables $(\tau,m)$ and transforms as:
\bea
Z_{k_{1}k_{2}\cdots k_{N-1}}(-\tfrac{1}{\tau},\tfrac{m}{\tau},\tfrac{\epsilon_{+}}{\tau},\tfrac{\epsilon_{-}}{\tau})=e^{\frac{2\pi i}{\tau}f_{\vec{k}}(m,\epsilon_{+},\epsilon_{-})}Z_{k_{1}k_{2}\cdots k_{N-1}}(\tau,m,\epsilon_{+},\epsilon_{-})\,.
\eea
As the function $f_{\vec{k}}(m,\epsilon_{+},\epsilon_{-})$ is quadratic in $k_{a}$, linear combinations of products of $Z_{k_{1}k_{2}\cdots k_{N-1}}$ with different charges $k_{a}$ will not transform with just an overall phase-factor. This implies that $\widetilde{F}^{(k_1,k_2,\cdots,k_{N-1})}$ given in (\ref{Fs}) will not in general transform nicely under the $S$-transformation of $SL(2, \mathbb{Z})$.  However, if we consider the expansion in $\epsilon_{1}$ and $\epsilon_{2}$ (the genus expansion), then coefficients of $\epsilon_{1}^{n_{1}}\epsilon_2^{n_{2}}$ will transform as Jacobi forms of weights $(n_{1}+n_{2})$ and index $K$ under $\Gamma_{0}(s)\subset SL(2,\mathbb{Z})$, where $s=\mbox{gcd}(k_{1},k_{2},\cdots,k_{N-1})$. Here, the subgroup $\Gamma_0(s)$ is defined as
\begin{align}
\Gamma_0(s)=\left\{\left(\begin{array}{cc}a & b \\ c & d\end{array}\right)\in SL(2,\mathbb{Z})\bigg|c \equiv 0\,\text{mod}\,s\right\}\,.
\end{align}

Index $K$ implies that we can decompose both $Z_{k_{1}\cdots k_{N}}$ and $\widetilde{F}^{(k_{1}\cdots k_{N-1})}$ in terms of index $K$ theta functions defined in section (\ref{Sec:IndexTheta}):
\bea
Z_{k_{1}\cdots k_{N-1}}(\tau,m,\epsilon_{+},\epsilon_{-})&=&\sum_{\ell=0}^{2K-1}R^{(k_{1}\cdots k_{N-1})}_{\ell}(\tau,\epsilon_{1},\epsilon_{2})\vartheta_{K,\ell}(\tau,m)\,,\\\nonumber
\widetilde{F}^{(k_{1}\cdots k_{N-1})}(\tau,m,\epsilon_{1},\epsilon_{2})&=&\sum_{\ell=0}^{2K-1}H^{(k_{1}\cdots k_{N-1})}_{\ell}(\tau,\epsilon_{1},\epsilon_{2})\vartheta_{K,\ell}(\tau,m)\,.
\eea
Since $Z_{k_{1}\cdots k_{N-1}}$ and $\widetilde{F}^{(k_{1}\cdots k_{N-1})}$ are both invariant under $m\mapsto -m$, it follows that
\bea
R^{(k_{1}\cdots k_{N-1})}_{\ell}(\tau,\epsilon_{1},\epsilon_{2})&=&R^{(k_{1}\cdots k_{N-1})}_{2K-\ell}(\tau,\epsilon_{1},\epsilon_{2})\,,\\\nonumber
H^{(k_{1}\cdots k_{N-1})}_{\ell}(\tau,\epsilon_{1},\epsilon_{2})&=&H^{(k_{1}\cdots k_{N-1})}_{2K-\ell}(\tau,\epsilon_{1},\epsilon_{2})\,.
\eea

Another basis of index $K$ theta functions is given by $\vartheta_{1,0}(\tau,m)^{a}\vartheta_{1,1}(\tau,m)^{K-a}$. In this basis,
\bea
\widetilde{F}^{(k_1,\cdots,k_{N-1})}(\tau,m,\epsilon_1,\epsilon_2)=\sum_{a=0}^{K}L^{(k_1,\cdots,k_{N-1})}_{a}(\tau,\epsilon_1,\epsilon_2)\,\vartheta_{1,0}(\tau,m)^{a}\,\vartheta_{1,1}(\tau,m)^{K-a} \ ,
\eea
where the $(K+1)$ coefficient functions $L^{(k_1,\cdots,k_{N-1})}_{a}$ are independent of each other.

In the following subsections, we will decode the structure of $\widetilde{F}^{(k_1,\cdots,k_{N-1})}(\tau, m,\epsilon_1,\epsilon_2)$ for several configurations with  lower $\{k_i\}$ charges. In all these cases, $\widetilde{F}$ is not modular invariant but holomorphic. We will also present the physical spin contents of a few low-lying states for each charge configuration discussed.

%------------------------------------------------------
\subsection{Single M2-Brane}
We begin with configurations in which a single M2-brane is stretched between every pair of consecutive M5-branes. Depending on the number of M5-branes, we have various possibilities.

%----------------------------------------------------------
\subsubsection{Configuration $(k_i)=(1)$}
The simplest configuration arise when a single M2-brane is stretched between two M5-branes. For this configuration,
\bea\label{f1}
\F^{(1)}(\tau,m,\epsilon_{1},\epsilon_{2}):=-\frac{\theta_{1}(\tau,m+\epsilon_{+})\theta_{1}(\tau,m-\epsilon_{+})}{\theta_{1}(\tau,\epsilon_{1})\theta_{1}(\tau,\epsilon_{2})}\,.
\eea
As we discussed before, $\widetilde{F}^{(1)}$ has index one with respect to $m$ and therefore it can be decomposed in the following form:
\begin{align}
\F^{(1)}(\tau,m,\epsilon_{1},\epsilon_{2})=H^{(1)}_0(\tau,\epsilon_1,\epsilon_2)\,\vartheta_{1,0}(\tau,m)+H^{(1)}_1(\tau,\epsilon_1,\epsilon_2)\,\vartheta_{1,1}(\tau,m)\,.\label{ThetaExpansionU21}
\end{align}
Here, $\vartheta_{1,0}$ and $\vartheta_{1,1}$ are index $1$ theta functions defined in appendix~\ref{Sec:IndexTheta}.
The coefficient functions $H^{(1)}_{0}(\tau,\epsilon_{1},\epsilon_{2})$ and $H^{(1)}_{1}(\tau,\epsilon_{1},\epsilon_{2})$ are residues of $\widetilde{F}^{(1)}$ and its first derivative \footnote{It was also noted \cite{Kim:2011mv} that the coefficients $H^{(1)}_\ell$ can be computed by a contour integration.}:
\begin{align}
&H_{0}^{(1)}=\oint \tfrac{dQ_{m}}{2\pi i} \  Q_{m}^{-1}\ \F^{(1)}\,,&&
H_{1}^{(1)}=Q_\tau^{-\frac{1}{4}}\oint \tfrac{dQ_{m}}{2\pi i}\ \F^{(1)}.
\end{align}
Using (\ref{f1}), we get
\begin{align}
H_{0}^{(1)}(\tau,\epsilon_1,\epsilon_2)=-\frac{\theta_{2}(2\tau,2\epsilon_{+})}{\theta_{1}(\tau,\epsilon_{1})\theta_{1}(\tau,\epsilon_{2})}, \qquad
H_{1}^{(1)}(\tau,\epsilon_1,\epsilon_2)=\frac{\theta_{3}(2\tau,2\epsilon_{+})}{\theta_{1}(\tau,\epsilon_{1})\theta_{1}(\tau,\epsilon_{2})}\,.\label{DefH01}
\end{align}

The pair  $(H^{(1)}_{0},H^{(1)}_{1})$ forms a vector-valued modular form of weight $-\frac{1}{2}$ which transforms as:
\bea
H^{(1)}_{0}(-\tfrac{1}{\tau},\tfrac{\epsilon_{1}}{\tau},\tfrac{\epsilon_{2}}{\tau})&=&\sqrt{\tfrac{i}{2\tau}}\,e^{-\frac{2\pi}{\tau} i \epsilon_{-}^2}(H^{(1)}_{0}(\tau,\epsilon_1,\epsilon_2)+H^{(1)}_{1}(\tau,\epsilon_{1},\epsilon_2))\\\nonumber
H^{(1)}_{1}(-\tfrac{1}{\tau},\tfrac{\epsilon_{1}}{\tau},\tfrac{\epsilon_{2}}{\tau})&=&\sqrt{\tfrac{i}{2\tau}}\,e^{-\frac{2\pi}{\tau} i \epsilon_{-}^2}(H^{(1)}_{0}(\tau,\epsilon_1,\epsilon_2)-H^{(1)}_{1}(\tau,\epsilon_{1},\epsilon_2)) \ .
\eea
These functions are the fundamental building blocks of distinct M-string configurations: We will soon find that degeneracies of M2-brane configurations of type $(k_{i})=(1,1,\cdots,1)$ are completely determined by $H^{(1)}_{0}$ and $H^{(1)}_{1}$.

We also extracted the spin contents {\it i.e.}, $\sum_{(j_{L},j_{R})}N^{(j_{L},j_{R})}_{\beta}(j_L,j_R)$ for some $\beta$.
\begin{tcolorbox}
\underline{Spin Contents from $H^{(1)}_{0}$:}

The function $H^{(1)}_{0}(\tau,\epsilon_1,\epsilon_2)$ contains the $SU(2)_{L}\times SU(2)_{R}$ spin contents of the states corresponding to $Q_{f}\,Q_{\tau}^{n}$. For some small values of $n$ we list $\sum_{(j_{L},j_{R})}N^{(j_{L},j_{R})}_{\beta}(j_L,j_R)$ below:
\bea\nonumber
n&=&0: (0,\tfrac{1}{2})\\\nonumber
n&=&1: (\tfrac{1}{2},1)+(\tfrac{1}{2},0)+2(0,\tfrac{1}{2})\\\nonumber
n&=&2: (1,\tfrac{3}{2})+(1,\tfrac{1}{2})+3(\tfrac{1}{2},1)+3(\tfrac{1}{2},0)+(0,\tfrac{3}{2})+5(0,\tfrac{1}{2})\\\nonumber
n&=&3: (\tfrac{3}{2},2)+(\tfrac{3}{2},1)+3(1,\tfrac{3}{2})
+4(1,\tfrac{1}{2})+9(\tfrac{1}{2},1)
+(\tfrac{1}{2},2)+
8(\tfrac{1}{2},0)+3(0,\tfrac{3}{2})+12(0,\tfrac{1}{2})
\eea
\end{tcolorbox}
\begin{tcolorbox}
\underline{Spin Contents from $H^{(1)}_{1}$:}

The function $Q_{\tau}^{\frac{1}{4}}\,H^{(1)}_{1}(\tau,\epsilon_1,\epsilon_2)$ contains the $SU(2)_{L}\times SU(2)_{R}$ spin contents of the states corresponding to $Q_{f}Q_{m}Q_{\tau}^{n}$:
\bea\nonumber
n&=&0: (0,0)\\\nonumber
n&=&1: (\tfrac{1}{2},\tfrac{1}{2})+(0,1)+(0,0)\\\nonumber
n&=&2: (1,1)+(\tfrac{1}{2},\tfrac{3}{2})+3(\tfrac{1}{2},\tfrac{1}{2})+2(0,1)+4(0,0)\\\nonumber
n&=&3: (\tfrac{3}{2},\tfrac{3}{2})+(1,2)+3(1,1)+2(1,0)+3(\tfrac{1}{2},\tfrac{3}{2})+9(\tfrac{1}{2},\tfrac{1}{2})+7(0,1)+7(0,0)
\eea

\end{tcolorbox}
%

%--------------------------
\subsubsection{Configuration $(k_i)=(1,1)$}
The next simpler configuration arises when there are 3 parallel M5 branes $(\text{M5}_1, \text{M5}_2, \text{M5}_3)$ and two M2 branes suspended between them: the first one stretches between $\text{M5}_1$ and $\text{M5}_2$ while the second one stretches between $\text{M5}_2$ and $\text{M5}_3$. The corresponding free energy is given by
\bea\label{f11}
\widetilde{F}^{(1,1)}&=&\frac{\theta_{1}(\tau,m+\epsilon_{+})
\theta_{1}(\tau,m-\epsilon_{+})\theta_{1}(\tau,m+\epsilon_{-})\theta_{1}(\tau,m-\epsilon_{-})}{\theta_{1}(\tau,\epsilon_1)^2 \theta_1(\tau,\epsilon_2)^2}\\\nonumber
&&-\frac{\theta_{1}(\tau,m+\epsilon_{+})^2\theta_{1}(\tau,m-\epsilon_{+})^2}{\theta_{1}(\tau,\epsilon_1)^2 \theta_1(\tau,\epsilon_2)^2} \ .
\eea

As $\widetilde{F}^{(1,1)}$ is of index 2, it must be decomposable as
\begin{align}
\F^{(1,1)}=\sum_{\ell=0}^3H_\ell^{(1,1)}(\tau,\epsilon_1,\epsilon_2)\,\vartheta_{2,\ell}(\tau,Q_m)\,,\label{ThetaExpansionU311}
\end{align}
The coefficients $(H_0^{(1,1)},H_1^{(1,1)},H_2^{(1,1)},H_3^{(1,1)})$ form a vector-valued modular form. They are given by
\bea
H_{0}^{(1,1)}&=&\oint \frac{dQ_{m}}{2\pi i} \ Q_{m}^{-1}\ \F^{(1,1)}\nonumber \\
H_{1}^{(1,1)}&=&Q_\tau^{-\frac{1}{8}}\oint \frac{dQ_{m}}{2\pi i}\ \F^{(1,1)} =H_3^{(1,1)} \nonumber \\
H_{2}^{(1,1)} &=& Q_\tau^{-\frac{1}{2}}\oint \frac{dQ_{m}}{2\pi i} \,Q_m \ \F^{(1,1)}\, .
\eea
These coefficients $H^{(1,1)}_{\ell}$ contain information for degeneracies of the states corresponding to $Q_{f_1}Q_{f_2}Q_{m}^{\ell}Q_{\tau}^n$ for $n\geq  0$. As asserted above, they are completely determined by $H^{(1)}_{0}$ and $H^{(1)}_{1}$ in (\ref{DefH01}). To see this, note from (\ref{f11})\footnote{For limiting values of $m, \epsilon_1, \epsilon_2$, this relation was also noted in \cite{Kim:2011mv} and more explicitly in \cite{Bak:2014xwa}.}
\bea\label{recursion0}
\widetilde{F}^{(1,1)}(\tau,m\epsilon_1,\epsilon_2)=\widetilde{F}^{(1)}(\tau,m,\epsilon_1,\epsilon_2)\,W(\tau,m,\epsilon_1,\epsilon_2)\, .
\eea
Here,
\bea\nonumber
W(\tau,m,\epsilon_1,\epsilon_2)&=&
\frac{\theta_{1}(\tau,m+\epsilon_{+})
\theta_{1}(\tau,m-\epsilon_{+})}{\theta_{1}(\tau,\epsilon_1) \theta_1(\tau,\epsilon_2)}-\frac{\theta_{1}(\tau,m+\epsilon_{-})\theta_{1}(\tau,m-\epsilon_{-})}{\theta_{1}(\tau,\epsilon_1) \theta_1(\tau,\epsilon_2)}\\
&=&-\widetilde{F}^{(1)}(\tau,m,\epsilon_1,\epsilon_2)-\widetilde{F}^{(1)}(\tau,m,\epsilon_1,-\epsilon_2)\\\nonumber
&=&W_{0}(\tau,\epsilon_1,\epsilon_2)\vartheta_{1,0}(\tau,m)+W_{1}(\tau,\epsilon_1,\epsilon_2)\vartheta_{1,1}(\tau,m)
\eea
where we introduced
%\bea
\begin{align}
W_{0}(\tau,\epsilon_1,\epsilon_2)&=H^{(1)}_{0}(\tau,\epsilon_1,\epsilon_2)+H^{(1)}_{0}(\tau,\epsilon_1,-\epsilon_2)\,,\nonumber\\
W_{1}(\tau,\epsilon_1,\epsilon_2)&=H^{(1)}_{1}(\tau,\epsilon_1,\epsilon_2)+H^{(1)}_{1}(\tau,\epsilon_1,-\epsilon_2).
\end{align}
Therefore, $\widetilde{F}^{(1,1)}$ can be written as
\bea\nonumber
\widetilde{F}^{(1,1)}=H^{(1)}_{0}W_{0}\,\vartheta_{1,0}(\tau,m)^2+(H^{(1)}_{0}W_{1}+H^{(1)}_{1}W_{0})\vartheta_{1,0}(\tau,m)\vartheta_{1,1}(\tau,m)+H^{(1)}_{1}W_{1}\vartheta_{1,1}(\tau,m)^2\,.
\eea

As claimed above, the coefficient functions $H^{(1,1)}_{\ell=0,1,2}$ are completely determined by $H^{(1)}_{\ell=0,1}$. Indeed, using the identities relating the index 2 and products of index 1 elliptic theta functions:
\bea
\vartheta_{1,0}(\tau,m)^2&=&\theta_{3}(4\tau,0)\vartheta_{2,0}(\tau,m)+\theta_{2}(4\tau,0)\vartheta_{2,2}(\tau,m)\,,\\\nonumber
\vartheta_{1,1}(\tau,m)^2&=&\theta_{2}(4\tau,0)\vartheta_{2,0}(\tau,m)+\theta_{3}(4\tau,0)\vartheta_{2,2}(\tau,m)\,,\\\nonumber
\vartheta_{1,0}(\tau,m)\vartheta_{1,1}(\tau,m)&=&\theta_{2}(\tau,0)(\vartheta_{2,1}(\tau,m)+\vartheta_{2,3}(\tau,m))\,,
\eea
we obtain
\bea
H^{(1,1)}_{0}&=&H^{(1)}_{0}W_{0}\theta_{3}(4\tau,0)+H^{(1)}_{1}W_{1}\theta_{2}(4\tau,0)\,,\\\nonumber
H^{(1,1)}_{1}&=&H^{(1,1)}_{3}=(H^{(1)}_{0}W_{1}+H^{(1)}_{1}W_{0})\theta_{2}(\tau,0)\,,\\\nonumber
H^{(1,1)}_{2}&=&H^{(1)}_{0}W_{0}\theta_{2}(4\tau,0)+H^{(1)}_{1}W_{1}\theta_{3}(4\tau,0)\,.
\eea
This is the beginning of an emergent recursive structure, which we will fully explore in the next subsection.

We extracted the spin contents of low-lying states, as encoded by $H^{(1,1)}_{\ell}$.\\

\begin{tcolorbox}

\underline{Spin contents from $H^{(1,1)}_{0}$:}
The function $H^{(1,1)}_{0}(\tau,\epsilon_1,\epsilon_2)$ contains the degeneracies of the states corresponding $Q_{f_1}Q_{f_2}\,Q_{\tau}^{n}$. For some small values of $n$ these are listed below:
\bea\nonumber
n&=&0: 0\\\nonumber
n&=&1:0\\\nonumber
n&=&2:3(1,\tfrac{3}{2})+17(\tfrac{1}{2},1)+5(0,\tfrac{3}{2})+9(1,\tfrac{1}{2})+21(\tfrac{1}{2},0)+31(0,\tfrac{1}{2})\\\nonumber
n&=&3: 4(\tfrac{3}{2}, 2)+ 14(\tfrac{3}{2}, 1) + 6(\tfrac{3}{2}, 0) + 10(\tfrac{1}{2}, 2) +
 28(1, \tfrac{3}{2}) + 60(1, \tfrac{1}{2}) + 98(\tfrac{1}{2}, 1)\\\nonumber
 &&\,\,\,\,\,\,\,+32(0, \tfrac{3}{2}) +
 128(0, \tfrac{1}{2}) + 100 (\tfrac{1}{2}, 0)
\eea
\end{tcolorbox}
\begin{tcolorbox}
\underline{Spin contents from $H^{(1,1)}_{2}$:}
The function $H^{(1,1)}_{2}(\tau,\epsilon_1,\epsilon_2)$ contains the degeneracies of the states corresponding $Q_{f_1}Q_{f_2}\,Q_{m}^{-2}\,Q_{\tau}^{n}$. For some small values of $n$ these are listed below:
\bea\nonumber
n&=&0: 0\\\nonumber
n&=&1:(\tfrac{1}{2},0)+3(0,\tfrac{1}{2})\\\nonumber
n&=&2:2(1, \tfrac{1}{2}) + 7(\tfrac{1}{2}, 1) +(0, \tfrac{3}{2}) + 9 (\tfrac{1}{2}, 0) + 14(0, \tfrac{1}{2})\\\nonumber
n&=&3: 3(\tfrac{3}{2}, 1) + 11(1, \tfrac{3}{2}) + 2 (\tfrac{1}{2}, 2) + 13 (0, \tfrac{3}{2}) +
 42 (\tfrac{1}{2}, 1) + 24 (1, \tfrac{1}{2}) +49 (\tfrac{1}{2}, 0) + 64 (0, \tfrac{1}{2})
\eea
\end{tcolorbox}
\begin{tcolorbox}
\underline{Spin content from $H^{(1,1)}_{1}$:}
The function $H^{(1,1)}_{2}(\tau,\epsilon_1,\epsilon_2)$ contains the degeneracies of the states corresponding $Q_{f_1}Q_{f_2}\,Q_{m}^{-1}\,Q_{\tau}^{n}$. For some small values of $n$ these are listed below:
\bea\nonumber
n&=&0: (0,0)\\\nonumber
n&=&1:3(\tfrac{1}{2}, \tfrac{1}{2}) + 2 (0, 1) + 5 (0, 0)\\\nonumber
n&=&2:4(\tfrac{1}{2}, \tfrac{3}{2}) + 5(1, 1) + 14 (0, 1) + 4 (1, 0) + 22 (\tfrac{1}{2}, \tfrac{1}{2}) + 22 (0, 0)\\\nonumber
n&=&3: 7(\tfrac{3}{2},\tfrac{3}{2})  + 8 (\tfrac{3}{2}, \tfrac{1}{2})+6(1,2) + 34 (\tfrac{1}{2}, \tfrac{3}{2}) +
 4 (0, 2) + 42 (1, 1) + 33 (1, 0)\\\nonumber&&\,\,\,\, + 71 (0, 1) + 110 (\tfrac{1}{2}, \tfrac{1}{2}) +
 86 (0, 0)
\eea

\end{tcolorbox}

%-------------------------------------------------------
\subsubsection{Configuration $(k_{i})=(1,1,\cdots,1)$}
The configuration $(1,1,\cdots,1)$ is the generalization of the configuration studied above in which a single M2-brane traverse through the $N$ many arrayed M5-branes. This should be thought of as a bound-state of a configuration of $(N-1)$ M2-branes with a single M2-brane per each two consecutive M5-branes, with additional winding of M-strings on $\mathbb{T}^2$ that each M5-brane wraps around. The corresponding BPS states are counted by $\F^{(1,1,\cdots,1)}$, as defined in (\ref{bound}). Using (\ref{mm}), we can see that it is given by
\bea
\F^{(1,1,\cdots,1)}:=\sum_{\ell=0}^{N-1}\sum_{(k_{1},\cdots,k_{\ell})\,\sum k_{i}=N-1}(-1)^{\ell-1}G_{k_{1}}G_{k_{2}}\cdots\,G_{k_{\ell}}\,,
\eea
where
\begin{align}
G_{k}&:=H_{01}\,(H_{11})^{k-1}
H_{10}\,,\nonumber
\end{align}
with the definitions~\footnote{
We remark that $H_{01}, H_{11}, H_{10}$ are also expressible in terms of the domain-wall partition function $D_{\lambda\mu}$ introduced in \cite{Haghighat:2013gba}.}
\begin{align}
&H_{01}:=\frac{\theta_{1}(\tau,m-\epsilon_{+})}{\theta_{1}(\tau,-\epsilon_{2})}\,,&&H_{11}:=\frac{\theta_{1}(\tau,m+\epsilon_{-})\theta_{1}(\tau,m-\epsilon_{-})}{\theta_{1}(\tau,\epsilon_{1})\theta_{1}(\tau,-\epsilon_{2})}\,, &&H_{10}:=\frac{\theta_{1}(\tau,m+\epsilon_{+})}{\theta_{1}(\tau,\epsilon_{1})}\,. \label{id}
\end{align}
Using (\ref{id}), we get
\bea\label{comb}
\F^{(1,1,\cdots,1)}:=\sum_{\ell=1}^{N-1}(-1)^{\ell-1}r_{N-1}(\ell)\,(H_{01})^{\ell}\,(H_{11})^{N-\ell-1}
(H_{10})^{\ell},
\eea
where $r_{N}(\ell)$ is the number of $\ell$-tuples $(k_{1},k_{2},\cdots,k_{\ell})$ such that $\sum k_{i}=(N-1)$ and is given by $r_{N}(\ell)=\frac{(N-1)!}{(\ell-1)!(N-\ell)!}$.  In fact, this is the defining form of the  free energy encoding the degeneracies for all ``single M-string states": it contains the combinatorics for placing one M-string in each of $(N-1)$ intervals.

The free energy $\F^{(1,1,\cdots,1)}$ obeys a number of remarkable recursive relations for any $(m, \epsilon_1, \epsilon_2)$. Indeed, simplifying (\ref{comb}), we get
\bea\nonumber
 \F^{(1,1,\cdots,1)}&:=&H_{01}H_{10}\,\sum_{\ell=1}^{N-1}(-1)^{\ell-1}r_{N-1}(\ell)\,(H_{01}H_{10})^{\ell-1}\,(H_{11})^{N-\ell-1}\label{comb2}\\\nonumber
&=&\F^{(1)}\,\sum_{\ell=0}^{N-2}(-1)^{\ell}r_{N-1}(\ell+1)\,(H_{01}H_{10})^{\ell}\,(H_{11})^{N-\ell-2}\label{rel}\\\label{recursion}
&=&\F^{(1)}\,W(\tau,m,\epsilon_1,\epsilon_2)^{N-2},
\eea
where we used the `boundary condition' $\F^{(1)}=H_{01}H_{10}$. Relation (\ref{recursion})  between $\F^{(1,1,\cdots,1)}$ and $\F^{(1)}$ is a generalization of a relation observed in \cite{Bak:2014xwa} for limiting situations to general nonzero values of $m, \epsilon_{1,2}$. Furthermore,  (\ref{recursion}) generalizes  (\ref{recursion0}) to the case of $N$ M5-branes  with $W(\tau,m,\epsilon_1,\epsilon_2)$ defined as
\begin{align}
W(\tau,m,\epsilon_1,\epsilon_2) &=H_{11}-H_{01}H_{10} \nonumber \\
&=\frac{\theta_{1}(\tau,m+\epsilon_{+})\theta_{1}(\tau,m-\epsilon_{+})
-\theta_{1}(\tau,m+\epsilon_{-})\theta_{1}(\tau,m-\epsilon_{-})}
{\theta_{1}(\tau,\epsilon_{1})\theta_{1}(\tau,\epsilon_{2})}\,.\label{DefZ0}
\end{align}

\begin{figure}[h]
  \centering
  \vskip1cm
  \includegraphics[width=5.5in]{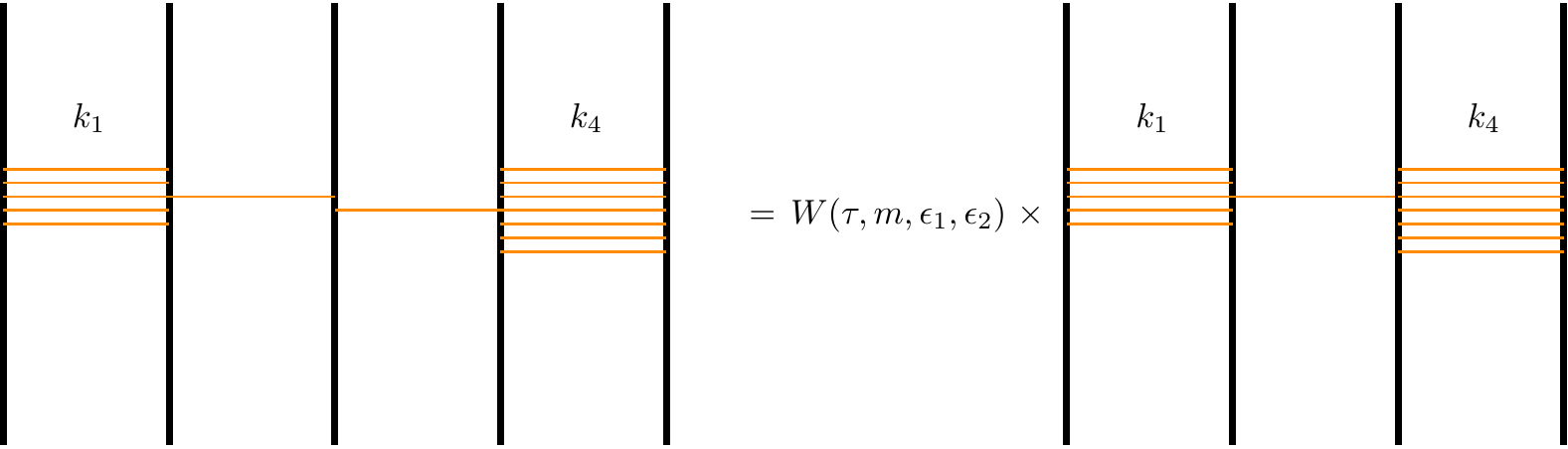}\\
  \vskip0.5cm
 \caption{An M5-brane is contractible whenever on both sides of it a single M2-brane ends. The contracted M5-brane contributes $W(\tau, m, \epsilon_1, \epsilon_2)$ to the free energy.}\label{RM5}
 \vskip0.5cm
\end{figure}

The observed recursive relations have a further generalization. Suppose an arbitrary number of M-strings is partitioned among the M5-brane intervals. If there are $s$ $(s \ge 2)$ consecutive intervals occupied by a single M-string, we conjecture that those intervals are further contractible down to a single interval.  In appendix A, we present evidence that supports our conjecture, generalizing (\ref{recursion}) to
\bea
\fbox{$\displaystyle
\F^{(k_{1},k_{2},\cdots,k_{r},1,1,1,\cdots,1,k_{r+s+1},\cdots,k_{N-1})}
=\F^{(k_{1},k_{2},\cdots,k_{r},1,k_{r+s+1},\cdots,k_{N-1})}\,(W(\tau,m,\epsilon_{1},\epsilon_{2}))^{s-1}$}
\eea
Algorithmically, if we have an M5-brane with a pair of single M2-branes ending on it on both sides, we can join the two M2-branes by removing the bridging M5-brane such that the partition function of the old configuration is  equal to the partition function of the new configuration times the factor $W(\tau,m,\epsilon_{1},\epsilon_{2})$ per each M5-brane removed, as indicated in \figref{RM5}.

Here again, we tabulate the spin contents of low-lying states.
\begin{tcolorbox}

\underline{Spin contents:}
The spin content of the states corresponding to $Q_{f_1}\cdots Q_{f_{N-1}}\,Q_{\tau}^{n}$ for some lower values of $n$ are listed below:
\bea\nonumber
n&=&0: (0,\tfrac{1}{2})\\\nonumber
n&=&1:(N - 1)(\tfrac{1}{2}, 1)+ (3 N - 5)(\tfrac{1}{2}, 0) + (4 N - 6)(0, \tfrac{1}{2})\\\nonumber
n&=&2:\tfrac{N(N - 1)}{2}(1, \tfrac{3}{2}) + (4N^2-12 N  + 9)(1,\tfrac{1}{2})+ (6N^2-16N+11)(\tfrac{1}{2}, 1) \\\nonumber
&&\,\,\,\,\,+ \tfrac{3 N^2 - 7 N + 4}{2} (0, \tfrac{3}{2}) + (15 N^2 + 43 - 49 N) (0, \tfrac{1}{2}) + (12 N^2 - 42 N + 39) (\tfrac{1}{2},0)\\\nonumber
n&=&3: \tfrac{N(N^2 - 1)}{6}(\tfrac{3}{2},2) + \tfrac{(N - 1) (15 N^2 - 39 N + 24)}{6}(\tfrac{3}{2}, 1) + (N - 1) (4 N^2 - 9 N + 5) (1, \tfrac{3}{2})\\\nonumber
&&\,\,\,\,\,+ \tfrac{(4 N^3 - 12 N^2 + 11 N - 3)}{3}(\tfrac{1}{2},2) + \tfrac{(10 N^3 - 54 N^2 + 98 N - 60)}{3} (\tfrac{3}{2}, 0) \\\nonumber
&&\,\,\,\,\,+ (20 N^3 - 101 N^2 + 181 N - 114) (1, \tfrac{1}{2})+ (26 N^3 - 123 N^2 + 210 N - 127) (\tfrac{1}{2}, 1)\\\nonumber
&&\,\,\,\,\,+ (8 N^3 - 36 N^2 + 57 N - 31)(0, \tfrac{3}{2}) + \tfrac{(110 N^3 - 609 N^2 + 1231 N - 882)}{3} (\tfrac{1}{2}, 0)\\\nonumber
&&\,\,\,\,\, + \tfrac{(124 N^3 - 663 N^2 + 1307 N - 918)}{3} (0, \tfrac{1}{2})
\eea
\end{tcolorbox}
\begin{tcolorbox}
\underline{Spin Contents:} The spin content of the states corresponding to $Q_{f_1}\cdots Q_{f_{N-1}}\,Q_{m}\,Q_{\tau}^{n}$ :
\bea\nonumber
n&=&0: (0,0)\\\nonumber
n&=&1:(2 N - 3)(\tfrac{1}{2}, \tfrac{1}{2})+(N - 1) (0, 1)+(4N-7)(0,0)\\\nonumber
n&=&2:(N - 1)^2 (\tfrac{1}{2}, \tfrac{3}{2}) + \tfrac{(3 N^2 - 7 N + 4)}{2} (1,1) + (3 N^2 - 11 N + 10) (1, 0)\\\nonumber
&&\,\,\,\,\,+ (11 N^2 - 36 N + 31)(\tfrac{1}{2}, \tfrac{1}{2}) + (6 N^2 - 18 N + 14) (0, 1) + \tfrac{(25 N^2 - 89 N + 86)}{2}(0,0)\\\nonumber
n&=&3: \tfrac{N (4 N^2 - 9 N + 5)}{6}(\tfrac{3}{2},\tfrac{3}{2})+\tfrac{N (N - 1)^2}{2}(1, 2)+ \tfrac{(6 N^3 - 24 N^2 + 30 N - 12)}{6}(0,2) \\\nonumber
&&\,\,\,\,\,+\tfrac{(44 N^3 - 183 N^2 + 265 N - 132)}{6} (\tfrac{1}{2}, \tfrac{3}{2}) +\tfrac{(63 N^3 - 279 N^2 + 432 N - 234)}{6}(1,1)\\\nonumber
&&\,\,\,\,\,+ \tfrac{(20 N^3 - 99 N^2 + 163 N - 90)}{6} (\tfrac{3}{2},\tfrac{1}{2}) + \tfrac{(132 N^3 - 675 N^2 + 1251 N - 816)}{6} (0, 1) \\\nonumber
&&\,\,\,\,\,+ \tfrac{(232 N^3 - 1242 N^2 + 2408 N - 1650)}{6} (\tfrac{1}{2}, \tfrac{1}{2}) + \tfrac{(30 N^3 - 169 N^2 + 337 N - 234)}{2} (1,0) \\\nonumber
&&\,\,\,\,\,+ \tfrac{(191N^3-1080 N^2 + 2245 N - 1656)}{6} (0, 0)
\eea

\end{tcolorbox}
%

%---------------
\subsubsection{Comparison with single-particle indices}

Dual to the M-string picture, the BPS degeneracies of the configuration $(1,1,\cdots,1)$ can also be computed from the five-dimensional $\mathcal{N}=1^{*}$ gauge theory. The multi-particle index $I_{SU(N)}$ of the $\mathcal{N}=1^*$ theory can be extracted in terms of the single particle index $z^{SU(N)}_{sp}$:
\begin{align}
I_{SU(N)}(Q_m,t,q)=\text{exp}\left[\sum_{n=1}^\infty\,\frac{1}{n}\,z^{SU(N)}_{sp}(Q_m^n,t^n,q^n)\right]\,.
\end{align}
In the limit $\epsilon_1=-\epsilon_2\to0$ and $Q_m=-1$ the single particle index $z_{sp}^{U(N)}$ was computed in \cite{Kim:2011mv}  in the form of $Q_\tau$-expansions
\begin{align}
z^{SU(2)}_{sp}&=1+8Q_\tau+40Q_\tau^2+160Q_\tau^3+552Q_\tau^4+1712Q_\tau^5+4896Q_\tau^6+\ldots\nonumber\\
z^{SU(3)}_{sp}&=1+24Q_\tau+264Q_\tau^2+2016Q_\tau^3+12264Q_\tau^4+63504Q_\tau^5+290976Q_\tau^6+\ldots\nonumber\\
z^{SU(4)}_{sp}&=1+40Q_\tau+744Q_\tau^2+8992Q_\tau^3+82344Q_\tau^4+618864 Q_\tau^5 + 4002336 Q_\tau^6+\ldots\nonumber\\
z^{SU(5)}_{sp}&=1 + 56 Q_\tau + 1480 Q_\tau^2 + 25184 Q_\tau^3 + 317288 Q_\tau^4 + 3207888 Q_\tau^5 +  27375520 Q_\tau^6+\ldots\nonumber
\end{align}
They just correspond to the genus-zero of free energy $F^{(1,1,\cdots,1)}(\tau, m, \epsilon_1, \epsilon_2)$:
\begin{align}
&z^{SU(2)}_{sp}=-\tfrac{1}{4}\,\lim_{\epsilon_{1,2}\mapsto 0}\epsilon_{1}\epsilon_{2}F^{(1)}(\tau,m=\tfrac{1}{2},\epsilon_1,\epsilon_2)\,,\nonumber\\
&z^{SU(3)}_{sp}=-\tfrac{1}{4}\,\lim_{\epsilon_{1,2}\mapsto 0}\epsilon_{1}\epsilon_{2}F^{(1,1)}(\tau,m=\tfrac{1}{2},\epsilon_1,\epsilon_2)\,,\nonumber\\
&z^{SU(4)}_{sp}=-\tfrac{1}{4}\,\lim_{\epsilon_{1,2}\mapsto 0}\epsilon_{1}\epsilon_{2}F^{(1,1,1)}(\tau,m=\tfrac{1}{2},\epsilon_1,\epsilon_2)\,,\nonumber\\
&z^{SU(5)}_{sp}=-\tfrac{1}{4}\,\lim_{\epsilon_{1,2}\mapsto 0}\epsilon_{1}\epsilon_{2}F^{(1,1,1,1)}(\tau,m=\tfrac{1}{2},\epsilon_1,\epsilon_2)\,.\nonumber
\end{align}
It was also observed in \cite{Kim:2011mv, Bak:2014xwa} that the single particle indices are related as
\bea
\frac{z^{SU(N)}_{sp}}{z^{SU(2)}_{sp}}=W(\tau,m,0,0)^{N-2}\ .
\eea
This also corresponds to the genus-zero limit of our recursion relation (\ref{recursion}) for the $\epsilon_{1,2}\mapsto 0$.

%--------------------------
\subsubsection{Properties of $W(\tau,m,\epsilon_{1},\epsilon_{2})$}
We showed that the function $W(\tau,m,\epsilon_1,\epsilon_2)$ defined in (\ref{DefZ0}) appears whenever an M5-brane, with a single M2-brane ending on it from both sides, is removed. In the next section, we will be identifying this function in the NS limit with the refined elliptic genus of the Taub-NUT space.

Here, we collect relevant properties of this function: under the $SL(2, \mathbb{Z})$ modular transformation $(\tau,m,\epsilon_1,\epsilon_2)\mapsto (-\tfrac{1}{\tau},\tfrac{m}{\tau},\tfrac{\epsilon_{1}}{\tau},\tfrac{\epsilon_2}{\tau})$, the function $W(\tau,m,\epsilon_1,\epsilon_2)$ transforms in the following way:
\bea\nonumber
W(-\tfrac{1}{\tau},\tfrac{m}{\tau},\tfrac{\epsilon_{1}}{\tau},\tfrac{\epsilon_2}{\tau})=e^{\frac{2\pi i }{\tau}(m^2-\epsilon_{-}^2)}\Big[\frac{\theta_{1}(\tau,m+\epsilon_{+})\theta_{1}(\tau,m-\epsilon_{+})
-e^{\frac{2\pi i }{\tau}(\epsilon_{-}^2-\epsilon_{+}^2)}\theta_{1}(\tau,m+\epsilon_{-})\theta_{1}(\tau,m-\epsilon_{-})}
{\theta_{1}(\tau,\epsilon_{1})\theta_{1}(\tau,\epsilon_{2})}\Big]
\eea
Due to the relative phase factor between the two terms in the numerator, the function $W(\tau, m, \epsilon_1, \epsilon_2)$ transforms as a weight-zero Jacobi form if and only if $\epsilon_{+}=\pm \epsilon_{-}$. This is precisely the NS limit, $\epsilon_1=0$ or $\epsilon_2=0$. In this NS limit $(\epsilon_{2}\mapsto 0$),  the function $W(\tau, m, \epsilon_1, \epsilon_2)$ is reduced to
\bea
W(\tau,m,\epsilon_1,0)=i\frac{\theta_{1}^{'}(\tau,m+\frac{\epsilon_{1}}{2})\theta_{1}(\tau,m-\frac{\epsilon_{1}}{2})-\theta_{1}^{'}(\tau,m-\frac{\epsilon_{1}}{2})\theta_{1}(\tau,m+\frac{\epsilon_{1}}{2})}{\theta_{1}(\tau,\epsilon_{1})\eta(\tau)^3}\,,
\eea
while in the genus-zero limit ($\epsilon_{1,2}\mapsto 0$), the function $W(\tau, m, \epsilon_1, \epsilon_2)$ is reduced to
\bea
W(\tau,m,0,0)&=&\frac{\varphi_{0,1}(\tau,m)}{24}+\frac{E_{2}(\tau)}{12}\,\varphi_{-2,1}(\tau,m)\,,\\\nonumber
&=&\frac{\theta_{1}^{''}(\tau,m)\theta_{1}(\tau,m)-\theta_{1}^{'}(\tau,m)^2}{\eta(\tau)^6},
\eea
where $\varphi_{-2,1}(\tau,z)$ and $\varphi_{0,1}(\tau,z)$ denote the weight $-2$ index 1 and weight $0$ index 1 Jacobi forms, respectively, defined in appendix (\ref{BasicJacForms}).
%-----------------------------------------------------------------------------------------------
\subsection{Two M2-branes}
More involved configurations arise when more than two M2-branes are stretched between any two M5-branes. Here, we consider the simplest such configuration, {\it i.e.} the configuration with $(k_i)=(2)$.

Following (\ref{bound}), we have
\begin{align}
\F^{(2)}&=\frac{\theta_1(\tau,m+\epsilon_{+})\theta_1(\tau,m-\epsilon_{+})}{\theta_1(\tau,\epsilon_{1})\theta_1(\tau,\epsilon_{2})\theta_1(\tau,\epsilon_{1}-\epsilon_{2})} \nonumber \\
& \times \Big[
\frac{\theta_1(\tau,m+\epsilon_{+}+\epsilon_{2})\theta_1(\tau,m-\epsilon_{+}-\epsilon_{2})}{\theta_1(\tau,2\epsilon_{2})}
+\frac{\theta_1(\tau,m+\epsilon_{+}+\epsilon_{1})\theta_1(\tau,m-\epsilon_{+}-\epsilon_{1})}{\theta_1(\tau,2\epsilon_{1})}
\Big]
\nonumber \\
& -\frac{\theta_1(\tau,m+\epsilon_{+})^2\theta_1(\tau,m-\epsilon_{+})^2}{2\theta_1(\tau,\epsilon_{1})^2\theta_1(\tau,\epsilon_{2})^2} +
\frac{\theta_1(2\tau,2m+2\epsilon_{+})\theta_1(2\tau,2m-2\epsilon_{+})}{2\theta_1(2\tau,2\epsilon_{1})\theta_1(2\tau,2\epsilon_{2})}\,.
\end{align}
Since this is of index 2 with respect to $m$,  it is expandable in terms of index $2$ theta functions $\vartheta_{2,\ell}(\tau,m)$ defined in (\ref{DefThetaGen}), with the explicit $Q_\tau$-expansion given in (\ref{VarthetaIndex2}):
\begin{align}
\F^{(2)}=\sum_{\ell=0}^3H^{(2)}_{\ell}(\tau,\epsilon_1,\epsilon_2)\,\vartheta_{2,\ell}(\tau, m)\, .
\end{align}
Here, the coefficient functions are defined by
{\allowdisplaybreaks
\begin{align}
&H_{0}^{(2)}:=\int \frac{dQ_{m}}{2\pi i} \ Q_{m}^{-1}\ F^{(2)} \,, \nonumber \\
&H_{1}^{(2)}:=Q_\tau^{-\frac{1}{8}}\int \frac{dQ_{m}}{2\pi i}F^{(2)} =H_3^{(2)} \,,\nonumber \\
&H_{2}^{(2)} =Q_\tau^{-\frac{1}{2}}\int \frac{dQ_{m}}{2\pi i}\,Q_m \ F^{(2)} \,.\nonumber
\end{align}}
In the genus-zero limit ($\epsilon_{1,2}\mapsto 0$), they have the $Q_\tau$-expansions as follows:
\begin{align}
&\lim_{\epsilon_{1,2}\to 0}\epsilon_{1}\epsilon_{2}H^{(2)}_0(\tau,\epsilon_1,\epsilon_2)=24 Q_\tau + 368 Q_\tau^2 + 3376 Q_\tau^3 + 23168 Q_\tau^4 + 131248 Q_\tau^5\nonumber\\
&\hspace{2cm}+ 645568 Q_\tau^6 + 2845536 Q_\tau^7 + 11477824 Q_\tau^8 + 43006152 Q_\tau^9 + 151352896 Q_\tau^{10}+\cdots,\nonumber\\
&\lim_{\epsilon_{1,2}\to 0}\epsilon_{1}\epsilon_{2}H^{(2)}_{1}(\tau,\epsilon_1,\epsilon_2)= -16 Q_\tau - 272 Q_\tau^2 - 2608 Q_\tau^3 - 18432 Q_\tau^4 - 106576 Q_\tau^5\nonumber\\
& \hspace{2cm}- 532480 Q_\tau^6 - 2376304 Q_\tau^7 - 9683120 Q_\tau^8 - 36592880 Q_\tau^9 - 129728864 Q_\tau^{10}+\cdots,\nonumber\\
&\lim_{\epsilon_{1,2}\to 0}\epsilon_{1}\epsilon_{2}H^{(2)}_{2}(\tau,\epsilon_1,\epsilon_2)= 4 Q_\tau + 104 Q_\tau^2 + 1168 Q_\tau^3 + 9104 Q_\tau^4 + 56276 Q_\tau^5\nonumber\\
&\hspace{2cm}+295608 Q_\tau^6 + 1372048 Q_\tau^7 + 5772688 Q_\tau^8 + 22406176 Q_\tau^9 + 81266232 Q_\tau^{10}+\cdots\nonumber .
\end{align}

We tabulate the spin contents of the BPS states extracted for this M2-brane configuration.
\hfill\break
\begin{tcolorbox}

\underline{Spin contents:}
The degeneracies of the states corresponding $Q_{f_1}^2\,Q_{\tau}^{n}$. For some lower values of $n$, they are listed below:
\bea\nonumber
n&=&0:0\\\nonumber
n&=&1:(\tfrac{1}{2},2)+(\tfrac{1}{2},1)+2(0,\tfrac{3}{2})\\\nonumber
n&=&2: (\tfrac{3}{2}, 3) + (\tfrac{1}{2}, 3) + 3(1, \tfrac{5}{2}) + (\tfrac{3}{2}, 2) + 3(0, \tfrac{5}{2}) +
 8(\tfrac{1}{2}, 2) + 4(1, \tfrac{3}{2}) + 10 (0, \tfrac{3}{2}) + 8 (\tfrac{1}{2}, 1)\\\nonumber
 &&\,\,\,\,\,+(1, \tfrac{1}{2}) + 5 (0, \tfrac{1}{2}) + (\tfrac{1}{2}, 0)
\eea
\end{tcolorbox}
\begin{tcolorbox}
\underline{Spin Contents for $Q_{f_1}^2\,Q_{m}\,Q_{\tau}^{n}$:}\\
\bea\nonumber
n&=&0:0 \\\nonumber
n&=&1:(\tfrac{1}{2},\tfrac{3}{2})+(0,2)+(0,1)\\\nonumber
n&=&2:(\tfrac{3}{2},\tfrac{5}{2})+(1, 3) + 3 (\tfrac{1}{2}, \tfrac{5}{2}) + 3 (1, 2) +6 (0, 2) + 8 (\tfrac{1}{2}, \tfrac{3}{2}) + 2 (1, 1) +7 (0, 1) \\\nonumber
&&\,\,\,\,\,+ 3 (\tfrac{1}{2}, \tfrac{1}{2}) + (0, 0)
\eea

\end{tcolorbox}

%--------------------------------------------------------------------------------
\subsection{Three M2-branes}
We can repeat the analysis for the case of three M2-branes suspended between two M5-branes, corresponding to the partition $(k_i)=(3)$. Due to the complexity of $\F^{(3)}$, however, here we only present the expression in the particular case $\epsilon_1=-\epsilon_2=\epsilon$,
\begin{align}
3\F^{(3)}(\tau,m,\epsilon,-\epsilon)
&= -\frac{3 \theta_1(\tau,m )^2 \theta_1\left(\tau,m +\epsilon \right) \theta_1(\tau,m -\epsilon )}{\theta_1(\tau,\epsilon )^4 \theta_1\left(\tau,2\epsilon\right)^2 \theta_1\left(\tau,3\epsilon\right)^2} \nonumber \\
& \hskip-2cm \times \left[ \theta_1(\tau,\epsilon )^2 \theta_1\left(\tau,m +2\epsilon\right) \theta_1\left(\tau,m-2\epsilon\right)+\theta_1\left(\tau,2\epsilon\right)^2 \theta_1\left(\tau,m+\epsilon\right) \theta_1(\tau,m-\epsilon)\right]
\nonumber \\
&+
\frac{\theta_1\left(3\tau,3m\right)^2}{\theta_1\left(3\tau,3\epsilon\right)^2}
+\frac{6 \theta_1(\tau,m )^4 \theta_1\left(\tau,m+\epsilon\right) \theta_1(\tau,m -\epsilon )}{\theta_1(\tau,\epsilon )^4 \theta_1\left(\tau,2\epsilon\right)^2}
-\frac{\theta_1(\tau,m )^6}{\theta_1(\tau,\epsilon )^6}\, .
\end{align}
Once again, this function is expandable in term of $\vartheta$ functions in the form
\begin{align}
\F^{(3)}=\sum_{\ell=0}^5H^{(3)}_{\ell}(\tau,\epsilon_1,\epsilon_2)\,\vartheta_{3,\ell}(\tau, Q_m)\,.
\end{align}

We also tabulate the spin content of low-lying BPS states for this M2-brane configuration.
\hfill\break
\begin{tcolorbox}

\underline{Spin contents:}
The degeneracies of the states corresponding $Q_{f_1}^3\,Q_{\tau}^{n}$. For some small values of $n$ are listed below:
\bea\nonumber
n&=&0:0\\\nonumber
n&=&1:(\tfrac{1}{2},3)+(\tfrac{1}{2},2)+2(0,\tfrac{5}{2})\\\nonumber
n&=&2: 4(\tfrac{3}{2}, 3) + 17(\tfrac{1}{2}, 3) + 10(1, \tfrac{5}{2}) + (\tfrac{3}{2}, 2) + 20(0, \tfrac{5}{2}) +
 17(\tfrac{1}{2}, 2) + 5(1, \tfrac{3}{2}) + 11(0, \tfrac{3}{2}) \\\nonumber
 &&\,\,\,\,\,+ 6 (\tfrac{1}{2}, 1)+(1, \tfrac{1}{2}) + 5 (0, \tfrac{1}{2}) + 3(\tfrac{1}{2},0)+(2, \tfrac{9}{2}) + (1, \tfrac{9}{2}) + 3(\tfrac{3}{2}, 4) + (2, \tfrac{7}{2}) + 3 (\tfrac{1}{2}, 4)\\\nonumber
&&\,\,\,\,\,+
 9 (1, \tfrac{7}{2}) + 6 (0, \tfrac{7}{2})
\eea
\end{tcolorbox}
\begin{tcolorbox}
\underline{Spin Contents for $Q_{f_1}^2\,Q_{m}\,Q_{\tau}^{n}$:}
\bea\nonumber
n&=&0:0 \\\nonumber
n&=&1:(\tfrac{1}{2},\tfrac{5}{2})+(0,3)+(0,2)\\\nonumber
n&=&2:(\tfrac{3}{2},\tfrac{9}{2})+3(2,4)+3(\tfrac{3}{2},\tfrac{7}{2})+2(\tfrac{3}{2},\tfrac{5}{2})+3(1,4)+9(1, 3) + 16 (\tfrac{1}{2}, \tfrac{5}{2})
+ 5(1,2)\\\nonumber
&&\,\,\,\,\,+15(0, 2) +8(\tfrac{1}{2},\tfrac{7}{2})+ 9 (\tfrac{1}{2}, \tfrac{3}{2}) + 2 (1, 1) +4(0, 1)+ 3 (\tfrac{1}{2}, \tfrac{1}{2}) + (0,4)+11(0,3)\\\nonumber
&&\,\,\,\,\,+4(0, 0)
\eea

\end{tcolorbox}

%%%%%%%%%%%%%%%%%%%%%%%%%%%%%%%%%%%%%%%%%%%%%%%%%%%%%%%%
\section{BPS Degeneracies of m-Strings}
Based on the information of the BPS degeneracies of M-strings, we now study the BPS degeneracies of m-strings.

%--------------------------------------------------------
\subsection{m-String Free Energies}
In section~\ref{Sect:DegMonopole}, we discussed the free energies $\F^{k_1,\ldots,k_{N-1}}$, which capture degeneracies of M-strings, for generic values of $\epsilon_{1,2}$ as well as $m$. However, as explained in section 2, in order to interpret them in terms of degeneracies of m-strings, it is necessary to take the NS-limit, sending $\epsilon_2\to 0$. This yields
\bea \nonumber
\lim_{\epsilon_{2}\mapsto 0}\epsilon_{2}\,\F^{(k_1,\ldots,k_{N-1})}(\tau,m,\epsilon_1,\epsilon_2),
\eea
where the parameter $\epsilon_1$ is kept finite. In this section, we shall study the leading term in their series expansions and learn about BPS states of m-strings. In particular, we aim to understand their modular properties in detail.

Before considering the limit $\epsilon_1=-\epsilon_2=0$ let us try to understand the modular properties of the NS limit of $\widetilde{F}^{(k_1,k_2,\cdots,k_{N-1})}$. Recall from section 4.2 that $\widetilde{F}^{(k_1,k_{2},\cdots,k_{N-1})}$ do not transform covariantly under the $SL(2,\mathbb{Z})$. Since $\widetilde{F}^{(k_1,k_{2},\cdots,k_{N-1})}$ is a sum of the product of different $Z_{r_{1}\cdots r_{N-1}}$, different pieces transform with different phase-factors. For example, consider $\widetilde{F}^{(1,2)}$,
\bea
\widetilde{F}^{(1,2)}=Z_{12}-Z_{1}Z_{11}-Z_{2}Z_{1}+Z_{1}^3\,.
\eea
Under $(\tau,m,\epsilon_{\pm})\mapsto (-\tfrac{1}{\tau},\tfrac{m}{\tau},\tfrac{\epsilon_{\pm}}{\tau})$,  $\widehat{F}^{(1,2)}$ transforms as
\bea \label{cov}
\widetilde{F}^{(1,2)}(-\tfrac{1}{\tau},\tfrac{m}{\tau},\tfrac{\epsilon_{\pm}}{\tau})=e^{\frac{2\pi i}{\tau}f_{12}}Z_{12}-e^{\frac{2\pi i}{\tau}(f_{1}+f_{11})}Z_{1}Z_{11}-e^{\frac{2\pi i}{\tau}(f_{1}+f_{2})}Z_{2}Z_{1}+e^{\frac{2\pi i}{\tau}3f_{1}}Z_{1}^3 \ ,
\eea
where
\bea
f_{12}(m,\epsilon_{+},\epsilon_{-})&=&3m^2-\tfrac{15}{2}\epsilon_{+}^2+
\tfrac{9}{2}\epsilon_{-}^2\,,\,\,\,f_{1}(m,\epsilon_{+},\epsilon_{-})=m^2-\epsilon_{+}^2\,,\\\nonumber
f_{11}(m,\epsilon_{+},\epsilon_{-})&=&2m^2-3\epsilon_{+}^2+
\epsilon_{-}^2\,,\,\,\,f_{2}(m,\epsilon_{+},\epsilon_{-})=2m^2-4\epsilon_{+}^2+2\epsilon_{-}^2\, .
\eea
One readily sees that $f_{12}$, $f_{1}+f_{11}$, $f_{1}+f_{2}$ and $3f_{1}$ are not  equal even pairwise. So the four terms in (\ref{cov}) have different phase factors. However, notice that for $\epsilon_{+}^2=\epsilon_{-}^2$ the phase factors are precisely the same and hence $\widetilde{F}^{(1,2)}$ transforms covariantly under $SL(2,\mathbb{Z})$. The condition $\epsilon_{+}^2=\epsilon_{-}^2$ is precisely the NS limit. This is essentially due to the fact that $f_{\vec{k}}(m,\epsilon_{+},\epsilon_{-})$ given in (\ref{fk}) which are quadratic in $k_a$ for generic $\epsilon_{1,2}$ become linear in $k_{a}$ in the NS limit.

Let's introduce
\bea
\boxed{
J_{k_1\cdots k_{N-1}}(\tau,m, \epsilon_1):=\lim_{\epsilon_{2}\mapsto 0}\frac{\widetilde{F}^{(k_1,k_2,\cdots,k_{N-1})}(\tau,m,\epsilon_1,\epsilon_2)}{\widetilde{F}^{(1)}(\tau,m,\epsilon_1,\epsilon_2)}
}.
\label{jfunction}
\eea
From the above discussion and (\ref{Zktrans}), it follows that for $\text{gcd}(k_1,k_2,\cdots,k_{N-1})=1$:
\bea \label{Jtrans}
J_{k_{1}\cdots k_{N-1}}(\tau+1,m,\epsilon_{1})&=&J_{k_{1}\cdots k_{N-1}}(\tau,m,\epsilon_{1})\,,\\\nonumber
J_{k_{1}\cdots k_{N-1}}(-\tfrac{1}{\tau},\tfrac{m}{\tau},\tfrac{\epsilon_{1}}{\tau})&=&e^{\frac{2\pi i}{\tau}(K-1)(m^2-\epsilon_{1}^2)}J_{k_{1}\cdots k_{N-1}}(\tau,m,\epsilon_{1})\,,\\\nonumber
J_{k_{1}\cdots k_{N-1}}(\tau,m+\ell\tau+r,\epsilon_{1})&=&e^{-2\pi i K\,\ell^2\tau+4\pi i m K}J_{k_{1}\cdots k_{N-1}}(\tau,m,\epsilon_{1})\,.
\eea
If we further consider the genus-zero limit $\epsilon_1\mapsto 0$, then from the above  equations it is clear that $J_{k_1\cdots k_{N-1}}(\tau,m,0)$ has the same modular transformation properties as the elliptic genus of a manifold with dimension $4(K-1)$.

We consider the properties of individual $\F^{(k_1,\ldots,k_{N-1})}(\tau,m,\epsilon_1,\epsilon_2)$ in the limit $\epsilon_1=-\epsilon_2= 0$ i.e., studying the leading order in the NS-limit. Since $\F^{(k_{1}k_{2}\cdots k_{N-1})}$ captures all single-string bound-states, by  extensiveness, it should be proportional to the volume of $\mathbb{R}^4$. This infinite volume is regularized by the $\Omega$-background parameters $\epsilon_{1,2}$\footnote{The first Chern class of $\mathbb{R}^4$ also gets deformed to $(\epsilon_{1}+\epsilon_{2})$.},
\bea
\mbox{Vol}(\mathbb{R}^4) \quad \rightarrow \quad {1 \over \epsilon_1 \epsilon_2}.
\eea
Details of proportionality constant does not matter us since we will be always taking ratios of free energies that are always regular in this limit.
Indeed, for $\epsilon_1=-\epsilon_2=0$, the residue of the free energy,
\begin{align}
\widehat{F}^{(k_1,\ldots, k_{N-1})}(\tau,m)=\lim_{{\epsilon_1\to 0} \atop {\epsilon_2\to 0}} \epsilon_{1}\epsilon_{2} \F^{(k_1,\ldots,k_{N-1})}(\tau,m,\epsilon_1,\epsilon_2)\,,\label{DefhatFgen}
\end{align}
is nothing but the genus-zero contribution to the partition functions defined in (\ref{bound}) in section~\ref{Sect:DegMonopole}. These residues can be written in the form
\begin{align}
 \widehat{F}^{(k_1,\ldots, k_{N-1})}(\tau,m)
= & \varphi_{-2,1}(\tau,m)\sum_{a=0}^{s-1} \frac{g_{2a}^{(k_1,\ldots,k_{N{-1}} )}(\tau)}{2^a\,12^{s-1}}\,\left(\varphi_{0,1}(\tau,m)\right)^{s-1-a}\left(\varphi_{-2,1}(\tau,m)\right)^{a}\,.\label{ExpandhatFGen}
\end{align}
Here, $\varphi_{-2,1}(\tau,m)$ and $\varphi_{0,1}(\tau,m)$ are the standard Jacobi forms of $SL(2,\mathbb{Z})$ with index $1$ and weights $-2$ and $0$, respectively, as introduced in (\ref{BasicJacForms}) in appendix~\ref{App:WeakJacForms}. Note that, because of the overall $\varphi_{-2,1}(\tau,m)$ factor, the residue vanishes in the limit the hypermultiplet mass is tuned to 0:
\begin{align}
\lim_{m\to 0}\widehat{F}^{(k_1,\ldots, k_{N-1})}(\tau,m)=0\,.
\end{align}

The functions $g_{2a}^{(k_1,\ldots k_{N-1})}$ in (\ref{ExpandhatFGen}) are anomalous modular forms. More precisely, they can be written as polynomials in the Eisenstein series (see appendix~\ref{App:DataFfunctions} for explicit examples) that include $E_2(\tau)$ as well. Upon replacing the latter by non-holomorphic $\widehat{E}_2(\tau,\bar{\tau})$, defined in (\ref{E2hat}), $g_{2a}^{(k_1,\ldots k_{N-1})}(\tau,\bar{\tau})$ transforms with weight $2a$ under modular transformations of a  congruence subgroup $\Gamma$ of $SL(2,\mathbb{Z})$. Finally, the numerical factors in (\ref{ExpandhatFGen}) are purely for convenience.
%%

%----------------------------------------------------
\subsection{Modular Transformations}
With the definitions given above, it can be seen that $\widehat{F}^{(k_1,\ldots, k_{N-1})}$ transforms as
\begin{align}
&\widehat{F}^{(k_1,\ldots, k_{N-1})}\left(\frac{a\tau+b}{c\tau+d},\frac{m}{c\tau+d}\right)=(c\tau+d)^w\,e^{2\pi i s\,\frac{cm^2}{c\tau+d}}\,\widehat{F}^{(k_1,\ldots, k_{N-1})}(\tau,m)\,, \nonumber\\
 &\widehat{F}^{(k_1,\ldots, k_{N-1})}(\tau,m+\ell\tau+\ell')=e^{-2\pi is(\ell^2\tau+2\ell m)}\,\widehat{F}^{(k_1,\ldots, k_{N-1})}(\tau,m)
\nonumber \\
%\end{align}
%for
%\begin{align}
&\hskip2cm \mbox{for} \quad \left(\begin{array}{cc}a & b \\ c & d\end{array}\right)\in\Gamma\subset SL(2,\mathbb{Z}) \quad \mbox{and} \quad \ell',\ell\in\mathbb{Z}\,,
\end{align}
with weight $w(k_1,\ldots,k_{N-1})$ and index $s(k_1,\ldots,k_{N-1})$ given by
\begin{align}
w(k_1,\ldots,k_{N-1})=-2 \qquad \mbox{and} \qquad s(k_1,\ldots,k_{N-1})=\sum_{a=1}^{N-1}k_a\,.
\end{align}

Due to the $\tau_2$ dependence of $\widehat{E}_2(\tau, \overline{\tau})$ induced by the replacement (\ref{E2hat}), $\widehat{F}^{(k_1,\ldots, k_{N-1})}$ is no longer holomorphic, but it is a so-called {\it quasi-holomorphic} modular object. However, this prescription is not the only way to obtain modular objects. We will discover in section~\ref{Sect:GrandCanonical} that, for a given index $s$, there always exist specific combinations of $\widehat{F}^{(k_1,\ldots, k_{N-1})}(\tau,m)$ (unique up to certain identities) for which the holomorphic anomaly cancels, thus yielding (holomorphic) weak Jacobi forms.

%-------------------------------------------------
\subsection{m-String Elliptic Genera from $\widehat{F}^{(k_{1}k_{2}\cdots k_{N-1})}$
}\label{Sect:EllipticGenera}
\subsubsection{Regularized elliptic genera}
In \cite{Bak:2014xwa}, it was shown that elliptic genera of the Atiyah-Hitchin and Taub-NUT space were captured by the five-dimensional \nonestar gauge theory. Since the M-strings point of view is natural for counting M2-branes, the M-string free energy captures the elliptic genera of monopole moduli spaces in the NS limit to all orders in $Q_\tau=e^{2\pi i \tau}$. Consequently, it must be that the reduced  free energy $\widehat{F}^{(k_{1}\cdots k_{N-1})}$ defined in the previous section are related to the elliptic genera of the m-string moduli space of charge $(k_{1},k_{2},\cdots, k_{N-1})$. In the subsequent sections, we provide evidence for this, along the lines of \cite{Bak:2014xwa}.

We  first recall a number of facts about elliptic genera on compact and non-compact hyperk\"ahler manifolds. For a compact manifold $\mathcal{M}$, the elliptic genus can be defined as
\begin{align}
\phi_{\mathcal{M}}(\tau,m)=\int_{\mathcal{M}}\,\prod_i\frac{x_i\,\theta_1(\tau,x_i+m)}{\theta_1(\tau,x_i)}\,,\label{DefEllGenInt}
\end{align}
where $x_i$ are the Chern roots of the tangent bundle on $\mathcal{M}$. Physically, the elliptic genus can be computed by the path integral over the loop space configurations:
\begin{align}
\phi_{\mathcal{M}}(\tau,m)=\sum_{\mathcal{H}_{RR}}(-1)^{F+\bar{F}}\,e^{2\pi imJ_0}\,Q_\tau^{L_0-\frac{c}{24}}\bar{Q}_\tau^{\bar{L}_0-\frac{\bar{c}}{24}}\,,\label{DefEllGenTrace}
\end{align}
where the sum is over the Hilbert space of the Ramond-Ramond sector of the two-dimensional supersymmetric sigma-model with target space $\mathcal{M}$. This Hilbert space consists of countably many normalizable states. Furthermore, $F$ ($\bar{F}$) is the left-(right) moving fermion number and $L_0$ and $J_0$ are generators of the $\mathcal{N}=2$ superconformal algebra of the sigma model. The elliptic genus encodes important information about the spectrum of the sigma-model which are intimately linked to topological properties and data of the target manifold $\mathcal{M}$. Moreover, as was discussed in \cite{Witten:1993jg,Eguchi:1988vra}, if the first Chern class of $\mathcal{M}$ vanishes ($c_1(\mathcal{M})=0$), the elliptic genus $\phi_{\mathcal{M}}$ is a weak Jacobi form of weight 0 and index $\text{dim}_{\mathbb{C}}(\mathcal{M})/2$. Physically, this is a {\color{green}consequence} of the $\mathcal{N}=2$ superconformal invariance of the sigma-model, as discussed in \cite{Kawai:1993jk}.

In the case that $\mathcal{M}$ is {\it non-compact}, the definitions (\ref{DefEllGenInt}) and (\ref{DefEllGenTrace}) need to be modified: from the geometric point of view, the integral in (\ref{DefEllGenInt}) becomes ill-defined and needs to be suitably regularized. For example, in \cite{Gritsenko:1999nm} it was proposed to perform the integration  equivariantly and it was argued that the corresponding {\it  equivariantly regularized elliptic genus} still transforms nicely under the modular transformations. Physically, besides well localized bulk states entering in (\ref{DefEllGenTrace}), sigma-models with non-compact target spaces generically also contain delocalized boundary  modes whose spectrum overlaps with the continuum scattering states, which also need to be taken into account (see for example \cite{Troost:2010ud}). In both cases, the modification requires introducing an additional parameter (which we call $\mu$ in the following), either in the form of a regularization parameter or in the form of the quantum numbers that label the delocalized states contributing to boundary part.

More specifically, for noncompact ${\cal M}$, we can define a {\it regularized} elliptic genus $\phi_{\text{reg}}(\tau,Q_m,\mu)$ with the following properties \cite{Gritsenko:1999nm} :
\begin{itemize}
\item For generic values of $\mu$, the regularized elliptic genus $\phi_{\text{reg}}(\tau,Q_m,\mu)$ transforms as a Jacobi form of weight $0$ under the full modular group $SL(2,\mathbb{Z})$.
\item Upon removing the parameter $\mu$, the genus $\phi_{\text{reg}}(\tau,Q_m,\mu=0)$ must be well-defined for $Q_m=\pm1$ and has to reproduce correctly the topological data of the target space manifold, {\it i.e.}
\begin{align}
&\phi_{\text{reg}}(\tau,Q_m=1,\mu=0)=\chi_\mathcal{M}\,,&&\phi_{\text{reg}}(\tau,Q_m=-1,\mu=0)=\sigma_\mathcal{M}\,,
\end{align}
where $\chi_{\mathcal{M}}$ is the Euler characteristic and $\sigma_{\mathcal{M}}$ the signature of $\mathcal{M}$.
\end{itemize}
%-------------------------------------------------------
\subsubsection{Comparison with other BPS bound-state problems}
In a variety of cases in string and field theories, it was observed that multi-instanton bound-state effects in $d$ dimensions encode {\sl part} of multi-particle bound-state effects in $(d+1)$-dimensions for reasons that have to do with non-compact configuration spaces and their continuous spectra~\cite{Imbimbo:1983dg}. Here, we recall some examples of this type and compare with the M \& m-string bound-state problem at hand.

One instance in string theory concerns the M-theory conjecture \cite{Witten:1995ex}  that multiple D0-particles in Type IIA string theory form a unique bound-state that builds the M-theory Kaluza-Klein tower. The bound-state is at threshold and so the relative moduli space is noncompact. The $L^2$-class Witten index for zero-energy, which counts BPS ground-states, is then calculated from the multi-D0-particle dynamics on $\mathbb{R}^9 \times \mathbb{S}^{1}_{\beta}$ in the limit of the radius $\beta \rightarrow 0$. It consists of two parts: a so-called {\it bulk part} and a so-called {\it boundary part} \cite{Sethi:1997pa}. If the IIA theory is compactified to $\mathbb{R}^{8, 1} \times \mathbb{S}^{1}_{R}$, the D0-particle circulating around $\mathbb{S}^{1}_{R}$ can be interpreted as an instanton in $\mathbb{R}^{8, 1}$. It was then observed \cite{Green:1997tn}  that the bulk part of the index can be extracted from the coefficient of an operator induced by the  D0-particle instanton.

Another instance from field theory concerns Sen's S-duality conjecture \cite{Sen:1993zi} that multiple monopoles in ${\cal N}=4$ super Yang-Mills theory~\footnote{The S-duality conjecture in string theory dates earlier and was first conjectured in~\cite{Rey:1989xj} and~\cite{Font:1990gx}.} form a unique bound-state that forms a unique bound-state that builds the Montonen-Olive \cite{Montonen:1977sn}
duality tower. Again, the relative moduli space is non-compact and the $L^2$-class Witten index is captured by the multi-monopole dynamics on $\mathbb{R}^3 \times \mathbb{S}^{1}_{\beta}$ in the limit $\beta \rightarrow 0$. Once more, it consists of a bulk part and a boundary part. Upon compactifying ${\cal N}=4$ super Yang-Mills theory on $\mathbb{R}^{2,1} \times \mathbb{S}^{1}_{R}$, the monopole circulating around $\mathbb{S}^{1}_{R}$ is interpretable as an instanton in $\mathbb{R}^{2,1}$. It was observed in \cite{Dorey:2000dt} that the bulk part of the index can be extracted from coefficient of an operator induced by the monopole instanton.

In both situations, the $\mathbb{S}^{1}_{R}$-compactification has the effect of converting bulk part of the $L^2$-class Witten index to the coefficient of the instanton-induced operator, while the boundary part of the index is not related to the compactified theory in any obvious way. Let us compare them with the situation at hand: on the one hand, an M-string bound-state wraps around $\mathbb{T}^2$ and behaves as a point like particle configuration on $ \mathbb{R}^4_\parallel$. On the other hand, the m-string lives on $\mathbb{T}^2 \times ( \mathbb{R}^3_\parallel \times \mathbb{S}^{1}_{R})$. We can view an m-string bound-state winding around the $\mathbb{S}^{1}_{R}$ as an Euclidean point like particle circulating around it. Therefore, drawing parallels to the above situations, one would expect that the BPS counting function for m-strings only accounts for the bulk contribution, whereas the BPS counting function for M-strings would contain both bulk and boundary contributions. It is interesting that the two counting problems are related by the NS limit. A seeming difference that the nature of the constituents, as particles (M-string) and instantons (m-string), are reversed compared to the above two examples. What is more important, however, is which constituents live in a space with $\mathbb{S}^{1}_{R}$ compactification and which ones live in space without. In this regard, our situation is essentially the same as the above two examples.

%-------------------------------------------------
\subsubsection{Elliptic genera of m-string moduli spaces}
We now would like to interpret the (refined) $\widehat{F}^{(k_{1}k_{2}\cdots k_{N-1})}$ as regularized elliptic genera for moduli spaces of m-strings with fixed charges. More precisely, we denote by ${\cal M}_{\vec{k}}$ the moduli space of monopoles of charge $\vec{k}=(k_{1},k_{2},\cdots,k_{N-1})$ and by $\widehat{\cal M}_{\vec{k}}$ the relative part of the monopole moduli space. Then, we propose
\begin{align}
J_{k_{1}k_{2}\cdots k_{N-1}}(\tau,m,0)=\phi_{\widehat{\cal M}_{\vec{k}}}(\tau,m)\,,&&\text{for} &&\mbox{gcd}(k_{1},\cdots,k_{N-1})=1\, ,
\end{align}
where the function $J_{k_{1}k_{2}\cdots k_{N-1}}(\tau,m,\epsilon_1)$ was defined in (\ref{jfunction}).

From (\ref{Jtrans}), it follows that
\begin{itemize}
\item $\phi_{\widehat{\cal M}_{\vec{k}}}(\tau,m)$ has zero weight under transformations with respect to full $SL(2,\mathbb{Z})$
\item the index of $\phi_{\widehat{\cal M}_{\vec{k}}}(\tau,m)$ is $K=(\sum_{a=1}^{N-1}k_{a})-1={1 \over 2} \mbox{dim}_{\mathbb{C}}\widehat{\cal M}_{\vec{k}}$.
\end{itemize}
We then expect that
\begin{align}
\phi_{\widehat{\cal M}_{\vec{k}}}(\tau,m,\epsilon_1)=J_{k_{1}k_{2}\cdots k_{N-1}}(\tau,m,\epsilon_1)\qquad \text{for} \qquad \mbox{gcd}(k_{1},\cdots,k_{N-1})=1\,,\label{PropEllGenRelMonopole}
\end{align}
is the regularized elliptic genus obtained by the insertion of $U(1)$ current  corresponding to the $U(1)$ symmetry with parameter $\epsilon_{1}$.

On the other hand, for $\mbox{gcd}(k_{a})>1$, $J_{k_{1}k_{2}\cdots k_{N-1}}(\tau,m,\epsilon_1)$ transforms covariantly not under the full $SL(2, \mathbb{Z})$ but only under a subgroup of $SL(2,\mathbb{Z})$. Therefore, we would expect that it only captures the universal (regularization independent) bulk part of the elliptic genus of the corresponding m-string moduli space. To restore covariance under the full $SL(2, \mathbb{Z})$, as discussed in subsection 5.1, we would need to add regularization-specific, boundary contribution coming from boundary contribution of delocalized states. Below, we will see this explicitly for the case of charge 2.

%%%%%%%%%%%%%%%%%%%%%%%%%%%%%%%%%
\subsection{Charge $(1,1,\cdots,1)$ Configurations}

Let us look at the simplest configuration with all distinct magnetic charges  equal to $1$ .
%-------------------------------------------------------
\subsubsection{$\widetilde{F}^{(1)}$ and $\mathbb{R}^3\times \mathbb{S}^{1}$ elliptic genus}
The moduli space of charge $1$ m-string in $SU(2)$ gauge group is given by $\mathbb{R}^3\times \mathbb{S}^{1}$. This factor is common in all m-string moduli spaces. So, to get the elliptic genus of the relative m-string moduli space, we quotient by the elliptic genus of this common factor. In the NS limit, we get
\bea
\lim_{\epsilon_2\mapsto 0}\epsilon_{2}\,\widetilde{F}^{(1)}(\tau,m,\epsilon_1,\epsilon_2)=\frac{\theta_{1}(\tau,m+\tfrac{\epsilon_1}{2})\theta_{1}(\tau,m-\tfrac{\epsilon_1}{2})}{\theta_1 (\tau,\epsilon_1)\eta(\tau)^3}.
\eea
As mentioned in \cite{Haghighat:2015coa}, the factor $\theta_{1}(\tau,\epsilon_1)\,\eta(\tau)$ in the denominator corresponds to four bosonic modes in which two of them are charged with charge $\pm \epsilon_{1}$. The remaining factor corresponds to the four fermionic zero modes. The left hand side above is the elliptic genus obtained after dividing by the volume of the transverse $\mathbb{R}^3$. Due to this regularization, the weight of the left hand side in the  equation above is $-1$.

%------------------------------------------------
\subsubsection{$\widetilde{F}^{(1,1)}$ and Taub-NUT elliptic genus}

The relative moduli space for the charge $(1,1)$ m-string in $SU(3)$ gauge group is the four-dimensional Taub-NUT space. The elliptic genus of the Taub-NUT space was calculated in \cite{Harvey:2014nha} and its dependence on the size of the asymptotic circle was studied in detail. The universal part of the elliptic genus of  the Taub-NUT space, which does not depend on the size of the Taub-NUT circle was shown to be
\begin{align}
\phi_{\widehat{M}_{1,1}}(\tau,m,\epsilon_{1})& :=\int_{0}^{1}\frac{\theta_{1}(\tau,m+\gamma)\theta_{1}(\tau,m-\gamma)}{\theta_{1}(\tau,\frac{\epsilon_{1}}{2}+\gamma)\theta_{1}(\tau,\frac{\epsilon_{1}}{2}-\gamma)}\,\nonumber\\
& \ =1+A_{1}(\tau,m,\epsilon_{1})Q_{\tau}+A_{2}(\tau,m,\epsilon_1)Q_{\tau}^2+A_{3}(\tau,m,\epsilon_1)Q_{\tau}^3+A_{4}(\tau,m,\epsilon_1)Q_{\tau}^4\cdots,
\label{tneg}
\end{align}
where
\bea
A_{1}(\tau,m,\epsilon_1)&=&q^{-1}(1-Q_{m}\sqrt{q})^2(1-Q_{m}^{-1}\sqrt{q})^2\,,\\\nonumber A_{2}(\tau,m,\epsilon_1)&=&(1-Q_{m}\sqrt{q})^2(1-Q_{m}^{-1}\sqrt{q})^2(1+4q^{-1}+q^{-2})Q_{\tau}^2\,,\\\nonumber
A_{3}(\tau,m,\epsilon_1)&=&(1-Q_{m}\sqrt{q})^2(1-Q_{m}^{-1}\sqrt{q})^2
\Big[(q+4+10q^{-1}+4q^{-2}+q^{-6})\\\nonumber
&&-2(Q_{m}+Q_{m}^{-1})(q^{-\frac{1}{2}}+q^{-\frac{3}{2}})\Big]\\\nonumber
A_{4}(\tau,m,\epsilon_1)&=&(1-Q_{m}\sqrt{q})^2(1-Q_{m}^{-1}\sqrt{q})^2\Big[q^2+4q+14+28q^{-1}+14q^{-2}+4q^{-3}+q^{-4}\\\nonumber &&-2(Q_{m}+Q_{m}^{-1})(q^{\frac{1}{2}}+4q^{-\frac{1}{2}}+4q^{-\frac{3}{2}}+q^{-\frac{5}{2}})+q^{-1}(Q_{m}^2+Q_{m}^{-2})\Big]
\eea
etc. In the genus-zero limit $\epsilon_{1}\mapsto 0$,  we can write the above as
\bea
\phi_{{\widehat{\cal M}}_{1,1}}(\tau,m,0)&=&\frac{\theta^{''}_{1}(\tau,m)\theta_{1}(\tau,m)-\theta^{'}_{1}(\tau,m)^2}{\eta(\tau)^6}\\\nonumber
&=&\phi_{-2,1}(\tau,m)\Big[\frac{\theta^{''}_{1}(\tau,m)}{\theta_{1}(\tau,m)}-\frac{\theta^{'}_{1}(\tau,m)^2}{\theta_{1}(\tau,m)^2}\Big].
\eea
Recall that, in section (4.2.2), we studied the M-string configuration $(1,1)$ and obtained
\bea
\widetilde{F}^{(1,1)}(\tau,m,\epsilon_1,\epsilon_2)=\widetilde{F}^{(1)}\,W(\tau,m,\epsilon_{1},\epsilon_2)\,,
\eea
where
\bea \label{ww}
W(\tau,m,\epsilon_1,\epsilon_2)=\frac{\theta_{1}(\tau,m+\epsilon_{+})\theta(\tau,m-\epsilon_{+})-\theta_{1}(\tau,m+\epsilon_{-})\theta_{1}(\tau,m-\epsilon_{-})}{\theta_{1}(\tau,\epsilon_1)\theta_{1}(\tau,\epsilon_2)}
\eea
It is straightforward to show that, in the limit $\epsilon_{1,2}\mapsto 0$, this is reduced to
\bea
W(\tau,m,0,0)=\frac{\theta^{''}_{1}(\tau,m)\theta_{1}(\tau,m)-\theta^{'}_{1}(\tau,m)^2}{\eta(\tau)^6}
\eea
and therefore
\bea
\phi_{{\widehat{\cal M}}_{1,1}}(\tau,m,0)=W(\tau,m,0,0).
\eea
While not evident from (\ref{tneg}) and (\ref{ww}), one can check that~\footnote{We have checked this up to order $Q_{\tau}^{10}$.}
\bea
\phi_{{\widehat{\cal M}}_{1,1}}(\tau,m,\epsilon_1)=\lim_{\epsilon_{2}\mapsto 0}W(\tau,m,\epsilon_1,\epsilon_2)=J_{1\,1}(\tau,m,\epsilon_1).\label{LimitW}
\eea
We thus confirm that the NS limit relates the M-string free energies to the elliptic genus of m-string moduli space, which in this case is the Taub-NUT space.

%%%%%%%%%%%%%%%%%%%%%%%
\subsubsection{$\widetilde{F}^{(1,1,\cdots,1)}$, bound-states of fundamental monopoles and Sen's S-duality}

Consider the gauge group $SU(N)$. The charge $(1,1,\cdots,1)$ monopole is the bound-state of $(N-1)$ distinct fundamental monopoles, which is S-dual to the bound-state of $(N-1)$ distinct W-bosons. In this case, we have
\bea
\phi_{\widehat{\cal M}_{1,1,\cdots,1}}(\tau,m,\epsilon_1)=J_{1\,1\cdots 1}(\tau,m,\epsilon_1)=J_{1\,1}(\tau,m,\epsilon_1)^{N-2}\, .
\eea
%where
%\bea
%J_{1\,1}(\tau,m,\epsilon_1)=\lim_{\epsilon_2\mapsto 0} (W(\tau,m,\epsilon_1,\epsilon_2))^{N-2}\,.
%\eea

Let us take the limit $\tau\mapsto i\infty$. In this limit, the elliptic genus is reduced to the $\chi_{y}$-genus, which is just to take the leading part of the $Q_\tau$ expansion. In this limit, it also follows that $W\mapsto 1$. Therefore, we find that $\chi_y$-genus is given by
\bea
\chi_{y}(\widehat{\cal M}_{1,1,\cdots,1})=1\, .
\eea
This then implies that
\bea
\sum_{q}(-1)^{q}\ \mbox{dim} H^{p,q}(\widehat{\cal M}_{1,1,\cdots,1})=
\begin{cases}
0\,&\text{for}\,\,\,\,p\neq  \frac{1}{2} \mbox{dim}_{\mathbb{C}}\widehat{\cal M}_{1,1,\cdots,1}\\
1& \text{for}\,\,\,\,p= \frac{1}{2} \mbox{dim}_{\mathbb{C}}\widehat{\cal M}_{1,1,\cdots,1}\,.
\end{cases}
\eea
We thus proved higher-rank generalization of the Sen's S-duality conjecture \cite{Sen:1993zi}
from the regularized elliptic genus, starting from the M-string  free energies and then taking the
NS limit.

%%%%%%%%%%%%%%%%%%%%%%
\subsection{$\widetilde{F}^{(2)}$ and Atiyah-Hitchin Elliptic Genus}

For the charge $(2)$ m-string in a setting with $N=2$ M5-branes, the relative part of the moduli space is the four-dimensional Atiyah-Hitchin space. In \cite{Bak:2014xwa}, the contribution of  bulk contribution from localized states to the elliptic genus of the Atiyah-Hitchin space was derived directly from the path integral over the Atiyah-Hitchin space. It takes the form
\bea
\phi_{\rm AH}(\tau,m,\mu)=\frac{1}{2}\Big[\frac{\theta_{3}(\tau,m+\mu)\theta_{3}(\tau,m-\mu)}{\theta_{3}(\tau,\mu)^2}+\frac{\theta_{4}(\tau,m+\mu)\theta_{4}(\tau,m-\mu)}{\theta_{4}(\tau,\mu)^2}\Big]\,,
\eea
where $\mu$ is a regularization parameter corresponding to the Cartan of the $SO(3)$ action on the Atiyah-Hitchin space, as discussed in section (5.2). The charge $2$ m-string moduli space has a $\mathbb{Z}_{2}$ grading associated with the parity action with respect to which the elliptic genus can be decomposed into irreducible building blocks \cite{Bak:2014xwa}. The even part of this is the elliptic genus of the moduli space of electrically neutral monopoles of charge $2$. This even part is given by \footnote{Incidentally, we can express it also in the form $\phi_{\rm AH,even}(\tau,m,\mu)=2\phi_{-2,1}(\tau,m)\,\frac{W(2\tau,2\mu+\tau,0,0)}{\phi_{-2,1}(2\tau,2\mu+\tau)}$.}
\begin{align}
\phi_{\rm AH,even}(\tau,m,\mu)&:=2\phi_{-2,1}(\tau,m)\Big[\frac{\theta^{''}_{1}(2\tau,\mu+\tau)}{\theta_{1}(2\tau,\mu+\tau)}-\frac{\theta^{'}_{1}(2\tau,\mu+\tau)^2}{\theta_{1}(2\tau,\mu+\tau)^2}\Big]\,.
\end{align}
It is straightforward to show in the $Q_{\tau}$ expansion that \cite{Bak:2014xwa}
\bea\label{ }
\phi_{\rm AH,even}(\tau,m,0)=J_{2}(\tau,m,0)\,.
\eea
This duality does not extend to non-zero $\mu$. As such, although both $\mu$ and $\epsilon_1$ regularization parameters retain the same Cartan of the SO(3) action on the Atiyah-Hitchin space, the grading provided by $\mu$ for $\phi_{\rm AH,even}(\tau,m,\mu)$ and the grading provided by $\epsilon_{1}$ for $\frac{\widetilde{F}^{(2)}}{\widetilde{F}^{(1)}}$ in the NS limit are different. Nevertheless, curiously, if we expand them in powers of $\mu$ and $\epsilon_{1}$, we found that
\bea
\phi_{\rm AH,even}(\tau,m,\mu)&=&\phi_{\rm AH,even}(\tau,m,0)+\mu^2\,R_{1}(\tau,m,0)+\cdots\\\nonumber
J_{2}(\tau,m,\epsilon_1)&=&J_2(\tau, m, 0)+\epsilon_{1}^2\,K_{1}(\tau,m)+\cdots
\eea
where it also turned out $R_{1}(\tau,0)=K_{1}(\tau,0)$. This leads us to conclude that perhaps
the duality in (\ref{ }) extends to non-zero $\mu$ but with the regularization parameters corresponding to the action of various $U(1)$'s  on both sides identified in some non-trivial way.

%%%%%%%%%%%%%%%%%%%%%%%%%
\subsection{ $\chi_{y}(\widehat{\cal M}_{k_1,k_2,\cdots,k_{N-1}})$ Genus from $\widetilde{F}^{(k_1,k_2,\cdots,k_{N-1})}$}
For arbitrary charge $(k_1,k_2,\cdots,k_{N-1})$, we found that the function $\widehat{F}^{(k_1,\cdots,k_{N-1})}$ vanishes in the limit $Q_{\tau}\mapsto 0$ if any of the $k_{i}>1$. From (\ref{PropEllGenRelMonopole}), it follows that for $\mbox{gcd}(k_1,\cdots,k_{N-1})=1$ and some $k_i>1$ the $\chi_y$ genus is given by
\bea
\chi_{y}(\widehat{\cal M}_{k_1,\cdots,k_{N-1}})=0\,,
\eea
Recalling the definition of the $\chi_{y}$ genus, this yields
\bea
\sum_{q}(-1)^{q} \ \mbox{dim} H^{p,q}(\widehat{\cal M}_{k_1,k_2,\cdots,k_{N-1}})=0 \quad \text{for \ all}\quad p \,.
\eea

\section{M5-brane Ensemble and Holomorphic Jacobi Forms}\label{Sect:GrandCanonical}
%%%%%%%%%%%%%%%%%%%%%%%%%%%%%%%%%%%%%%%%%%%%%%%%%%%%%%%%%

The free energies $\F^{(k_{1}k_{2}\cdots k_{N-1})}$ we discussed in the previous sections behave very similar to multi-variable Jacobi forms under transformations with respect to congruent subgroups of $SL(2,\mathbb{Z})$. In the last section, we saw that the NS limit of these free energies is related to the elliptic genera of m-string moduli spaces. If we further take the genus-zero limit $\epsilon_{1}\mapsto 0$, then we are considering the genus-zero part of the free energy, which suffers from the so-called {\it modular anomaly}. We explained that they can be made into covariant objects by using the $\widehat{E}_2 (\tau, \overline{\tau})$ function at the expense of rendering them non-holomorphic functions. In the following section, we will however show that there exist unique linear combinations of various $\F^{(k_i)}$ (in the genus-zero limit $\epsilon_1=-\epsilon_2=0$) which are holomorphic and Jacobi forms of a particular congruence subgroup of $SL(2,\mathbb{Z})$. In other words, the modular anomaly cancels out in these linear combinations, which are unique, all the while retaining the holomorphy as well.

%%%%%%%%%%%%%%%%%%%%%%%
\subsection{What Is Special of  Equal K\"ahler Parameters?}

Before explaining the details of this observation, we would like to point out that, in general, linear combinations of different free energies $\F^{(k_i)}$  do not make sense. Firstly, although $K = \sum_i k_i$ is held fixed, different m-string configurations necessitate different number of M5-branes and hence different gauge groups. So, roughly speaking, summing over different free energies amount summing over different rank of the gauge group.     Secondly, these free energies are the coefficients of different monomials of the K\"ahler parameters $Q_{f_a}$, as can be seen from the expansion (\ref{bound}), and hence ought not to be bundled together in any straightforward manner in any sensible BPS state counting. However, at the particular point in the K\"ahler parameter space where
\bea
Q_{f_1} = Q_{f_2} = \cdots := Q_f\,,\label{IdentModuli}
\eea
it is meaningful to consider a linear combination of all possible $\F^{(k_i)}$ of fixed $K = \sum_{i}k_i$. We can view them as m-string configurations in the M5-brane ensemble, in which the number of M5-branes is freely varied or freely adjusted to fit to the m-string configurations of fixed $K$. This is the prescription we shall consider hereafter. Here we explain why (\ref{IdentModuli}) is in fact imperative to interpret the $\F^{(k_i)}$ (or their linear combinations) precisely as the elliptic genera of the relative moduli space of m-strings.

The special limit (\ref{IdentModuli}) corresponds to a configuration in which all M5-branes are separated by  equal distances. Furthermore, since all the K\"ahler parameters are  equal, the $\F^{(k_i)}$ only count the total number of M-strings, irrespective of the M5-branes they are attached to. We can gain a very intuitive picture of this setup by first compactifying the $x_6$-direction of the brane configuration on a circle with radius $R_6$ and then take the decompactification limit $R_6\to \infty$ in the end.  On the circle, the M5-branes are spread out at  equal distances. This corresponds to the configuration  (\ref{IdentModuli}).
Due to the compactification, this configuration can be interpreted as the Dynkin diagram of the affine extension $\mathfrak{a}_{N-1}^+$ of the Lie algebra $\mathfrak{a}_{N-1}$ and indeed, the M5-branes can be thought of being dual to Dynkin roots of $\mathfrak{a}^+_{N-1}$. The M-strings are distributed with multiplicities $K = (k_1, k_2, \cdots, k_N)$ associated with these roots. Note that here we consider all configurations of $k_a \ge 0$. The decompactification limit is obtained from removing any one of the Dynkin roots by making the distance between any two adjacent M5-branes infinitely large.  As the M5-branes are symmetrically distributed around the circle,  equivalently, as the distance between two adjacent M5-branes are all  equal according to (\ref{IdentModuli}), we can decompactify democratically any one of the intervals.
Although there are $N$ independent ways of doing this, all of them reproduce the Dynkin diagram of the Lie algebra $\mathfrak{a}_{N-1}$. From the M-strings point of view, we obtain all possible configurations over the remaining $(N-1)$ intervals (up to appropriate Weyl reflections), {\it i.e.} the remaining $(N-1)$ Dynkin nodes. Here, we make no distinction between M-strings at different Dynkin nodes and the only meaningful quantity is the total M-string number. For this arrangement to function as desired, it is necessary to start first with M5-branes as many as the total number of M2-branes under consideration. This then also explains why, after the decompactification, brane configurations with different number of M5-branes are taken all at  equal footings.

Let us now consider this configuration from the point of view of m-strings by studying the simplest non-trivial case: we take $N=3$ with three M5-branes separated by distances $a_{1,2}$ respectively, with a single M2-brane stretched between each of them ({\it i.e.} $K=(1,1)$). The monopole moduli space can be separated into  a center-of-mass and a relative parts,
\bea
{\cal M}_{\rm com} \times {\cal M}_{\rm rel} = \mathbb{R}^4 \times {\cal M}_{TN}\,,\label{ModSpaceRelCenter}
\eea
which represents 2 magnetic monopoles of distinct U(1) charges \cite{taub-nut}. We are interested in their electric charge excitations, corresponding to putting F1 strings $(n_1, n_2)$ on top of the M2-branes~\footnote{These F1 strings are additional M2-branes stretched along another orthogonal direction.} .The F1 string charge is quantized in the Dynkin basis discussed above, and should be interpreted as "momentum" for rotational excitations around the $\mathbb{S}^1$ part of the moduli space. However, from the viewpoint of (\ref{ModSpaceRelCenter}), we expect the interpretation to be more subtle, since the Taub-NUT space is a non trivially curved manifold, {\it i.e.} its sigma model is an interacting two-dimensional conformal field theory. Indeed, the $(n_1, n_2)$ are quantized F1 string charges and hence correspond to momenta conjugate to the $\mathbb{S}^1$'s of a single monopole moduli space $\mathbb{R}^3 \times \mathbb{S}^1$. The corresponding Hamiltonian is given by
\bea
{\cal H}
&=& a_1 \sqrt{g^{-2} + n_1^2} + a_2 \sqrt{g^{-2} + n_2^2} + E_{\rm int}(n_1, n_2) \nn
&=& \left[ {a_1 \over g} + {a_2 \over g} \right] + \left[{1 \over 2} (g a_1) n_1^2 + {1 \over 2} (g a_2) n_2^2 \right] + E_{\rm int}(n_1 - n_2) + \cdots
\eea
The first bracket is the sum of two monopole masses, while the second bracket is the kinetic energy of electric charge excitations, where $g a_1 = m_{W_1}$ and $g a_2 = m_{W_2}$ are the W-boson masses for two independent Cartan subalgebras. Note that, modulo the gauge coupling constant $g$, they are proportional to the M5-brane separations $(a_1,a_2)$ . The interaction energy between the two M2-branes depends only on the relative orientation of F1-strings attached to the middle M5-brane. This explains the dependence of $E_{\rm int}$ on $(n_1 - n_2)$.

The key idea is now that the electric charge excitations cannot be separated into a center of mass and a relative motion component, {\it unless} we set the masses of the two distinct W-bosons to be  equal.
To see this, let us quantize the charge excitations. The relevant quantum Hamiltonian is
\bea
H_{\rm total} = {1 \over 2} m_{W_1} n_1^2 + {1 \over 2} m_{W_2} n_2^2 + H_{\rm rel}(n_1 - n_2)\,,\label{HtotHam}
\eea
where $n_1, n_2$ are momenta conjugate to $\mathbb{S}^1(\phi_1), \mathbb{S}^1(\phi_2)$ of $(\mathbb{R}^3 \times \mathbb{S}^1)^2$:
\begin{align}
& n_1 := p_{\phi_1} &\text{and}&& n_2 := p_{\phi_2} &&\text{for} && 0 \le \phi_{1,2} \le 2 \pi.
\end{align}
The novel feature of (\ref{HtotHam}) is that the masses $m_{W_1}, m_{W_2}$, not their inverses, appear in front of the squares of the momenta. In order to decompose the Hamiltonian into the center-of-mass and the relative motion part, we define
\begin{align}
N \equiv {m_{W_1} n_1 + m_{W_2} n_2 \over m_{W_1} + m_{W_2}} := P_{\Phi_{\rm COM}}\qquad \mbox{and} \qquad
n \equiv {1 \over 2} (n_1 - n_2) := p_{\varphi_{\rm rel}}\, ,
\end{align}
which satisfy
\begin{align}
n_1 = N + {m_{W_2} \over m_{W_1} + m_{W_2}} n\qquad \mbox{and} \qquad n_2 = N - {m_{W_1} \over m_{W_1} + m_{W_2}} n\,.
\end{align}
In terms of the moduli coordinates of electric charge excitation, we have the relations
\begin{align}
\Phi_{\rm COM} = \phi_1 + \phi_2\qquad \text{and} \qquad
\varphi_{\rm rel} = 2 {m_{W_2} \phi_1 - m_{W_1} \phi_2 \over m_{W_1} + m_{W_2}}
\end{align}
as well as
\begin{align}
\phi_1 = {m_{W_1} \over m_{W_1} + m_{W_2}} \Phi_{\rm COM} + {1 \over 2} \varphi_{\rm rel}\qquad \text{and}
\qquad
\phi_2 = {m_{W_2} \over m_{W_1} + m_{W_2}} \Phi_{\rm COM} - {1 \over 2} \varphi_{\rm rel}\,.
\end{align}
These relations are very different from the standard situation, due to the reason stressed already -- the W-boson masses appear in the numerator of the charge excitation kinetic energies, which also affects the charge lattices $(N, n)$. The moduli coordinates $\phi_1, \phi_2$ take values over $[0, 2 \pi]$. The momenta $n_1, n_2$ conjugate to them are integrally quantized, {\it i.e.} $n_1, n_2 \in \mathbb{Z}$. However, when computing the elliptic genus of the {\it relative} moduli space, we are required to take the decoupling conditions, $N = 0$ and $ n \in \mathbb{Z}$. We now would like to see under what conditions these conditions are satisfied.

Consider first the shift
\begin{align}
\phi_1 \quad \rightarrow \quad \phi_1 + 2 \pi \mathbb{Z}\qquad \mbox{and} \qquad \phi_2 \quad \rightarrow \quad \phi_2 - 2 \pi \mathbb{Z}\,,
\end{align}
which corresponds to
\begin{align}
\Phi_{\rm COM} \quad \rightarrow \quad \Phi_{\rm COM}
\qquad \mbox{and} \qquad
\varphi_{\rm rel} \quad \rightarrow \quad\varphi_{\rm rel} + 4 \pi \mathbb{Z},
\end{align}
under which the spectrum of each individual electric charge excitations is invariant. This implies that the momentum $n$ conjugate to $\varphi_{\rm rel}$ must be $\mathbb{Z}/2$-quantized.

Consider next the situation that we shift
\begin{align}
\phi_1 \quad \rightarrow \quad \phi_1 + 2 \pi \mathbb{Z} \qquad \mbox{and} \qquad \phi_2 \quad \rightarrow \quad \phi_2.
\end{align}
This amounts to
\bea
\Phi_{\rm COM} \quad \rightarrow \quad
\Phi_{\rm COM} + 2 \pi \mathbb{Z} \qquad \mbox{and} \qquad
\varphi_{\rm rel} \quad \rightarrow \quad
\varphi_{\rm rel} + 4 \pi {m_{W_2} \over m_{W_1} + m_{W_2}} \mathbb{Z}.
\eea
Therefore, the moduli space is not quite factorized. The charge excitation part is given by
\bea
{\cal M}_{\rm charge} = {[\mathbb{R}_{\rm COM} \times \mathbb{S}^1({\rm Taub-NUT})] / 2 \pi \mathbb{Z}},
\eea
and we see that the decomposition is problematic. For generic $m_{W_1}, m_{W_2}$ we require
\begin{align}
0 = N= m_{W_1} n_1 + m_{W_2} n_2\qquad \mbox{and} \qquad
n = {1 \over 2} (n_1 - n_2) \in\mathbb{Z}
\end{align}
These conditions cannot be satisfied for generic $m_{W_1}, m_{W_2}$ since%
\begin{align}
n_1 = 2{m_{W_2} \over m_{W_1} + m_{W_2}}\, \mathbb{Z}\qquad \text{and}
\qquad n_2 = 2{m_{W_1} \over m_{W_1} + m_{W_2}}\,\mathbb{Z}\,.
\end{align}
They are integer-valued only for $m_{W_1} = m_{W_2} \ne 0$~\footnote{The possibility $m_{W_1} = 0$ or $m_{W_2}$ would imply that gauge symmetry is restored and the m-strings are replaced by magnetic charge cloud.}. The upshot of this intuitive analysis is that, in order to be able to interpret the counting functions $\F^{(k_i)}$ in terms of elliptic genera of the relative moduli spaces of m-strings, we are forced to take $m_{W_1}=m_{W_2}$, which corresponds to configurations in which the M5-branes are separated by  equal distances. But then, by the argument given at the beginning of this section, one needs to sum over all possible configurations of m-strings in so far as they all have the same value of $K = \sum_i k_i$.

%%%%%
\subsection{Explicit Examples}\label{Sect:ExforT}
It now remains to identify the pertinent M5-brane ensembles once a total number $K = \sum_i k_i$ of M-string is given. In this subsection, we will present the unique combinations which lead to holomorphic Jacobi forms in the genus-zero limit. We tabulate $\widehat{F}^{(k_1,\ldots, k_{N-1})}(\tau,m)$ ordered by their index $K=\sum_{i}k_{i}$.
\subsubsection{Index $K=1$}
In the configuration of index $K=1$, there is only a single $\widehat{F}^{(k_1,\ldots, k_{N-1})}(\tau,m)$
\begin{align}
\widehat{F}^{(1)}(\tau,m)=\varphi_{-2,1}(\tau,Q_m)\,,\label{ResF1}
\end{align}
which indeed is a Jacobi form of weight $w=-2$ and index $1$ under the full group $SL(2,\mathbb{Z})$. In this case, we do not encounter an anomaly.
The Fourier expansion of $\widehat{F}^{(1)}$ is given by
{\allowdisplaybreaks\begin{align}
&\widehat{F}^{(1)}=\sum_{n=0}^\infty\sum_{\ell\in\mathbb{Z}}c^{(1)}(n,\ell) Q_\tau^n Q_m^\ell=\sum_{n=0}^\infty\sum_{\ell\in\mathbb{Z}}c^{(1)}(4n-\ell^2)\,Q_\tau^nQ_m^\ell\nonumber\\
&=2-Q_m -\frac{1}{Q_m }+Q_\tau \left(2Q_m ^2+\frac{2}{Q_m ^2}-8 Q_m -\frac{8}{Q_m }+12\right)\nonumber\\
&\hspace{0.3cm}+Q_\tau^2 \left(-Q_m ^3-\frac{1}{Q_m ^3}+12Q_m ^2+\frac{12}{Q_m ^2}-39 Q_m -\frac{39}{Q_m }+56\right)\nonumber\\
&\hspace{0.3cm}-Q_\tau^3 \left(-8 Q_m ^3-\frac{8}{Q_m ^3Q_m }+56 Q_m ^2+\frac{56}{Q_m ^2}-152Q_m -\frac{152}{Q_m }+208\right)\nonumber\\
&\hspace{0.3cm}+Q_\tau^4 \left(2 Q_m ^4+\frac{2}{Q_m ^4}-39 Q_m ^3-\frac{39}{Q_m ^3}+208Q_m ^2+\frac{208}{Q_m ^2}-513 Q_m -\frac{513}{Q_m }+684\right)\nonumber\\
&\hspace{0.3cm}+\mathcal{O}(Q_\tau^5)\label{FourierF1}
\end{align}}

As for the theta-function decomposition (\ref{ThetaExpansionU21}), the functions $H_{0,1}$ defined in (\ref{DefH01}) behave in the following way in the genus-zero limit $\epsilon_1,\epsilon_2\to 0$:
\begin{align}
\lim_{\epsilon_{1,2}\mapsto 0}\epsilon_{1}\epsilon_{2}\,H_{0}(\tau,\epsilon_{1},\epsilon_{2})&=2+12Q_{\tau}+56Q_{\tau}^2+\cdots=-\sum_{m=0}^{\infty}c^{(1)}(4m)Q_\tau^{m}\nonumber\\
\lim_{\epsilon_{1,2}\mapsto 0}\epsilon_{1}\epsilon_{2}\,H_{1}(\tau,\epsilon_{1},\epsilon_{2})&=-1-8\,Q_{\tau}-39\,Q_{\tau}^2+\cdots
=-Q_{\tau}^{\frac{1}{4}}\frac{\eta(2\tau)^5}{\eta(\tau)^{8}\eta(4\tau)^2}=-\sum_{m=0}^{\infty}c^{(1)}(4m-1)Q_\tau^{m}
\end{align}

%------------------------------------------------
\subsubsection{Index $K=2$}\label{Sect:Index2}
In the configurations of $K=2$, we have two different $\widehat{F}^{(k_1,\ldots, k_{N-1})}(\tau,m)$
\begin{align}
&\widehat{F}^{(2)}(\tau,m)\,,&&\widehat{F}^{(1,1)}(\tau,m)\,.\label{Index2Functs}
\end{align}
Their explicit forms are given in (\ref{ExplicitIndex2}) in appendix~\ref{App:Index2}.

Concerning their modular properties of (\ref{Index2Functs}), we stress that both $\widehat{F}^{(2)}(\tau,m)$ and $\widehat{F}^{(1,1)}(\tau,m)$ are holomorphic, however, suffer from an anomaly under modular transformations.\footnote{As we already remarked, in both cases, this anomaly can be removed by the replacement (\ref{E2hat}), at the cost of turning $\hat{F}^{(2)}(\tau,m)$ and $\hat{F}^{(1,1)}(\tau,m)$ into quasi-holomorphic objects.} However, we found that there is a unique combination of these two objects, for which the anomaly cancels. Indeed, upon forming the sum
\begin{align}
T^{(2)}(\tau,m)=\widehat{F}^{(2)}(\tau,m)+\widehat{F}^{(1,1)}(\tau,m)=\frac{\varphi_{-2,1}}{12}\left[\varphi_{0,1}-\left(E_2(\tau)-2E_2(2\tau)\right)\varphi_{-2,1}\right]\,,\label{T2holfunc}
\end{align}
we notice that the Eisenstein series $E_2$ only appear in the combination $E_2(\tau)-2E_2(2\tau)$, which is the particular case $N=2$ of the generalized Eisenstein series introduced in (\ref{GeneralisedEisensteinDef})
\begin{align}
\psi^{(2)}(\tau)=E_2(\tau)-2E_2(2\tau)\,.
\end{align}
This transforms covariantly under the congruence subgroup $\Gamma_0(2)$.\footnote{More precisely, $\psi^{(2)}(\tau)$ is a holomorphic function which transforms with weight $2$ under $\Gamma_0(2)$.} Therefore, $T^{(2)}$ in (\ref{T2holfunc}) is a (holomorphic) Jacobi form of weight $-2$ and index $2$ under $\Gamma_0(2)$. We also remark that (\ref{T2holfunc}) can also be written as
\begin{align}
T^{(2)}(\tau,m)=\frac{\varphi_{-2,1}}{2}\left[\left(\frac{\theta_3(\tau,m)}{\theta_3(\tau,0)}\right)^2+\left(\frac{\theta_4(\tau,m)}{\theta_4(\tau,0)}\right)^2\right]\,.
\end{align}

We now display another interesting property of $T^{(2)}$. Comparing the Fourier expansion
\begin{align}
&T^{(2)}(\tau,Q_m)=\sum_{n=0}^\infty\sum_{\ell}c^{(2)}(n,\ell) Q_\tau^n Q_m^\ell\nonumber\\
&=2-Q_m -\frac{1}{Q_m }+Q_\tau\left(-Q_m ^3-\frac{1}{Q_m ^3}+12 Q_m ^2+\frac{12}{Q_m ^2}-39Q_m -\frac{39}{Q_m }+56\right)\nonumber\\
&+Q_\tau^2 \left(2 Q_m ^4+\frac{2}{Q_m ^4}-39 Q_m ^3-\frac{39}{Q_m ^3}+208
   Q_m ^2+\frac{208}{Q_m ^2}-513 Q_m -\frac{513}{Q_m }+684\right)+\mathcal{O}(Q_\tau^3)
\end{align}
with (\ref{FourierF1}), we notice that
\begin{align}
&c^{(2)}(n,\ell)=c^{(1)}(2n,\ell) \qquad \mbox{for} \quad n\in\mathbb{N}\quad \mbox{and} \quad \forall \ell\in\mathbb{Z}\,.
\end{align}
This means all the information encoded in $T^{(2)}$ can already be extracted from $T^{(1)}$.

%-------------------------------------------------
\subsubsection{Index $K=3$}\label{Sect:10Index3}
For the configurations of $K=3$, there are three different $\widehat{F}^{(k_1,\ldots, k_{N-1})}(\tau,m)$:
\begin{align}
&\widehat{F}^{(3)}(\tau,m)\,,&&\widehat{F}^{(2,1)}(\tau,m)\,,&&\widehat{F}^{(1,1,1)}(\tau,m)\,,
\end{align}
where we used $\widehat{F}^{(2,1)}(\tau,m)=\widehat{F}^{(1,2)}(\tau,m)$. The explicit expressions are written in (\ref{Fsindex3}) in appendix~\ref{App:Index3}. Each of these functions suffers from a modular anomaly. However, we would expect that there are again possible combinations for which the anomalies cancel out. We will now show that there is indeed (up to overall normalization) a unique such combination. To this end, we replace each $E_2(n)$ for $n>1$ in (\ref{Fsindex3}) by
\begin{align}
E_2(n)=\frac{E_2(1)-\psi^{(n)}}{n} \quad \mbox{for all} \quad  n>1\,,\label{E2Replace}
\end{align}
and form the combination
\begin{align}
&a_1\,\widehat{F}^{(1,1,1)}(\tau,m)+ a_2\,\widehat{F}^{(2,1)}(\tau,m)+a_3\, \widehat{F}^{(3)}(\tau,m)=\frac{\varphi _{-2,1}}{2880} \nonumber \\
&\times \left[20 a_1 (\varphi _{0,1})^2+2 (\varphi _{-2,1})^2 \left(15 a_2\, E_4(1) +7 a_3\,E_4(1) -27a_3\, E_4(3) \right)-20 a_3\, \psi^{(3)}
   \varphi _{0,1} \varphi _{-2,1}\right]
\nonumber\\
&+\frac{a_1-a_3}{72} E_2(1) (\varphi _{-2,1})^2 \varphi _{0,1} +\frac{2 a_1-3 a_2+4 a_3}{288} E_2(1)^2 (\varphi_{-2,1})^3\label{T3Expand}
\end{align}
for some numerical coefficients $a_{1,2,3}$. The only source of anomaly in this expression are the $E_2(1)$ in the last line~\footnote{The first line in (\ref{T3Expand}) only contains holomorphic modular forms, which are also anomaly-free.}.  Since the two terms are linearly independent, in order for the anomalies to cancel, we have to impose
\begin{align}
a_1-a_3=0 \qquad \mbox{and} \qquad 2a_1-3a_2+4a_3=0\,.
\end{align}
The solution is $a_2=2a_1$ and $a_3=a_1$. Therefore, up to a overall normalization, the unique anomaly-free combination is
\begin{align}
T^{(3)}&=\widehat{F}^{(3)}(\tau,m)+2\,\widehat{F}^{(2,1)}(\tau,m)+\widehat{F}^{(1,1,1)}(\tau,m)\nonumber\\
&=\frac{1}{2880} \left( \varphi _{-2,1} \left[2 \left(37 E_4(1)-27 E_4(3)\right)\, (\varphi _{-2,1})^2-20 \psi^{(3)}\, \varphi _{0,1}\, \varphi _{-2,1}+20 \,(\varphi _{0,1})^2\right]\right) \, .\label{T3holfunc}
\end{align}
This is a (holomorphic) Jacobi form of weight $-2$ and index $3$ under $\Gamma_0(3)$.

We now analyze the Fourier expansion of $T^{(3)}$, along with the first few terms
\begin{align}
T^{(3)}&=\sum_{n=0}^\infty\sum_{\ell}c^{(3)}(n,\ell) Q_\tau^n Q_m^\ell\nonumber\\
&=\left(2-Q_m-\frac{1}{Q_m }\right) +Q_\tau \left(-8 Q_m ^3-\frac{8}{Q_m ^3}+56 Q_m ^2+\frac{56}{Q_m ^2}-152
   Q_m -\frac{152}{Q_m }+208\right)
\nonumber \\
&+\mathcal{O}(Q_\tau^2)\, .
\end{align}
Comparing the coefficients $c^{(3)}$ with (\ref{FourierF1}), we find the relation
\begin{align}
&c^{(3)}(n,\ell)=c^{(1)}(3n,\ell) \qquad \mbox{for all} \qquad  n\in\mathbb{N}\quad \mbox{and} \quad \forall \ell\in\mathbb{Z}\, .
\end{align}
This again indicates that $T^{(3)}$ can be fully reconstructed from $T^{(1)}$.
%------------------------------------------------------
\subsubsection{Index $K=4$}
For the configurations of $K=4$, we have six distinct $\widehat{F}^{(k_1,\ldots, k_{N-1})}(\tau,m)$ with $\sum_{i}k_i=4$:
\begin{align}
\widehat{F}^{(1,1,1,1)}\,,&&\widehat{F}^{(2,1,1)}\,,&&\widehat{F}^{(1,2,1)}\,,&&\widehat{F}^{(3,1)}\,,&&\widehat{F}^{(2,2)}\,,&&\widehat{F}^{(4)}\,,
\end{align}
where we have already made use of relations of the form $\widehat{F}^{(3,1)}(\tau,m)=\widehat{F}^{(1,3)}(\tau,m)$, etc. The explicit expressions are given in (\ref{Fsindex4}) in appendix~\ref{App:Index4}. Each of these functions suffers from a modular anomaly, however, we expect that there are again possible combinations for which the latter cancel out.

Following a strategy parallel to subsection~\ref{Sect:10Index3}, we consider the most general linear combination of these six functions
{\allowdisplaybreaks
\begin{align}
& a_1\widehat{F}^{(1,1,1,1)}+a_2\,\widehat{F}^{(2,1,1)}+a_3\,\widehat{F}^{(1,2,1)}+a_4\,\widehat{F}^{(3,1)}+a_5\,\widehat{F}^{(2,2)}+a_6\,\widehat{F}^{(4)} =-\frac{\varphi _{-2,1}}{1451520} \nonumber \\
&\times \big[\! -840 a_1 \varphi _{0,1}^3 \! -84 \varphi _{0,1} \varphi _{-2,1}^2 \big((15 a_2+25 a_4+18 a_5 +28a_6) E_4(1)-48 (a_5+a_6) E_4(2)\big)\nonumber\\
&\hspace{0.5cm}+8 \varphi _{-2,1}^3 \big((420 a_3+280 a_4 +181 a_5+174 a_6)  E_6(1)-608 a_6  E_6(2)-832 a_5  E_6(2)\big)\nonumber\\
&\hspace{0.5cm}+42 \psi^{(2)} \varphi _{-2,1} \big(64 E_4(2)
   \varphi _{-2,1}^2 (a_6 -a_5 )+20 \varphi _{0,1}^2 (2 a_6 +a_5 )\big)+1680 (\psi^{(2)})^2 \varphi _{0,1} \varphi _{-2,1}^2
   (a_6 -a_5 )\big]\nonumber\\
&-\frac{\varphi _{-2,1}^2E_2(1)}{34560}\big[20 \varphi _{0,1}^2 (-3 a_1 +a_5+2 a_6  )+2 \varphi _{-2,1}^2 \big((-15 a_2-60 a_3+15 a_4+8 a_5+52 a_6)E_4(1)\nonumber\\
&\hspace{0.5cm}+32 (a_5-a_6)
   E_4(2)\big)-80 \psi^{(2)} \varphi _{0,1} \varphi _{-2,1} (a_6 -a_5 )\big]\nonumber\\
&+\frac{1}{3456} \left(  E_2(1){}^2 \varphi _{0,1} \varphi _{-2,1}^3 (6 a_1 -3 a_2 -5 a_4 -6
   a_5+16 a_6  ) \right)\nonumber\\
&+\frac{1}{10368} \left( E_2(1){}^3 \varphi _{-2,1}^4 (6 a_1 -9 a_2 -12 a_3 +25 a_4 +6 a_5 -32
   a_6 ) \right) \ . \nonumber
\end{align}}
We have replaced all $E_2(n)$ with $n>1$ by (\ref{E2Replace}). In order to form an anomly-free combination ({\it i.e.} a holomorphic modular form), we need to make sure that all terms proportional to (a power of) $E_2(1)$ vanish. Since $E_4(1)$ and $E_4(2)$ as well as $\varphi_{0,1}$ and $\varphi_{-2,1}$ are linearly independent, we find the following five conditions on the coefficients $a_{i=1,2,3,4,5,6}$
\begin{align}
-3a_1+a_5+2a_6&=0\,,\nonumber\\
-15a_2-60a_3+15a_4+8a_5+52a_6&=0\,,\nonumber\\
a_5-a_6&=0\,,\nonumber\\
6a_1-3a_2-5a_4-6a_5+16a_6&=0\,,\nonumber\\
6a_1-9a_2-12a_3+25a_4+6a_5-32a_6&=0\,.
\end{align}
The solution is
\begin{align}
&a_2=2a_1\,,&&a_3=a_1\,,&&a_4=2a_1\,,&&a_5=a_1\,,&&a_6=a_1\,.
\end{align}
Therefore, modulo overall normalization, we find a unique linear combination of the $\widehat{F}^{(k_1,\ldots, k_{N-1})}(\tau,m)$ with index $4$ which is a holomorphic modular form of $\Gamma_0(2)$ with weight $-2$ and index $4$
\begin{align}
T^{(4)}&=\widehat{F}^{(1,1,1,1)}+2\,\widehat{F}^{(2,1,1)}+\widehat{F}^{(1,2,1)}+2\,\widehat{F}^{(3,1)}+\widehat{F}^{(2,2)}+\widehat{F}^{(4)}\nonumber\\
&=\frac{\varphi _{-2,1}}{483840} \big[40 \left(96 E_6(2)-89 E_6(1)\right) \varphi _{-2,1}^3+84 \left(21 E_4(1)-32 E_4(2)\right) \varphi _{0,1} \varphi_{-2,1}^2\nonumber\\
   &\hspace{2cm}-840 \psi^{(2)} \varphi _{0,1}^2 \varphi _{-2,1}+280 \varphi _{0,1}^3\big]\label{T4holfunc}
\end{align}

Again, comparing the coefficient $c^{(4)}$ in the Fourier expansion
\begin{align}
T^{(4)}&=\sum_{n=0}^\infty\sum_{\ell}c^{(4)}(n,\ell) Q_\tau^n Q_m^\ell \nonumber \\
& = \left( 2-Q_m -\frac{1}{Q_m }\right) \nonumber \\
& +Q_\tau \left(2 Q_m ^4+\frac{2}{Q_m ^4}-39 Q_m ^3-\frac{39}{Q_m ^3}+208 Q_m ^2+\frac{208}{Q_m ^2}-513
   Q_m -\frac{513}{Q_m }+684\right)
\nonumber \\
& +\mathcal{O}(Q_\tau^2),
\end{align}
with (\ref{FourierF1}), we find the relation
\begin{align}
&c^{(4)}(n,\ell)=c^{(1)}(4n,\ell) \qquad \mbox{for all} \qquad n\in\mathbb{N}\quad \mbox{and} \quad \ell\in\mathbb{Z}\, .
\end{align}
This means that $T^{(4)}$ can be fully reconstructed from $T^{(1)}$.

%%------------------------------------------------------
\subsubsection{Index $K=5$}
For the configurations of $K=5$, we have ten distinct $\widehat{F}^{(k_1,\ldots, k_{N-1})}(\tau,m)$ with $\sum_{i}k_i=5$:
\begin{align}
\widehat{F}^{(1,1,1,1,1)}\,,\widehat{F}^{(2,1,1,1)}\,,\widehat{F}^{(1,2,1,1)}\,,\widehat{F}^{(3,1,1)}\,,\widehat{F}^{(1,3,1)}\,,\widehat{F}^{(2,2,1)}\,,\widehat{F}^{(2,1,2)}\,,\widehat{F}^{(4,1)}\,,\widehat{F}^{(3,2)}\,,\widehat{F}^{(5)}\, . \label{Index5Functions}
\end{align}
Here, we have already used relations of the form $\widehat{F}^{(2,1,1,1)}(\tau,m)=\widehat{F}^{(1,1,1,2)}(\tau,m)$, etc. The explicit expressions are given in (\ref{Fsindex5}) in appendix~\ref{App:Index5}. In contrast to $K<4$, however, we find additional relations~\footnote{We have checked that these relations are an accident at the genus-zero limit and do not hold for the full ($\epsilon_-$-dependent) $F^{(\{k_i\})} (\tau,m,\epsilon_-)$.} among the functions (\ref{Index5Functions}):
\begin{align}
&3\widehat{F}^{(1,3,1)}+6\widehat{F}^{(2,1,1,1)}=4\widehat{F}^{(2,1,2)}+6\widehat{F}^{(2,2,1)}\,,\nonumber\\
&3\widehat{F}^{(1,3,1)}=6\widehat{F}^{(1,2,1,1)}+16\widehat{F}^{(2,1,2)}\,,\nonumber\\
&\widehat{F}^{(1,3,1)}=20\widehat{F}^{(2,1,2)}+2\widehat{F}^{(2,2,1)}+34\widehat{F}^{(3,1,1)}-36\widehat{F}^{(3,2)}+16\widehat{F}^{(4,1)}\,.\label{IdentitiesIndex5}
\end{align}

Let us analyze the modular properties. Each of the functions (\ref{Index5Functions}) suffers from a modular anomaly.  However, we expect that there are again possible combinations for which the anomaly cancels out. Indeed, following the pattern discussed for $K<5$, we find that the combination
\begin{align}
T^{(5)}=&\widehat{F}^{(1,1,1,1,1)}+2\widehat{F}^{(2,1,1,1)}+2\widehat{F}^{(1,2,1,1)}+2\widehat{F}^{(3,1,1)}+\widehat{F}^{(1,3,1)}+2\widehat{F}^{(2,2,1)}+\widehat{F}^{(2,1,2)}\nonumber\\
&+2\widehat{F}^{(4,1)}+2\widehat{F}^{(3,2)}+\widehat{F}^{(5)}\label{T5holfunc}
\end{align}
is a holomorphic modular form of weight $-2$ and index $5$ of $\Gamma_0(5)$. This combination is unique up to the identities (\ref{IdentitiesIndex5}) and an overall normalization.

From the Fourier expansion of $T^{(5)}$
\begin{align}
T^{(5)}=\sum_{n=0}^\infty\sum_{\ell}c^{(5)}(n,\ell) Q_\tau^n Q_m^\ell \ ,
\end{align}
we again notice the relation
\begin{align}
c^{(5)}(n,\ell)=c^{(1)}(5n,\ell) \qquad \mbox{for all} \qquad  n\in\mathbb{N}\quad \mbox{and} \quad \ell\in\mathbb{Z}\, .
\end{align}
Since $c^{(1)}$ is given by the expansion of $T^{(1)}$ in (\ref{FourierF1}), this relation implies that $T^{(5)}$ is reconstructable entirely from $T^{(1)}$.
%%%%%%%-
%%%%%%%
\subsubsection{Index $K=6$}
For the configurations of $K=6$, we have the following 20 distinct $\widehat{F}^{(k_1,\ldots, k_{N-1})}(\tau,m)$ for which $\sum_{i}k_i=6$:
\begin{align}
&\widehat{F}^{(1,1,1,1,1,1)}\,,&&\widehat{F}^{(2,1,1,1,1)}\,,&&\widehat{F}^{(1,2,1,1,1)}\,,&&\widehat{F}^{(1,1,2,1,1)}\,,&&\widehat{F}^{(3,1,1,1)}\,,&&\widehat{F}^{(1,3,1,1)}\,,&&\widehat{F}^{(2,2,1,1)}\,,\nonumber\\
&\widehat{F}^{(2,1,2,1)}\,,&&\widehat{F}^{(2,1,1,2)}\,,&&\widehat{F}^{(1,2,2,1)}\,,&&\widehat{F}^{(3,2,1)}\,,&&\widehat{F}^{(2,3,1)}\,,&&\widehat{F}^{(3,1,2)}\,,&&\widehat{F}^{(2,2,2)}\,,\nonumber\\
&\widehat{F}^{(4,1,1)}\,,&&\widehat{F}^{(1,4,1)}\,,&&\widehat{F}^{(3,3)}\,,&&\widehat{F}^{(4,2)}\,,&&\widehat{F}^{(5,1)}\,,&&\widehat{F}^{(6)}\,,\label{Index6Functions}
\end{align}
where we have already made use of relations of the form $\widehat{F}^{(2,1,1,1,1)}(\tau,m)=\widehat{F}^{(1,1,1,1,2)}(\tau,m)$, etc. The explicit expressions are given in (\ref{Fsindex6}) in appendix~\ref{App:Index6}. As in the case $K=5$, we find relations among the functions (\ref{Index6Functions})
\begin{align}
&3\widehat{F}^{(1,3,1,1)}+6\widehat{F}^{(2,1,1,1,1)}=4\widehat{F}^{(2,1,1,2)}+6\widehat{F}^{(2,2,1,1)}\,,\nonumber\\
&6\widehat{F}^{(1,2,1,1,1)}+16\widehat{F}^{(2,1,1,2)}=3\widehat{F}^{(1,3,1,1)}\,,\nonumber\\
&\widehat{F}^{(1,1,2,1,1)}=\widehat{F}^{(1,2,1,1,1)}\,,\nonumber\\
&18\widehat{F}^{(1,3,1,1)}+6\widehat{F}^{(1,4,1)}+64\widehat{F}^{(2,1,1,2)}=24\widehat{F}^{(2,1,2,1)}+27\widehat{F}^{(3,2,1)}\,,\nonumber\\
&\widehat{F}^{(2,1,1,2)}+9\widehat{F}^{(2,3,1)}=9\widehat{F}^{(1,3,1,1)}+3\widehat{F}^{(1,4,1)}+6\widehat{F}^{(2,1,2,1)}\,,\nonumber\\
&9\widehat{F}^{(1,3,1,1)}+24\widehat{F}^{(2,1,2,1)}+54\widehat{F}^{(3,1,2)}=6\widehat{F}^{(1,4,1)}+10\widehat{F}^{(2,1,1,2)}\,.\label{IdentitiesIndex6}
\end{align}
As in the previous cases, each individual function in  (\ref{Index6Functions}) suffers from a modular anomaly.  However, repeating the above constructions, we find that the combination
\begin{align}
T^{(6)}=&\widehat{F}^{(1,1,1,1,1,1)}+2\widehat{F}^{(2,1,1,1,1)}+2\widehat{F}^{(1,2,1,1,1)}+\widehat{F}^{(1,1,2,1,1)}+2\widehat{F}^{(3,1,1,1)}+2\widehat{F}^{(1,3,1,1)}\nonumber\\
&+2\widehat{F}^{(2,2,1,1)}+2\widehat{F}^{(2,1,2,1)}+\widehat{F}^{(2,1,1,2)}+\widehat{F}^{(1,2,2,1)}+2\widehat{F}^{(3,2,1)}+2\widehat{F}^{(2,3,1)}+2\widehat{F}^{(3,1,2)}\nonumber\\
&+\widehat{F}^{(2,2,2)}+2\widehat{F}^{(4,1,1)}+\widehat{F}^{(1,4,1)}+\widehat{F}^{(3,3)}+2\widehat{F}^{(4,2)}+2\widehat{F}^{(5,1)}+\widehat{F}^{(6)}\label{T6holfunc}
\end{align}
is a holomorphic modular form of weight $-2$ and index $6$ of $\Gamma_0(6)$. This combination is unique up to the identities (\ref{IdentitiesIndex6}) and an overall rescaling.

From the Fourier expansion of $T^{(6)}$
\begin{align}
T^{(6)}=\sum_{n=0}^\infty\sum_{\ell}c^{(6)}(n,\ell) Q_\tau^n Q_m^\ell \, ,
\end{align}
we also found the relation
\begin{align}
&c^{(6)}(n,\ell)=c^{(1)}(6n,\ell)\, , \qquad \forall n\in\mathbb{N}\quad \mbox{and} \quad \forall \ell\in\mathbb{Z}\,,
\end{align}
where $c^{(1)}$ is again given by the expansion of $T^{(1)}$ in (\ref{FourierF1}). We can reconstruct $T^{(6)}$ entirely from $T^{(1)}$.

 %%%%%%%%%%%%%%%%%
\subsection{Conjecture for the General Structure}

Built upon the emerging patterns we discovered in the previous subsections, we now put forward the following conjecture:\\

\noindent
\begin{tcolorbox}
The unique combination
\begin{align}
&T^{(K)}(\tau,m)=\sum_{\{k_i\},\sum k_i = K} \widehat{F}^{(\{k_i\})}\,,\label{DefTnGen}
\end{align}
%
%%
%\begin{align}
%&T^{(K)}(\tau,m)=\sum_{\{k_i\},\sum k_i=K}a_{\{k_i\}}\,\widehat{F}^{(\{k_i\})}\,,\label{DefTnGen}
%\end{align}
%%
%with the coefficients $a_{\{k_i\}}$:
%%
%\begin{align}
%a_{\{k_i\}}=\left\{\begin{array}{lcl} 1\hspace{0.5cm} & \text{if} & (k_1,k_2,\ldots,k_{N-1})=(k_{(N-1)},\ldots,k_2,k_1) \\ 2\hspace{0.5cm} & \text{if} & (k_1,k_2,\ldots,k_{N-1})\neq  (k_{(N-1)},\ldots,k_2,k_1)  \end{array}\right.
%\end{align}
%%
summed over all possible positive-integer partitions of $K$, can be expressed in terms of Hecke transforms (see (\ref{ExpHeckeExtend}) below for the definition) as
\begin{align}
T^{(K)}(\tau,m)=\sum_{a|K}\frac{\mu(a)}{a^3}\,\mathcal{T}_{\frac{K}{a}}(\varphi_{-2,1}(a\tau,am))\ .
\nonumber
\end{align}
Therefore, they transforms as a weak Jacobi form of index $K$ and weight $-2$ under a congruent subgroup $\Gamma$ of $SL(2, \mathbb{Z})$:
\begin{align}
&T^{(K)}\left(\frac{a\tau+b}{c\tau+d},\frac{m}{c\tau+d}\right)=(c\tau+d)^{-2}\,e^{2\pi i K\,\frac{cm^2}{c\tau+d}}\,T^{(K)}(\tau,m)\,, \nonumber\\
 &T^{(K)}(\tau,m+\ell\tau+\ell')=e^{-2\pi iK(\ell^2\tau+2\ell m)}\,T^{(K)}(\tau,m)
\nonumber \\
%\end{align}
%for
%\begin{align}
&\hskip2cm \mbox{for} \quad \left(\begin{array}{cc}a & b \\ c & d\end{array}\right)\in\Gamma\subset SL(2,\mathbb{Z}) \quad \mbox{and} \quad \ell',\ell\in\mathbb{Z}\,,
\end{align}
\end{tcolorbox}
%\\
We note that the summation in (\ref{DefTnGen}) is over \emph{all} configurations $\{k_i\}$ with $\sum_{i}k_i=K$ in a \emph{democratic} fashion. To reproduce (\ref{T2holfunc}), (\ref{T3holfunc}), (\ref{T4holfunc}), (\ref{T5holfunc}) and (\ref{T6holfunc}), we recall that not all such $\widehat{F}^{\{k_i\}}$ are independent and in particular $\widehat{F}^{(k_1,k_2,\ldots,k_{N-1})}=\widehat{F}^{(k_{N-1},\ldots,k_2,k_1)}$. Furthermore, denote the Fourier expansion of $T^{(K)}$ as
\begin{align}
T^{(K)}(\tau,m)=\sum_{n=0}^\infty\sum_{\ell}c^{(K)}(n,\ell) Q_\tau^n Q_m^\ell \ .
\end{align}
Then we have the relation
\begin{align}
&c^{(K)}(n,\ell)=c^{(1)}(nK,\ell)\, \qquad \mbox{for all} \qquad n\in\mathbb{N}\quad \mbox{and} \quad \ell\in\mathbb{Z}\,,
\end{align}
where the $c^{(1)}$ are given by the expansion of $T^{(1)}$ in (\ref{FourierF1}). This implies that we can express $T^{(K)}$ in terms of $T^{(1)}$ as
\begin{align}
T^{(K)}(\tau,m)=\frac{1}{K}\sum_{r=0}^{K-1}T^{(1)}(\tfrac{\tau+r}{K},m)\,.
\end{align}

The modular transformation properties of $T^{(K)}(\tau,Q_{m})$ can be determined by expressing it in terms of the Hecke transform of $T^{(1)}$. The Hecke transform of a weak Jacobi form $\phi(\tau,m)$ of weight $w$ is defined as
\begin{align}
\mathcal{T}_K(\phi(\tau,m))\equiv K^{w-1}\sum_{{ad=K}\atop{b\,\text{mod}\,d}}{1 \over d^{w}} \,\phi\left(\frac{a\tau+b}{d},am\right)\,.\label{ExpHeckeExtend}
\end{align}
So, $\mathcal{T}_K$ maps a weak Jacobi form of $SL(2,\mathbb{Z})$ of index $r$ into a weak Jacobi form of $SL(2,\mathbb{Z})$ of index $Kr$. In terms of the Hecke transform, $T^{(K)}$ is given by
\bea\label{hecketk}
T^{(K)}(\tau,m)=\sum_{a|K}\frac{\mu(a)}{a^3}\,\mathcal{T}_{\frac{K}{a}}(T^{(1)}(a\tau,am))\,.
\eea
Given a prime factor decomposition
\begin{align}
K=\prod_{i=1}^rp_i^{m_i}\qquad \mbox{where} \qquad
m_i\geq  1\,,
\end{align}
we introduce the congruence subgroup
\begin{align}
\Gamma=\Gamma_0(p)\subset SL(2,\mathbb{Z})\qquad \text{with} \qquad
p=\prod_{i=1}^rp_i\,.\label{DefCongSubgroup}
\end{align}
As $T^{(1)}$ transforms covariantly under $\Gamma_{0}(1)$ and the largest $a$ that occurs in (\ref{hecketk}) is $p$,   $T^{(K)}$ transforms covariantly under $\Gamma_{0}(p)$.

%--------------------------------------------------------
\subsection{$T^{(K)}$ and m-String Moduli Spaces}

In the previous section,  we found that the genus-zero part of the free energy for various m-string configurations can be combined to form holomorphic Jacobi forms that can be expressed in terms of Hecke transforms of $\varphi_{-2,1}(\tau,m)$.

These combinations are not arbitrary. They arise when we consider the grand canonical ensemble summing over the number of M5-branes in the  equal K\"ahler parameter configurations (whose special physical properties were  explained in 6.1)
\bea
\boxed{
{\cal G}(\tau,m,\epsilon_1,\epsilon_2,Q)=1+\sum_{N=2}^{\infty}{\cal Z}_{N}(\tau,m,t,\epsilon_1,\epsilon_2)
} \,,
\eea
where we have taken $t_{f_{a}}=t$ for all $a$ and $Q=e^{-t}$. The free energy associated with ${\cal G}(\tau,m,\epsilon_1,\epsilon_2)$ naturally combines $\widetilde{F}^{(k_{1}k_{2}\cdots k_{N-1})}$ for various $(k_{1},k_{2},\cdots,k_{N-1})$ in exactly such a way that the genus-zero part is a holomorphic Jacobi form as discussed in the last subsection.

Recall that the free energy, after subtracting multi-coverings, is given by
\bea
{\cal F}(\tau,m,\epsilon_1,\epsilon_2,Q)=\sum_{\ell=1}^{\infty}\frac{\mu(\ell)}{\ell}{\cal G}(\ell\tau,\ell\,m,\ell \epsilon_{1},\ell \epsilon_{2},Q^{\ell})\,.
\eea
In terms of ${\cal F}$, we can write $T^{(K)}$ as
\bea
\boxed{
\sum_{K=1}^{\infty}Q^{K}T^{(K)}=\lim_{\epsilon_1, \epsilon_2\mapsto 0}\epsilon_{1}\epsilon_{2}{\cal F}(\tau,m,\epsilon_1,\epsilon_2,Q)} \, ,
\eea
where by $T^{(1)}$ we mean the elliptic genus of $\mathbb{R}^3\times S^{1}$ in the limit $\epsilon_{1}\mapsto 0$. This is not surprising given that $\widetilde{F}^{(1)}$ in the NS limit is the elliptic genus of $\mathbb{R}^3\times S^{1}$. However, what is surprising is that $T^{(2)}$ is also related to the elliptic genus of the Atiyah-Hitchin space.

Recall from the discussion of the last section that the contribution of bulk states to the elliptic genus of the Atiyah-Hitchin space is given by \cite{Bak:2014xwa}
\bea
\phi_{\text{AH}}(\tau,m)&:=&\frac{1}{2}\Big[\left(\frac{\theta_{3}(\tau,m)}{\theta_{3}(\tau,0)}\right)^2+\left(\frac{\theta_{4}(\tau,m)}{\theta_{4}(\tau,0)}\right)^2\Big]\,.
\eea
Note that we refer to the full elliptic genus, not just the even part. It was also observed in \cite{Bak:2014xwa} that the elliptic genus can be decomposed (in our notations) as
\bea
\phi_{\text{AH}}(\tau,m)=
J_{(1,1)}(\tau, m, 0) + J_{(2)}(\tau, m, 0).
%\mbox{lim}_{\epsilon_{1},\epsilon_{2}\mapsto 0}\Big(\frac{\widetilde{F}^{(1,1)}+\widetilde{F}^{(2)}}{\widetilde{F}^{(1)}}\Big)\,.
\eea
Notice that $T^{(2)}$ is precisely the genus-zero limit of $\widetilde{F}^{(1,1)}+\widetilde{F}^{(2)}$ and therefore
\bea
\phi_{\text{AH}}(\tau,m)=\frac{T^{(2)}(\tau,m)}{T^{(1)}(\tau,m)}\,.
\eea
Thus $T^{(2)}$ is the elliptic genus of the magnetic charge-2 m-string for $N=2$.

We believe the above relation is not just a coincident and that higher $T^{(K)}$, being holomorphic Jacobi forms, are also related to higher monopole charge m-string moduli spaces. Indeed, a natural guess would be that they capture the elliptic genus of charge-$K$ m-string moduli spaces for $N=2$. If this holds for any $K$ and $N$, then the $\chi_y$ genus would be
\bea
\chi_{y}(\widehat{\cal M}_{K})=1 \qquad \mbox{for all} \qquad K>1\,.
\eea

Attentive readers might have noticed that the above considerations left out m-string configurations with mixed (i.e. multiple identical plus multiple distinct) magnetic charges for which gcd$(k_1, \cdots, k_{N-1})$ is greater than unity. For those, we have a natural extrapolation of the constructions we have taken so far: build a new class of holomorphic Jacobi forms by taking multiple products of $J_{k_1, \cdots, k_{N-1}}(\tau, m, \epsilon_1)$ functions.
We conjecture that suitable linear combinations of them capture the elliptic genus of m-string moduli space for the situations gcd$(k_1, \cdots, k_{N-1}) > 1$. Since the combinatorics are more involved and since they have further distinguishing features, we will relegate their detailed construction to \cite{toappear}.

%%%%%%%%%%%%%%%%%%%%%%%%%%%%%%%%
\section{Summary and Further Remarks}
In this paper,  we have studied the correspondence between M-strings and m-strings. We proposed that the degeneracies of BPS bound-states of M-strings for certain configurations of M2-branes (denoted as $(k_1,\ldots,k_{N-1})$) capture the regularized elliptic genus of the relative moduli space $\widehat{\mathcal{M}}_{k_1,\ldots,k_{N-1}}$ of m-strings of magnetic charges $(k_1,\ldots,k_{N-1})$. Specifically, we proposed (see  equation (\ref{PropEllGenRelMonopole}))
\begin{align}
\phi_{\widehat{\mathcal{M}}_{k_1,\ldots,k_{N-1}}}(\tau,m,\epsilon_1) = \lim_{\epsilon_2\to 0}\frac{\widetilde{F}^{(k_1,\ldots,k_{N-1})}(\tau,m,\epsilon_1,\epsilon_2)}{\widetilde{F}^{(1)}(\tau,m,\epsilon_1,\epsilon_2)}\qquad \text{for} \qquad \text{gcd}(k_1,\ldots,k_{N-1})=1\,.
\end{align}
The NS limit ($\epsilon_2\to 0$) is crucial in this correspondence, since it restores the {\color{red}requisite} $ISO(2)$ boost isometry of the m-strings in this setup. Furthermore, the parameter $\epsilon_1$, from the point of view of the elliptic genus, corresponds to an  equivariant regularization using an $U(1)$ isometry of the relative moduli space $\widehat{\mathcal{M}}_{k_1,\ldots,k_{N-1}}$. In the simplest non-trivial case, corresponding to the charge configuration $(1,1)$, the relative moduli space $\widehat{\mathcal{M}}_{1,1}$ is the Taub NUT space. Its elliptic genus was recently computed in \cite{Harvey:2014nha} and the universal part of their result ({\it i.e.} the contribution independent of the size of the asymptotic circle) agrees with our (\ref{PropEllGenRelMonopole}).

Concerning the M-strings free energies $\widetilde{F}^{(k_1,\ldots,k_{N-1})}$ for generic configurations with $\text{gcd}(k_1$ $,\ldots,k_{N-1})\neq  1$, we have conducted an in-depth analysis of their (modular) properties. We have studied a number of interesting iterative relations among different $\widetilde{F}^{(k_1,\ldots,k_{N-1})}$ corresponding to configurations containing M5-branes that only have one M2-brane ending and beginning on them. Furthermore, we have extracted the explicit spin contents for the M-string BPS-states. In the limit $\epsilon_1\to 0$, we gave their explicit forms for all configurations up to $\sum_ik_i=6$ and expressed them in a way which allows to study their modular properties: while generically individual $\widetilde{F}^{(k_1,\ldots,k_{N-1})}$ have a modular anomaly, a unique combination $T^{(K)}$, defined in (\ref{DefTnGen}), of all configurations with $\sum_i k_i=K$, is a weak Jacobi form of weight $-2$ and index $K$ of the congruence subgroup $\Gamma_0(p)$ defined in (\ref{DefCongSubgroup}). While combinations of $\widetilde{F}^{(k_1,\ldots,k_{N-1})}$ in general do not make sense from a physics point of view, they are admissible at the point in moduli space where all K\"ahler moduli take an  equal value. We gave a physical interpretation of this fact from the viewpoint of m-strings, arguing that only at this point in the moduli space the factorization of electric excitations over the total moduli space into that of center-of-mass and of relative parts become possible.

It would be fruitful to further study and compare properties of the M\&m-string partition functions. Firstly, it is an interesting problem to elucidate the parallels of the BPS state counting in M\& m-strings with a variety of BPS bound-state counting problems in field and string theories. We recalled two situations in section 5.3.2. A new aspect of M\&m-strings, as compared to those situations, is that the BPS counting functions must exhibit modular covariance and that the modularity would impose additional constraints on the functions. Indeed, we were able to construct holomorphic Jacobi forms at least under particular congruence subgroup of $SL(2, \mathbb{Z})$. There is a priori no reason why the equivariantly regularized elliptic genus exhibit such modularity. While the parallels with other BPS bound-state problems suggest that this is the best we could get, it would still be useful to try to construct other modular covariant functions and, if not possible, to understand more precisely why the equivariantly regularised elliptic genus exhibits so. In \cite{Bak:2014xwa} it was suggested that a refined version of this quotient also captures additional contribution that would restore the full modular covariance under the $SL(2, \mathbb{Z})$. It would be very interesting to understand the refinement of \cite{Bak:2014xwa} from the viewpoint of the $\Omega$-deformations we used for equivariant regularization.
Secondly, the M-string configurations in which a direction transverse to M5-branes is compactified to a circle are related to m-string configurations in which calorons and Kaluza-Klein monopoles also contribute as new constituents. This will certainly entail new features to the BPS bound-state counting of M\&m-strings and poses an interesting new direction for building additional holomorphic Jacobi forms and corresponding elliptic genera.  We will report our results on these research programs  %corresponding elliptic genera
in a separate work \cite{toappear}.

%%%%%%%%%%%%%%%%%%%%%%%%%%%%%%%%
\section*{Acknowledgment}
We would like to thank Jinbeom~Bae, Dongsu~Bak, Michele Del Zotto, Andreas~Gustavsson, Babak Haghighat, Can Kozcaz, Guglielmo~Lockhart, Sameer Murthy, Dario Rosa and Cumrun~Vafa for many helpful discussions. SH is grateful to the Asia-Pacific Center for Theoretical Physics and Seoul National University for warm hospitality and for creating a stimulating research environment while part of this work was being done. AI thanks the Center for Mathematical Sciences and Applications at Harvard university for support and a stimulating environment. AI acknowledges the support of Higher Education Commission through grant HEC-2487. SJR acknowledges the support of the National Research Foundation of Korea(NRF) grant funded by the Korea government(MSIP) through Seoul National University with grant numbers 2005-0093843, 2010-220-C00003 and 2012K2A1A9055280.

%%%%%%%%%%%%%%%%%%%%%%%%%%%%%%%%%%%%%%%%%%%%%%%%%%%%%%%%
\appendix

%%%%%%%%%%%%%%%%%%%%%%%%%%%%%%%%%%%%%%%%%%%%%%%%%%%%%%%%
\section{Relevant Monopole Physics}
For ${\cal N}=4, 2$ super Yang-Mills with gauge group $G = SU(N)$, the Coulomb branch is parametrized by the asymptotic value of the Higgs field. Take the diagonal gauge in which all off-diagonal entries of the Higgs field are zero. Set the Cartan basis
\begin{align}
&{\mathtt H}_1 = (1, 0, 0, \cdots, 0)\,, && {\mathtt H}_2 = (0, 1, 0, \cdots, 0)\,, && \ldots && {\mathtt H}_N = (0, 0, 0, \cdots, 1)\,.
\end{align}
In this basis, the Higgs field reads
\bea
{\boldsymbol \phi} = \mbox{diag}(v_1, \cdots, v_N) = \sum_{a=1}^N v_a {\mathtt H}_a
\eea
where the asymptotic value of the Higgs field $v$'s are subject to the SU(N) condition $v_1 + \cdots + v_N = 0$. By Weyl symmetry, we can always order the asymptotic Higgs fields in the positive Weyl chamber as
\bea
v_1 \le v_2 \le \cdots \le v_N.
\eea
The 2nd homotopy group of the coset $SU(N)/(U(1))^{N-1}$ yields $(N-1)$ species of magnetic monopoles. In the Cartan basis, the asymptotic magnetic field reads
\begin{align}
&{\bf B}_a = {\bf g} {\hat{\bf r}_a \over 4 \pi r^2}\,,&&\text{where}&&{\bf g} = \sum_{a=1}^N g_a {\mathtt H}_a.
\end{align}
The $g_1, \cdots, g_N$ are magnetic charges subject to the SU(N) condition  $g_1 + \cdots + g_N = 0$. The SU(N) condition is automatically satisfied in the Weyl basis
\begin{align}
&{\boldsymbol \alpha}_1 = (1, -1, 0,0,  \cdots, 0) \,,&& {\boldsymbol \alpha}_2 = (0, 1, -1, 0, \cdots, 0) \,,&& \ldots\,, && {\boldsymbol \alpha}_{N-1} = (0, \cdots, 0, 1, -1)\,.
\end{align}
The asymptotic Higgs field and the magnetic charge are expanded as
\begin{align}
&{\boldsymbol \phi} = \sum_{a=1}^{N-1} \mu_a {\boldsymbol \alpha}_a
\,,&&\text{and} && {\bf g} = \sum_{a=1}^{N-1} k_a {\boldsymbol \alpha}_a.
\end{align}
The magnetic charge components can be related between the two bases:
\bea
&& (v_1, \cdots, v_N) = (\mu_1, \mu_2 - \mu_1, \cdots, \mu_{N-1} - \mu_{N-2}, - \mu_{N-1})  \nn
&& (g_1, \cdots, g_N) = (k_1, k_2 - k_1, \cdots, k_{N-1} - k_{N-2}, - k_{N-1}).
\eea
The BPS configuration has the mass
\bea
M_m = |{\bf g} \cdot {\boldsymbol \phi}| = |\sum_{a=1}^{N-1} n_a \mu_a|.
\eea

The total moduli space ${\cal M}_{\bf g}$ of magnetic charge ${\bf g}$ monopoles is a noncompact hyperk\"ahler space, whose asymptotic geometry is given by
\bea
\otimes_{a=1}^{N-1} (\mathbb{R}^3 \times \mathbb{S}_\phi^1)^{k_a}/\Gamma_{\bf g}.
\eea
Here, $\Gamma_{\bf g}$ is the permutation group of $(k_1, \cdots, k_{N-1})$.
It has the real dimension
\bea
\mbox{dim} {\cal M}_{\bf g} = 4 \sum_{a=1}^{N-1} k_a.
\eea
%

%%%%%%%%%%%%%%%%%%%%%%%%%%%%%%%%%%%%%%%%%%%%%%%%%%%%%%%%
\section{Noncompact Hyperk\"ahler Geometry}
In this appendix we summarize some basics on hyperk\"ahler geometry, relevant for the discussions in the main part of this paper.  We first recall that the holonomy group $H$ of a simply connected manifold $M$ must belong to the following Berger's classification:
\bea
\begin{matrix}
H & \mbox{Dim}(M) & \mbox{manifold class} \\
\hline
SO(n) & \quad n \ (n \ge 1) \quad & \mbox{Riemannian} \\
U(n) & \quad 2n \ (n \ge 1) \quad & \mbox{K\"ahler} \\
SU(2n) & \quad 2n \ (n \ge 1) \quad & \mbox{Calabi-Yau} \\
Sp(n) & \quad 4n \ (n \ge 1) \quad & \mbox{hyperk\"ahler} \\
Sp(n) \times Sp(1)/\mathbb{Z}_2 & \quad 4n  \ (n \ge 2) \quad & \mbox{quaternionic K\"ahler} \\
G_2 & \qquad 7 \qquad & G_2 \\
\mbox{Spin}(7) & \qquad 8 \qquad & \mbox{Spin}(7)
\end{matrix}
\eea
in which $M$ is assumed to be a non-symmetric and irreducible space. This means that the holonomy group $h$ acts as an irreducible representation on tangent bundle $TM$.
%--------------------------------------------------------
\subsection{Hyperk\"ahler Manifolds}
A hyperk\"ahler manifold is a Riemannian manifold $(M, g)$ with three complex structures $I_a: TM \rightarrow TM$,  $(a = 1, 2, 3, \ I_a^2 = - 1)$ that commute with parallel transport. They satisfy
\begin{align}
I_a I_b = \epsilon_{abc} I_c\,.
\end{align}
Accordingly, at any point on $M$, there is an $SO(3)$ family of skew-symmetric and closed K\"ahler 2-forms,  $(\omega_a, a = 1, 2, 3)$:
%\omega_J, \omega_K$:
%
\bea
\omega_{a} (u, v) = g ({I_a} u, v) \quad \mbox{for all} \quad
u, v \in TM \ .
\eea
%
%and similarly for $\omega_J, \omega_K$.

The holonomy group of hyperk\"ahler manifold is contained in  $Sp(n)$, {\it i.e.} the group of orthogonal transformation of $\mathbb{R}^{4n} = \mathbb{H}^n$. They are linear with respect to $I_a$ and
$I_a$'s are parallel and make $TM\big\vert_x$ a quaternionic vector space. Conversely, if a $4n$-dimensional manifold $M$ has holonomy group contained in $Sp(n)$, the complex structures $I_a\big\vert_x$ can be chosen on $TM\big\vert_x$ and render $TM\big\vert_x$ a quaternionic vector space. Parallel transport of $I_a\big\vert_x$ furnishes three complex structures on $M$, so $M$ is a hyperk\"ahler manifold.

From the viewpoint of K\"ahler geometry, we can think of the hyperk\"ahler manifold $M$ as a holomorphic symplectic manifold. Namely, choosing $I_1$ as the complex structure, $(M, g, I_1)$ is a K\"ahler manifold  uipped with an additional holomorphic symplectic form (viz. a closed and everywhere nondegenerate holomorphic 2-form) $\omega := \omega_2 + I_1 \omega_3$. Conversely, Yau's theorem asserts that a holomorphic symplectic manifold $M$ admits a Ricci flat metric for which the holomorphic symplectic form commutes with parallel transports. This implies
that the holonomy group is contained in $Sp(n)$ and hence $M$ is a hyperk\"ahler manifold.

The minimal dimension for a hyperk\"ahler manifold is 4. Since $Sp(1) \simeq  SU(2)$, it is also a CY2fold. If $M_4$ is compact and simply connected, it is actually an irreducible symplectic manifold, {\it i.e.} a K3 surface. If not simply connected, $M_4$ could be a complex 2-torus $\mathbb{T}_{\mathbb{C}}^2$ as well.

Hereafter, we summarize several constructions of noncompact hyperk\"ahler manifolds that are relevant for the present work.
%---------------------------------------------------------
\subsection{Cotangent Bundle of K\"ahler Manifold}
A class of noncompact hyperk\"ahler manifold is cotangent bundle $T^\star M_K$ of a K\"ahler manifold $M_K$. This is because the cotangent bundle can be canonically decomposed to Lagrangian subspaces $T^\star M_K \sim  V \oplus V^\star$ and the obvious pairing furnishes a holomorphic symplectic form $\omega$. This implies that $T^\star M$ is holomorphic symplectic. Its holomorphic form $\omega$ is in general defined patch wise with well-defined transition functions. 
Furthermore, it is known that, in an open neighborhood of the zero section, $T^\star M_K $ is a noncompact hyperk\"ahler manifold \cite{kaledin}.

%--------------------------------------------------------
\subsection{Hilbert Scheme}
The Hilbert schemes $X^{[K]}$ of $K (\ge 2)$ points on a four-dimensional hyperk\"ahler manifold $X$ are also hyperk\"ahler. Blow-ups by deleting a suitable codimension-2 sets provides the Hilbert-Chow morphism $X^{[K]} \rightarrow S^K X = (X)^K/S_K$, the $K$-th symmetric product of $X$, and guarantees the existence of a holomorphic symplectic form $\omega$. If $X$ is (non)compact, $X^{[K]}$ is also (non)compact.

In case $X = K3$, the moduli space $M_X(N, c_1, c_2)$ of rank-$N$ sheaves with Chern class $(c_1, c_2)$ is an irreducible symplectic manifold (assuming that the moduli space is compact). Via the Fourier-Mukai transformation, the moduli space is diffeomorphic to the Hilbert scheme $X^{[K]}$ of the same dimension. For example, by the result of Vafa and Witten \cite{Vafa:1994tf}
\bea
\chi_E [M_{K3} (2, 0, 2K)] = {\cal E}[4K-3] + {1 \over 4} {\cal E}[K],
\eea
where ${\cal E}[K]$ is the Euler characteristic of the $X^{[K]}$ of $K$ points on K3 manifold $X$.

%---------------------------------------------------------
\subsection{Monopole Moduli Space}
The noncompact hyperk\"ahler space we consider as the target space of the m-string is the moduli space of magnetic monopoles on $\mathbb{R}^4$. It can be described by the data $(A,\Phi)$ that satisfies the BPS  equation
\bea
\{ (A, \Phi)\vert F_A = \star_3 \rmd_A \Phi, \ F_A = \rmd_A + A^2, \ \rmd_A = \rmd + A\}/G.
\eea
Here $A$ is a connection on a principal $G = A_{N-1}$-bundle on $\mathbb{R}^3$ and $\Phi$ is a Lie algebra valued holomorphic Higgs form, both with appropriate fall-off conditions at spatial infinity. The magnetic charge is defined by the second Chern class of the data. The moduli space $\mathcal{M}_m(N, K)$ of BPS magnetic
monopoles of charge $K$ is the space of in equivalent data $(A, \Phi)$ modulo gauge equivalence. According to Donaldson's theorem \cite{Donaldson:1985id}, this moduli space is isomorphic to the space of rational maps $h: \mathbb{P}^1 \rightarrow \mathbb{P}^{N-1}$ of degree-$K$ with the boundary condition $h(\infty) = 0$. For example, for $G = A_1$,
\bea
\mathcal{M}(2, K) =
\left\{
{a_0 + a_1 z + \cdots + a_{K-1}z^{K-1}
\over
b_0 + b_1z + \cdots + b_{K-1} z^{K-1} + z^K}
\Big\vert \Delta \ne 0 \right\}	
\subset \mathbb{C}^{2K} \simeq  \mathbb{H}^K,
\eea
where $\Delta$ is the resultant of the numerator and the denominator. Being an open
subset of $\mathbb{H}^K$,  the moduli space $\mathcal{M}(2, K)$ is a noncompact hyperk\"ahler manifold. One of spin-offs of this paper is that, utilizing the free energy $T^{(K)}$, we were able to extract topological information of the multi-monopole moduli space ${\cal M}(N, K)$.

%--------------------------------------------------------
\subsection{Instanton Moduli Space}
The hyperk\"ahler manifold taken as the target space of the M-string is the moduli space of instantons on $\mathbb{R}^4$. It can be described by the data $A$ that satisfies the anti-self-duality condition
\bea
\{ A \vert F_A = - \star_4 F_A, F_A = \rmd A + A^2 \}/G.
\eea
Here, $A$ is a connection of $G = A_{N-1}$ bundle on $\mathbb{R}^4$, with appropriate fall-off
conditions at spacetime infinity. The instanton charge is defined by the second Chern class of $A$. The moduli space $M_i (N, K)$ of anti-self-dual instantons of charge $K$ is the space of in equivalent data $A$ modulo gauge  equivalence. This moduli space is diffeomorphic to the moduli space of rank $N$ torsion-free sheaves $E$ on $\mathbb{P}^2$ with the second Chern class $K$. Explicitly,
\bea
M_i (N, K) = \{ (B_1, B_2, P, Q) \vert [B_1, B_2] + P^TQ = 0\}/GL(K, \mathbb{C})
\eea
where the matrices $B_1, B_2$ are $(K \times K)$ and $P, Q$ are $(N \times K)$. So, $M_i (N, K)$ is the hyperk\"ahler quotient by the $GL(K, \mathbb{C})$ action of the cotangent bundle $T^\star {\cal M}$ of ${\cal M} = $ Hom$(\mathbb{C}^K, \mathbb{C}^K) \times$ Hom $(\mathbb{C}^N, \mathbb{C}^K)$.

%-------------------------------------------------------
\section{Relations among $\widetilde{F}^{(k_1,\cdots,k_{N-1})}$}
In this appendix, we explicitly show relations among different $\widetilde{F}^{(k_1,k_2,\cdots,k_{N-1})}$ whose indices $(k_1,\cdots,k_{N-1})$ contain several consecutive entries of $1$. Indeed, the upshot of our analysis is that these factors can be 'compressed' at the expense of additional factors of $W(\tau,m,\epsilon_1,\epsilon_2)$.
%%-------------------------------------------------------
\subsection{$\widetilde{F}^{(1,1,\cdots,1,2)}$ and $\widetilde{F}^{(1,2,1,\cdots,1)}$}

We start by considering $\widetilde{F}^{(1,1,\cdots,1,2)}$, {\it i.e.}
\begin{align}
k_i=1\quad \mbox{for} \quad i=1,\cdots, N-2\qquad \text{and} \qquad k_{N-1}=2\,.
\end{align}
For this configuration, we have
\begin{align}
\widetilde{F}^{(1,1,\cdots,1,2)}&=(-1)^{N}\sum_{\ell=1}^{N}(-1)^{\ell}\sum_{{k^{i}_{1},\cdots,k^{i}_{N-1}\geq  0}\atop{\sum_{i=1}^{\ell}k^{i}_{a}=1+\delta_{a,N-1}}}\prod_{i=1}^{
\ell}Z_{k^{i}_{1}\,k^{i}_{2}\,\cdots\,k^{i}_{N-1}}\nonumber\\
%&=-(H_{01}H_{10})^{2}Z_{0}^{N-2}+(Z_{2}H_{01}H_{10}-Z_{12})Z_{0}^{N-3}\nonumber\\
&=\Big[\underbrace{-(H_{01}H_{10})^{2}W+(Z_{2}H_{01}H_{10}-Z_{12})}_{\widetilde{F}^{(1,2)}}\Big]W^{N-3}
\end{align}
Since the term in the bracket is precisely $\widetilde{F}^{(1,2)}$ we have,
\bea
\widetilde{F}^{(1,1,\cdots,1,2)}&=&\widetilde{F}^{(1,2)} \,W^{N-3}
\eea
In a similar fashion we can treat
\begin{align}
&\widetilde{F}^{(1,2,1,\cdots,1)}=-2(H_{01}H_{10})^{2}\,W^{N-2}+Z_{2}(H_{01}H_{10})^{2}\,W^{N-4}-Z_{12}H_{01}H_{10}W^{N-4}\nonumber\\
&\hspace{1cm}-Z_{21}H_{01}H_{10}W^{N-4}+Z_{121}W^{N-4}\nonumber\\
&\hspace{0.5cm}=\Big[\underbrace{-2(H_{01}H_{10})^{2}\,W^{2}+Z_{2}(H_{01}H_{10})^{2}\,-Z_{12}H_{01}H_{10}-Z_{21}H_{01}H_{10}+Z_{121}}_{\widetilde{F}^{(1,2,1)}}\Big]W^{N-4}\nonumber
\end{align}
Therefore, we find the relation
\begin{align}
\widetilde{F}^{(1,2,1,\cdots,1)}=\widetilde{F}^{(1,2,1)}\,W^{N-4}\,.
\end{align}
In the same fashion we can treat any combination of $(k_i)$ which has only a single entry 2 and else only $1$'s.
%%-------------------------------------------------------
\subsection{$\widetilde{F}^{(2,2,1,\cdots,1)}$ and $\widetilde{F}^{(2,1,\cdots,1,2)}$}
The next class of examples contains sets of $(k_i)$ with two entries  equal to 2 and the remaining ones $1$. {\it i.e.} the simplest example is
\begin{align}
&\widetilde{F}^{(2,2,1,\cdots,1)}=-(H_{01}H_{10})^2(3H_{01}H_{10}-H_{11})W^{N-2}-Z_{21}H_{01}H_{10}(3H_{01}H_{10}-H_{11})W^{N-4}\nonumber\\
&\hspace{0.5cm}+Z_{121}H_{01}H_{10}W^{N-4}+Z_{2}(H_{01}H_{10})^2(3H_{01}H_{10}-H_{11})W^{N-4}+Z_{2}Z_{21}W^{N-4}\nonumber\\
&\hspace{0.5cm}-Z_{2}^2(H_{01}H_{10})W^{N-4}+Z_{22}H_{01}H_{10}W^{N-4}-Z_{221}W^{N-4}-Z_{12}\,W^{N-4}=\widetilde{F}^{(2,2,1)}\,W^{N-4}\nonumber
\end{align}
In a similar fashion we can consider the case where the first and the last entry are 2 while the remaining ones are $1$
\begin{align}
\widetilde{F}^{(2,1,\cdots,1,2)}&=\Big[-Z_{1}^{3}W^{3}-Z_{1}Z_{12}W^{2}+ 2Z_{1}^{2}Z_{2}W^{2}+Z_{2}Z_{12}W- Z_{2}^{2}Z_{1}W-Z_{1}Z_{21}W^{2}\nonumber\\
&\hspace{1cm}- \frac{Z_{12} Z_{21}}{Z_1}W +Z_{2}Z_{21}W\Big]W^{N-4}=\widetilde{F}^{(2,1,2)}W^{N-4}
\end{align}

%------------------------------------------------------
\subsection{$\widetilde{F}^{(3,1,\cdots,1)}$}
The next non-trivial example is to have $k_1=3$ and the remaining $k_i=1$
\begin{align}
\widetilde{F}^{(3,1,\cdots,1)}&=-(H_{01}H_{10})^{3}W^{N-2}+Z_{2}H_{01}H_{10}(2H_{01}H_{10}-H_{11})W^{N-3}-Z_{21}H_{01}H_{10}W^{N-3}\nonumber\\
&-Z_{3}H_{01}H_{10}W^{N-3}+Z_{31}W^{N-3}=\widetilde{F}^{(3,1)}\,W^{N-3}
\end{align}
\subsection{$\widetilde{F}^{(3,1,1,2)}$}
The final example we consider is the case $\widetilde{F}^{(3,1,1,2)}$. As a preparation, we compute $\widetilde{F}^{(3,1,2)}$
\begin{align}
\widetilde{F}^{(3,1,2)}&=-H_{11}^2 Z_{1}^4 + 2 H_{11} Z_{1}^5 - Z_{1}^6 + H_{11} Z_{1}^2 Z_{12} - Z_{1}^3 Z_{12} +
 H_{11}^2 Z_{1}^2 Z_{2} - 4 H_{11} Z_{1}^3 Z_2\nonumber\\
 & + 3 Z_{1}^4 Z_2 - H_{11} Z_{12}Z_2 +
 2 Z_1 Z_{12} Z_2 + H_{11}Z_{1} Z_{2}^2 - 2 Z_{1}^2 Z_{2}^2 + H_{11} Z_{1}^2 Z_{21}\nonumber\\
 & - Z_{1}^3 Z_{21} -
 Z_{12} Z_{21} + Z_{1} Z_{2} Z_{21} + H_{11} Z_{1}^2 Z_{3}
 - Z_{1}^3 Z_{3} - Z_{12} Z_{3} + Z_{1} Z_{2} Z_3\nonumber\\
 & -
 H_{11} Z_1 Z_{31} + Z_{1}^2 Z_{31} +
 \frac{Z_{12} Z_{31}}{Z_1} - Z_{2} Z_{31}\nonumber
\end{align}
We compare this expression to
\begin{align}
 \widetilde{F}^{(3,1,1,2)}=&H_{11}^3 Z_1^4 - 3 H_{11}^2 Z_1^5 + 3 H_{11} Z_1^6 - Z_1^7 - H_{11}^2 Z_1^2 Z_{12} +
 2 H_{11} Z_1^3 Z_{12} - Z_1^4 Z_{12}\nonumber\\
 & - H_{11}^3 Z_1^2 Z_2 + 5 H_{11}^2 Z_1^3 Z_2 -
 7 H_{11} Z_1^4 Z_2 + 3 Z_1^5 Z_2 + H_{11}^2 Z_{12} Z_2 - 3 H_{11} Z_1 Z_{12} Z_2\nonumber\\
 & +
 2 Z_1^2 Z_{12} Z_2 - H_{11}^2 Z_1 Z_2^2 + 3 H_{11} Z_1^2 Z_2^2 - 2 Z_1^3 Z_2^2 -
 H_{11}^2 Z_1^2 Z_{21} + 2 H_{11} Z_1^3 Z_{21} \nonumber\\
 &- Z_1^4 Z_{21} + H_{11} Z_{12} Z_{21} -
 Z_1 Z_{12} Z_{21} - H_{11} Z_1 Z_2 Z_{21} + Z_1^2 Z_2 Z_{21} - H_{11}^2 Z_1^2 Z_3\nonumber\\
 & +
 2 H_{11} Z_1^3 Z_3 - Z_1^4 Z_3 + H_{11} Z_{12} Z_3 - Z_1 Z_{12} Z_3 - H_{11} Z_1 Z_2 Z_3 +
 Z_1^2 Z_2 Z_3 \nonumber\\
&+ H_{11}^2 Z_1 Z_{31} - 2 H_{11} Z_1^2 Z_{31} + Z_1^3 Z_{31} + Z_{12} Z_{31} - (
 H_{11} Z_{12} Z_{31})/Z_1 + H_{11} Z_2 Z_{31}\nonumber\\
& - Z_1 Z_2 Z_{31}=\widetilde{F}^{(3,1,2)}\,W(\tau,m,\epsilon_1,\epsilon_2)
\end{align}

%%%%%%%%%%%%%%%%%%%%%%%%%%%%%

\section{Modular Building Blocks}
In this section, we compile a number of relevant definitions and useful relations of modular objects,  which we will use throughout the paper. Our conventions follow mostly \cite{zagierbook}.
%------------------
\subsection{Jacobi Theta Functions}
A class of functions used for the M-strings partition functions are the {\it Jacobi theta functions}, which are defined as follows:
{\allowdisplaybreaks\begin{align}
&\theta_1(\tau,m)=-iQ_\tau^{1/8}Q_m^{1/2}\prod_{n=1}^\infty(1-Q_\tau^n)\,(1-Q_mQ_\tau^n)\,(1-Q_m^{-1}Q_\tau^{n-1})\,,\nonumber\\
&\theta_2(\tau,m)=2Q_\tau^{1/8}\,\cos(\pi m)\,\prod_{n=1}^\infty(1-Q_\tau^n)\,(1+Q_mQ_\tau^n)\,(1+Q_m^{-1}Q_\tau^{n})\,,\nonumber\\
&\theta_3(\tau,m)=\prod_{n=1}^\infty(1-Q_\tau^n)\,(1+Q_mQ_\tau^{n-1/2})\,(1+Q_m^{-1}Q_\tau^{n-1/2})\,,\nonumber\\
&\theta_4(\tau,m)=\prod_{n=1}^\infty(1-Q_\tau^n)\,(1-Q_mQ_\tau^{n-1/2})\,(1-Q_m^{-1}Q_\tau^{n-1/2})\,.\label{JacobiTheta}
\end{align}}
Here, we use the notation
\begin{align}
&Q_\tau=e^{2\pi i \tau}\qquad \mbox{and} \qquad
Q_m=e^{2\pi im}\,.
\end{align}
Furthermore, we also introduce the Dedekind eta function
\begin{align}
\eta(\tau)=Q_\tau^{{1}/{24}}\,\prod_{n=1}^\infty (1-Q_\tau^n)\,.
\end{align}
%------------------------------------------------------
\subsection{Weak Jacobi Forms}\label{App:WeakJacForms}
In studying the M- and m-string partition functions, we encountered weak Jacobi forms of $SL(2,\mathbb{Z})$ and its subgroups. Here we outline the most important properties of these objects (a more complete treatment can be found in \cite{zagierbook}). A weak Jacobi form $\phi_{w,s}$ of weight-$w$ and index-$s$ of $SL(2,\mathbb{Z})$ is the mapping function
\begin{align}
\phi_{w,s}: \quad \,&\mathbb{H}\times \mathbb{C} \quad \longrightarrow \qquad \mathbb{C}\nonumber\\
& \ (\tau,m) \quad \longmapsto \quad \phi_{w,s}(\tau,m),
\end{align}
where $\mathbb{H}$ is the upper half-plane. It satisfies
\begin{align}
\phi_{w,s}\left(\frac{a\tau+b}{c\tau+d},\frac{m}{c\tau+d}\right) &=(c\tau+d)^w\,e^{2\pi i s\,\frac{cm^2}{c\tau+d}}\,\phi_{w,s}(\tau,m)\,,&&\left(\begin{array}{cc}a & b \\ c & d\end{array}\right)\in  SL(2,\mathbb{Z})\nonumber\\
\phi_{w,s}(\tau,m+\ell\tau+\ell')&=e^{-2\pi is(\ell^2\tau+2\ell m)}\,\phi_{w,s}(\tau,m)\,,&&\qquad \ell,\ell'\in\mathbb{Z} \ .
\end{align}
It can be Fourier-expanded
\begin{align}
\phi_{w,s}(\tau,m)=\sum_{n\geq  0}\sum_{\ell\in\mathbb{Z}}c(n,\ell)Q_\tau^n\,Q_m^\ell\,,
\end{align}
with the coefficients $c(n,\ell)=(-1)^w c(n,-\ell)$.

 The standard weak Jacobi-forms of $SL(2,\mathbb{Z})$ of index $1$ and weight $0$ and $-2$, respectively, are given by
\begin{align}
\varphi_{0,1}(\tau,m)=4\sum_{i=2}^4\frac{\theta_i(\tau,m)^2}{\theta_i(\tau,0)}
\qquad \text{and} \qquad
\varphi_{-2,1}(\tau,m)=-\frac{\theta_1^2(\tau,m)}{\eta(\tau)^6}\,.\label{BasicJacForms}
\end{align}
In fact, we have the following structure theorem: every weak Jacobi form of index $1$ and even weight $w$ (of a congruence subgroup $\Gamma\subset SL(2,\mathbb{Z})$) can be expressed as a linear combination \cite{zagierbook}
\begin{align}
\phi_{w,1}(\tau,m)=g_w(\tau)\,\varphi_{0,1}(\tau,m)+g'_{w+2}(\tau)\,\varphi_{-2,1}(\tau,m)\,,
\end{align}
where $g_w(\tau)$ and $g'_{w+2}(\tau)$ are modular forms of $\Gamma$ with weights $w$ and $w+2$, respectively.
%%%%%%%%------------------------------------------
\subsection{Theta Functions of index $k$}\label{Sec:IndexTheta}
We also define the following theta-functions of index $k$:
\begin{align}
\vartheta_{k,\ell}(\tau,m):=\sum_{n\in\mathbb{Z}}Q_\tau^{k\left(n+\frac{\ell}{2k}\right)^2}\,Q_m^{\ell+2kn}  %\ell \in\mathbb{Z}/2k\mathbb{Z}\,
,\label{DefThetaGen}
\end{align}
where $\ell$ takes values $\ell=0,\ldots, 2k-1$. They exhibit the property
\bea
\vartheta_{k,\ell}(\tau,m)=\vartheta_{k,2k-\ell}(\tau,-m)\,.
\eea
Explicitly, we find the series expansions for $k=1$
\begin{align}
&\vartheta_{1,0}(\tau, m)=1+Q_\tau\left(Q_m^2+Q_m^{-2}\right)+Q_\tau^4\left(Q_m^4+Q_m^{-4}\right)+Q_\tau^9\left(Q_m^6+Q_m^{-6}\right)+\ldots\nonumber\\
&\vartheta_{1,1}(\tau, m)=Q_\tau^{{1}/{4}}\left[Q_m+Q_m^{-1}+Q_\tau^2\left(Q_m^3+Q_m^{-3}\right)+Q_\tau^6\left(Q_m^5+Q_m^{-5}\right)\right]+\ldots\,,\label{DefThetaIndex1}
\end{align}
and for $k=2$
\begin{align}
&\vartheta_{2,0}(\tau,m)=1+Q_\tau^2(Q_m^4+Q_m^{-4})+Q_\tau^8(Q_m^8+Q_m^{-8})+\ldots\,,\nonumber\\
&\vartheta_{2,1}(\tau,m)=Q_\tau^{1/8}\left[Q_m+Q_\tau\,Q_m^{-3}+Q_\tau^3\,Q_m^5+Q_\tau^6\,Q_m^{-7}+\ldots\right]\,,\nonumber\\
&\vartheta_{2,2}(\tau,m)=Q_\tau^{1/2}\left[(Q_m^2+Q_m^{-2})+Q_\tau^4(Q_m^6+Q_m^{-6})+\ldots\right]\,,\nonumber\\
&\vartheta_{2,3}(\tau,m)=Q_\tau^{1/8}\left[Q_m^{-1}+Q_\tau\,Q_m^{3}+Q_\tau^3\,Q_m^{-5}+Q_\tau^6\,Q_m^{7}+\ldots\right]\,,\label{VarthetaIndex2}
\end{align}

%%%%%%%%
\subsection{Modular Forms for $SL(2,\mathbb{Z})$ and Its Congruence Subgroups}
In order to express weak Jacobi forms of congruence subgroups, we need a basis for modular forms of congruence subgroups of $SL(2,\mathbb{Z})$. Here we will only compile the forms relevant for us -- essentially the Eisenstein series -- and refer the interested reader to the original mathematics literature for the complete basis \cite{Lang,Stein} (see also \cite{Gaberdiel:2010ca} for a review).
%-------------------------------------------------------
\subsubsection{Eisenstein Series of $SL(2,\mathbb{Z})$}
The Eisenstein series of $SL(2,\mathbb{Z})$ are defined as
\begin{align}
E_{2k}(\tau):=1+\frac{(2\pi i)^{2k}}{(2k-1)! \zeta(2k)}\sum_{n=1}^\infty \sigma_{2k-1}(n)\, Q_\tau^n\,,\label{DefEisenstein}
\end{align}
where $\sigma_k(n)$ is the divisor function. For $k>1$ the function $E_{2k}$ is a modular form of weight $2k$. Furthermore, every $E_{2k}$ with $k>3$ can be written as a polynomial in $E_4$ and $E_6$.

For $k=2$ the function $E_2(\tau)$ is not a modular form, but transforms with an additional shift term. More precisely, only the combination
\begin{align}
\widehat{E}_2(\tau,\bar{\tau})=E_2(\tau)-\frac{3}{\pi\tau_2}\,,\label{E2hat}
\end{align}
transforms with weight $2$ under transformations of $SL(2,\mathbb{Z})$. However, the latter is no longer a holomorphic function, but is called a {\it quasi-holomorphic form}.
%----------------------------------------------------
\subsubsection{Modular Forms of $\Gamma_0(N)$}
In this section, we recall important modular forms for congruence subgroups $\Gamma_0(N)$ os $SL(2,\mathbb{Z})$. Our main references are \cite{Lang,Stein} (see also \cite{Gaberdiel:2010ca} for an overview).

The space $\mathcal{M}_{2k}(\Gamma_0(N))$ of weight $2k$ modular forms for $\Gamma_0(N)$ has the structure
\begin{align}
\mathcal{M}_{2k}(\Gamma_0(N))=\mathcal{E}_{2k}(\Gamma_0(N))\oplus \mathcal{S}_{2k}(\Gamma_0(N))\,,
\end{align}
where $\mathcal{E}_{2k}(\Gamma_0(N))$ is the subspace that is invariant under all Hecke operators, while $\mathcal{S}_{2k}(\Gamma_0(N))$ is the space of cusp forms. The latter will not be important for our current work and we therefore focus exclusively on the former. A basis for $\mathcal{E}_k(\Gamma_0(N))$ is given by (generalized) Eisenstein series of weight $2k$. This comprises the following objects
\begin{itemize}
\item {\it standard Eisenstein series of weight $2k$:}\\
If $k>1$ this comprises
\begin{align}
&E_{2k}(n\tau)\,,&&\text{for} &&n|N \,,
\end{align}
with $E_{2k}$ defined as in (\ref{DefEisenstein}). For $k=1$ we also have the combination
\begin{align}
\psi^{(N)}(\tau)=Q_\tau\,\frac{\partial}{\partial Q_\tau}\,\log\frac{\eta(N\tau)}{\eta(\tau)}=E_2(\tau)-NE_2(N\tau)\label{GeneralisedEisensteinDef}
\end{align}
which is holomorphic, since the shift-term (\ref{E2hat}) precisely cancels out.
\item {\it generalized Eisenstein series:}\\
If $N=m^2$, we can define the generalized Eisenstein series as follows
\begin{align}
E_{2k}^{\chi_m}(\tau)=\sum_{n=1}^\infty\left(\sum_{d|n}\overline{\chi_m(d)}\,\chi_m(n/d)\,d^{2k-1}\right)Q_\tau^n
\end{align}
where $\chi_m$ is a non-trivial Dirichlet character of modulus $m$. We will not need these objects in the main part of this paper.
\end{itemize}

%%%%%%%%%%%%%%%%%%%%%%%%%%
%%%%%%%%%%%%%%%%%%%%%%%%%%
\section{Explicit Examples of $\widehat{F}^{(k_1,\ldots, k_{N-1})}$}\label{App:DataFfunctions}
In this appendix we compile explicit expressions for the functions $\widehat{F}^{(k_1,\ldots, k_{N-1})}$ introduced in (\ref{DefhatFgen}). We recall that they can be written in the form (\ref{ExpandhatFGen})
\begin{align}
\widehat{F}^{(k_1,\ldots, k_{N-1})}(\tau,m)=\varphi_{-2,1}(\tau,m)\sum_{a=0}^{K} g_{2a}^{(k_1,\ldots,k_{N{-1}} )}(\tau)\,\left(2\varphi_{0,1}(\tau,m)\right)^{K-a}\left(\varphi_{-2,1}(\tau,m)\right)^{a}\,.\nonumber
\end{align}
In the following we will give explicit expressions for the modular forms $g_a^{(k_1,\ldots,k_{N-1})}$ for $K\geq  2$.
%%%
\subsection{Index $K=2$}\label{App:Index2}
As explained in section~\ref{Sect:Index2}, for $K=\sum_{a=1}^{N-1}k_a=2$, there are two functions $\widehat{F}^{(k_1,\ldots, k_{N-1})}(\tau,m)$, written in (\ref{Index2Functs}). Each of them can be written in the form
\begin{align}
&\widehat{F}^{(K=2)}(\tau,m)=\varphi_{-2,1}(\tau,m)\,\left[\frac{g^{(k_i)}_0}{12}\,\varphi_{0,1}(\tau,m)+\frac{g^{(k_i)}_2(\tau)}{24}\,\varphi_{-2,1}(\tau,m)\right]\,,\,
\end{align}
where $\sum k_i=2$ and $g^{(k_i)}_0$ are constants and $g^{(k_i)}_2(\tau)$ are modular objects subject to an anomaly. More precisely, when replacing
\begin{align}
E_2(\tau)\longrightarrow \hat{E}_2(\tau,\bar{\tau})=E_2(\tau)-\frac{3}{\pi\tau_2}\,,
\end{align}
$g^{(k_i)}_2(\tau,\bar{\tau})$ is a quasi-holomorphic modular form of weight $2$ under $\Gamma_0(2)\subset SL(2,\mathbb{Z})$. Specifically we find
\begin{align}
&g^{(2)}_0=0\,,&&g^{(2)}_2(\tau)=4(E_2(2\tau)-E_2(\tau))\,,\nonumber\\
&g^{(1,1)}_0=1\,,&&g^{(1,1)}_2(\tau)=2E_2(\tau)\,.
\end{align}
and thus
\begin{align}
&\widehat{F}^{(2)}(\tau,m)=\left(\varphi_{-2,1}(\tau,m)\right)^2\,\frac{E_2(2\tau)-E_2(\tau)}{6}\,,\nonumber\\
&\widehat{F}^{(1,1)}(\tau,m)=\frac{\varphi_{-2,1}(\tau,m)}{12}\left[\varphi_{0,1}(\tau,m)+E_2(\tau)\varphi_{-2,1}(\tau,m)\right]\,.\label{ExplicitIndex2}
\end{align}
%%%
\subsection{Index $K=3$}\label{App:Index3}
  The general form of the functions $\widehat{F}^{(k_i)}(\tau,m)$ with $\sum k_a=K=3$ is
\begin{align}
\widehat{F}^{(K=3)}(\tau,m)=\frac{\varphi_{-2,1}}{24^2}\left[g^{(k_i)}_0\,(2\varphi_{0,1})^2+2g^{(k_i)}_2\,\varphi_{0,1}\,\varphi_{-2,1}+g^{(k_i)}_4\,(\varphi_{-2,1})^2\right]
\end{align}
where $\sum k_i=3$ and $g^{(k_i)}_0$ is a constant, while $g^{(k_i)}_{2}$ and $g^{(k_i)}_{4}$ are anomalous modular quantities, {\it i.e.} under the change (\ref{E2hat}) they are quasi-holomorphic modular forms of weight $2$ and $4$ respectively, under $\Gamma_0(3)$. Specifically, we find
\begin{align}
&g^{(3)}_0=0\,,&&g^{(3)}_2=6(E_2(3\tau)-E_2(\tau))\,,&&g^{(3)}_4=\frac{2}{5}(20E_2(\tau)^2+7E_4(\tau)-27E_4(3\tau))\,,\nonumber\\
&g^{(2,1)}_0=0\,,&&g^{(2,1)}_2=0\,,&&g^{(2,1)}_4=6(E_4(\tau)-E_2(\tau)^2)\,,\nonumber\\
&g^{(1,1,1)}_0=1\,,&&g^{(1,1,1)}_2=4E_2(\tau)\,,&&g^{(1,1,1)}_4=4E_2(\tau)^2\,.
\end{align}
And thus we have
\begin{align}
\widehat{F}^{(3)}&=\varphi_{-2,1}\left[\frac{E_2(3)-E_2(\tau)}{48}\varphi_{0,1}+\frac{20E_1(1)^2+7E_4(1)-27E_4(3)}{1440}\,\varphi_{-2,1}\right]\,,\nonumber\\
\widehat{F}^{(2,1)}&=(\varphi_{-2,1})^3\,\frac{E_4(\tau)-E_2(\tau)^2}{96}\,,\nonumber\\
\widehat{F}^{(1,1,1)}&=\varphi_{-2,1}\left[\frac{(\varphi_{0,1})^2}{144}+\frac{E_2(1)}{72}\,\varphi_{-2,1}\,\varphi_{0,1}+\frac{E_2(1)^2}{144}\,(\varphi_{-2,1})^2\right]\,.\label{Fsindex3}
\end{align}
where we introduced the shorthand notation $E_m(n):=E_m(n\tau)$.
%%%
\subsection{Index $K=4$}\label{App:Index4}
The general form of the functions $\widehat{F}^{(k_i)}(\tau,m)$ with $\sum k_i=K=4$ is
\begin{align}
\widehat{F}^{(K=4)}=\frac{\varphi_{-2,1}}{24^3}\left[g^{(k_i)}_0\,(2\varphi_{0,1})^3+g^{(k_i)}_2\,(2\varphi_{0,1})^2\,\varphi_{-2,1}+2g^{(k_i)}_4\,\varphi_{0,1}(\varphi_{-2,1})^2+g^{(k_i)}_6\,(\varphi_{-2,1})^3\right]\label{Fsindex4}
\end{align}
where $\sum k_i=4$ and $g^{(k_i)}_0$ is a constant, while $g^{(k_i)}_{2,4,6}$ are anomalous modular quantities, {\it i.e.} under the change (\ref{E2hat}) they are quasi-holomorphic modular forms of $\Gamma_0(4)$ with weight $2,4,6$ respectively. The explicit expressions we find are given in table~\ref{Tab:Index4}
\begin{table}[h!tbpp]
\begin{center}
\scalebox{.85}{\rotatebox{90}{\parbox{16cm}{
\begin{align}
&g^{(4)}_0=0\,,&&g^{(4)}_2=8[E_2(2)-E_2(1)]\,,&&g^{(4)}_4=\frac{8}{5}\left[25E_1(1)^2-20E_2(2)^2+7E_4(1)-12E_4(2)\right]\,,\nonumber\\[6pt]
&g_6^{(4)}=\frac{16}{105}\left[-280 E_2(1)^3-273 E_2(1)E_4(1)+336 E_2(2)E_4(2)-87 E_6(1)+304 E_6(2)\right]\,,\hspace{-14cm} &&{}&&\nonumber\\[24pt]
&g^{(2,2)}_0=0\,,&&g^{(2,2)}_2=4[E_2(2)-E_2(1)]\,,&&g^{(2,2)}_4=\frac{4}{5}\left[-25E_1(1)^2+40E_2(2)^2+9E_4(1)-24E_4(2\tau)\right]\,,\nonumber\\[6pt]
&g_6^{(2,2)}=\frac{8}{105}\left[105 E_2(1)^3-84 E_2(1)E_4(1)-672 E_2(2)E_4(2)-181 E_6(1)+832 E_6(2)\right]\,,\hspace{-16cm} && {} &&\nonumber\\[24pt]
&g^{(3,1)}_0=0\,,&&g^{(3,1)}_2=0\,,&&g^{(3,1)}_4=10[E_4(1)-E_2(1)^2]\,,\nonumber\\[6pt]
&g_6^{(3,1)}=\frac{4}{3}\left[25 E_2(1)^3-9 E_2(1)E_4(1)-16E_6(1)\right]\,,\hspace{-16cm}&& {} &&\nonumber\\[24pt]
&g^{(1,2,1)}_0=0\,,&&g^{(1,2,1)}_2=0\,,&&g^{(1,2,1)}_4=0\,,\nonumber\\[6pt]
&g_6^{(1,2,1)}=16\left[-2 E_2(1)^3+3 E_2(1)E_4(1)-2E_6(1)\right]\,,\hspace{-16cm} &&{}&&\nonumber\\[24pt]
&g^{(2,1,1)}_0=0\,,&&g^{(2,1,1)}_2=0\,,&&g^{(2,1,1)}_4=6[E_4(1)-E_2(1)^2]\,,\nonumber\\[6pt]
&g_6^{(2,1,1)}12[E_2(1)(E_4(1)-E_2(1)^2)]\,,\hspace{-16cm} &&{}&&\nonumber\\[24pt]
&g^{(1,1,1,1)}_0=1\,,&&g^{(1,1,1,1)}_2=6E_2(1)\,,&&g^{(1,1,1,1)}_4=12E_2(1)^2\,,\nonumber\\[6pt]
&g_6^{(1,1,1,1)}=8E_2(1)^3\,,\hspace{-16cm} && {} &&\nonumber
\end{align}}}}
\caption{Coefficients for $\widehat{F}^{(k_i)}(\tau,m)$ with $\sum k_i=K=4$.}\label{Tab:Index4}
\end{center}
\end{table}
where we again used the shorthand notation $E_m(n)=E_m(n\tau)$.
%%%
\subsection{Index $K=5$}\label{App:Index5}
The general form of the functions $\widehat{F}^{(k_i)}(\tau,m)$ with $\sum k_i=K=5$ is
\begin{align}
\widehat{F}^{(K=5)}=\frac{\varphi_{-2,1}}{24^4}\sum_{a=0}^4g^{(k_i)}_{2a}\,(2\varphi_{0,1})^{4-a}(\varphi_{-2,1})^a\label{Fsindex5}
\end{align}
where $\sum k_i=5$ and $g^{(k_i)}_0$ is a constant, while $g^{(k_i)}_{2,4,6,8}$ are anomalous modular quantities, {\it i.e.} under the change (\ref{E2hat}) they are quasi-holomorphic modular forms of $\Gamma_0(5)$ with weight $2,4,6,8$ respectively. The explicit expressions are given in table~\ref{Tab:Index5}.
\begin{table}[h!tbp]
\begin{center}
\scalebox{.77}{\rotatebox{90}{\parbox{15cm}{
\begin{align}
&g^{(5)}_0=0\,,&&g^{(5)}_2=10[E_2(5)-E_2(1)]\,,&&g^{(5)}_4=10 \left[7 E_2(1){}^2+10 E_2(5) E_2(1)-25 E_2(5){}^2+3 E_4(1)+5 E_4(5)\right]\,,\nonumber\\
&g_6^{(5)}=-\frac{8}{105} \left(3500 E_2(1){}^3+3633 E_4(1) E_2(1)-1365 E_2(5) E_4(1)+482 E_6(1)-6250 E_6(5)\right)\,, \hspace{-20cm} &&{} && \nonumber\\
&g_8^{(5)}=\frac{10}{21} \left[560 E_2(1){}^4+1008 E_4(1) E_2(1){}^2+304 E_6(1) E_2(1)+3 \left(83 E_4(1){}^2-98 E_4(5) E_4(1)-625 E_4(5){}^2+16
   E_2(5) E_6(1)\right)\right]\,,\hspace{-20cm} && {} && \nonumber\\[12pt]
&g^{(3,2)}_0=0\,,&&g^{(3,2)}_2=0\,,&&g^{(3,2)}_4=16 \left[E_4(1)-E_2(1){}^2\right]\,,\hspace{1cm}g^{(3,2)}_6=\frac{16}{3} \left[13 E_2(1){}^3-3 E_4(1) E_2(1)-10 E_6(1)\right]\,,\hspace{-20cm} && {} && \nonumber\\
&g^{(3,2)}_8=-\frac{32}{3} \left[2 E_2(1){}^4+9 E_4(1) E_2(1){}^2-2 E_6(1) E_2(1)-9 E_4(1){}^2\right]\,,\hspace{-20cm} && {} &&\nonumber\\[12pt]
&g^{(4,1)}_0=0\,,&&g^{(4,1)}_2=0\,,&&g^{(4,1)}_4=14 \left[E_4(1)-E_2(1){}^2\right]\,,\hspace{1cm} g^{(4,1)}_6=\frac{56}{3} \left[7 E_2(1){}^3-3 E_4(1) E_2(1)-4 E_6(1)\right]\,,\hspace{-20cm} && {} &&\nonumber\\
&g^{(4,1)}_8=\frac{2}{3} \left[-343 E_2(1){}^4-126 E_4(1) E_2(1){}^2+208 E_6(1) E_2(1)+261 E_4(1){}^2\right]\,,\hspace{-20cm} && {} &&\nonumber\\[12pt]
&g^{(2,1,2)}_0=0\,,&&g^{(2,1,2)}_2=0\,,&& g^{(2,1,2)}_4=0\,,\hspace{1cm}g^{(2,1,2)}_6=0\,,\hspace{1cm}g^{(2,1,2)}_8=36 \left(E_2(1){}^2-E_4(1)\right)^2\,,\nonumber\\[12pt]
&g^{(2,2,1)}_0=0\,,&&g^{(2,2,1)}_2=0\,,&&g^{(2,2,1)}_4=6 \left[E_4(1)-E_2(1){}^2\right]\,,\hspace{1cm} g^{(2,2,1)}_6=-8 \left[5 E_2(1){}^3-9 E_4(1) E_2(1)+4 E_6(1)\right]\,,\hspace{-20cm} && {} &&\nonumber\\
&g^{(2,2,1)}_8=8 \left[2 E_2(1){}^4-3 E_4(1) E_2(1){}^2-8 E_6(1) E_2(1)+9 E_4(1){}^2\right]\,,\hspace{-20cm} && {} &&\nonumber\\[12pt]
&g^{(1,3,1)}_0=0\,,&&g^{(1,3,1)}_2=0\,,&&g^{(1,3,1)}_4=0\,,\hspace{1cm} g^{(1,3,1)}_6=-32 \left[E_2(1){}^3-3 E_4(1) E_2(1)+2 E_6(1)\right]\,,\hspace{-20cm} && {} &&\nonumber\\
&g^{(1,3,1)}_8=64 \left[2 E_2(1){}^4-3 E_4(1) E_2(1){}^2-2 E_6(1) E_2(1)+3 E_4(1){}^2\right]\,,\hspace{-20cm} && {} &&\nonumber\\[12pt]
&g^{(3,1,1)}_0=0\,,&&g^{(3,1,1)}_2=0\,,&&g^{(3,1,1)}_4=10 \left[E_4(1)-E_2(1){}^2\right]\,,\hspace{1cm} g^{(3,1,1)}_6=\frac{8}{3} \left[5 E_2(1){}^3+3 E_4(1) E_2(1)-8 E_6(1)\right]\,,\hspace{-20cm} && {} &&\nonumber\\
&g^{(3,1,1)}_8=\frac{8}{3} E_2(1) \left[25 E_2(1){}^3-9 E_4(1) E_2(1)-16 E_6(1)\right]\,,\hspace{-20cm} && {} &&\nonumber\\[12pt]
&g^{(1,2,1,1)}_0=0\,,&&g^{(1,2,1,1)}_2=0\,,&&g^{(1,2,1,1)}_4=0\,,\hspace{1cm} g^{(1,2,1,1)}_6=-16 \left[E_2(1){}^3-3 E_4(1) E_2(1)+2 E_6(1)\right]\,,\hspace{-20cm} && {} &&\nonumber\\
&g^{(1,2,1,1)}_8=-32 E_2(1) \left[E_2(1){}^3-3 E_4(1) E_2(1)+2 E_6(1)\right]\,,\hspace{-20cm} && {} &&\nonumber\\[12pt]
&g^{(2,1,1,1)}_0=0\,,&&g^{(2,1,1,1)}_2=0\,,&&g^{(2,1,1,1)}_4=6 \left[E_4(1)-E_2(1){}^2\right]\,,\hspace{1cm} g^{(2,1,1,1)}_6=24 E_2(1) \left[E_4(1)-E_2(1){}^2\right]\,,\hspace{-20cm} && {} &&\nonumber\\
&g^{(2,1,1,1)}_8=24 E_2(1){}^2 \left[E_4(1)-E_2(1){}^2\right]\,,\hspace{-20cm} && {} &&\nonumber\\[12pt]
&g^{(1,1,1,1,1)}_0=1\,,&&g^{(1,1,1,1,1)}_2=8 E_2(1)\,,&&g^{(1,1,1,1,1)}_4=24 E_2(1){}^2\,,\hspace{1cm} g^{(1,1,1,1,1)}_6=32 E_2(1){}^3\,,\hspace{1cm}g^{(1,1,1,1,1)}_8=16 E_2(1){}^4\,.\nonumber
\end{align}}}}
\caption{Coefficients for $\widehat{F}^{(k_i)}(\tau,m)$ with $\sum k_i=K=5$.}\label{Tab:Index5}
\end{center}
\end{table}
\newpage
%%%%%%%%%%%%%%%%%
%%%
\subsection{Index $K=6$}\label{App:Index6}
The general form of the functions $\widehat{F}^{(k_i)}(\tau,m)$ with $\sum k_i=K=6$ is
\begin{align}
\widehat{F}^{(K=6)}=\frac{\varphi_{-2,1}}{24^5}\sum_{a=0}^5g^{(k_i)}_{2a}\,(2\varphi_{0,1})^{5-a}(\varphi_{-2,1})^a\label{Fsindex6}
\end{align}
where $\sum k_i=6$ and $g^{(k_i)}_0$ is a constant, while $g^{(k_i)}_{2,4,6,8,10}$ are anomalous modular quantities, {\it i.e.} under the change (\ref{E2hat}) they are quasi-holomorphic modular forms of $\Gamma_0(6)$ with weight $2,4,6,8,10$ respectively. The explicit expressions are given in tables~\ref{Tab:Index6a}, \ref{Tab:Index6b} and \ref{Tab:Index6c}.

\begin{table}[h!tbp]
\begin{center}
\scalebox{.63}{\rotatebox{90}{\parbox{15cm}{
\begin{align}
&g^{(6)}_0=0\,,&&g^{(6)}_2=-12[E_2(1)-E_2(2)-E_2(3)+E_2(6)]\,,&&g^{(6)}_4=\frac{4}{5} \left[260 E_2(1){}^2-180 E_2(6) E_2(1)-200 E_2(2){}^2-135 E_2(3){}^2+540 E_2(6){}^2+46 E_4(1)-88
   E_4(2)-135 E_4(3)-108 E_4(6)\right]\,,\nonumber\\
&g_6^{(6)}=-\frac{8}{35} \left[4480 E_2(1){}^3+3318 E_4(1) E_2(1)-2240 E_2(2){}^3-1134 E_2(3) E_4(1)+1134 E_2(6) E_4(1)-4032 E_2(2) E_4(2)-5103 E_2(3) E_4(3)+842 E_6(1)-2368 E_6(2)-2673 E_6(3)+7776 E_6(6)\right]\,, \hspace{-30cm} &&{} && \nonumber\\
&g_8^{(6)}=\frac{16}{5775} \big[831600 E_2(1){}^4+105 \left(12309 E_4(1)+26132 E_4(2)\right) E_2(1){}^2-3499200 E_6(6)
   E_2(1)+428811 E_4(1){}^2-4188528 E_4(2){}^2-14486688 E_4(6){}^2-280665 E_2(3){}^2 E_4(1)-11862480
   E_2(2){}^2 E_4(2)\,,\hspace{-100cm} && {} && \nonumber\\
&\hspace{1cm}   -631071 E_4(1) E_4(3)-142884 E_4(1) E_4(6)+733220 E_2(2) E_6(1)-137700 E_2(6)
   E_6(1)+10104160 E_2(2) E_6(2)-180 E_2(3) \left(2537 E_6(1)+8019 E_6(3)\right)+20995200 E_2(6)
   E_6(6)\big]   E_2(2){}^2 E_4(2)\,,\hspace{-100cm} && {} && \nonumber\\
&g_10^{(6)}=-\frac{32}{4375} \big(252000 E_2(1){}^5+10635000 E_2(6) E_6(1) E_2(1)+5538414 E_2(3) E_4(1){}^2-9914847 E_2(6)
   E_4(1){}^2+42408576 E_2(3) E_4(2){}^2+255927552 E_2(6) E_4(2){}^2-2460375 E_2(6) E_4(3){}^2\hspace{-100cm} && {} && \nonumber\\
&\hspace{1cm}+5040000
   E_2(2){}^3 E_4(1)-14 E_2(2) \left(1893 E_4(1){}^2+7674112 E_4(2){}^2\right)+101250 E_2(3){}^2
   E_6(1)-31905000 E_2(6){}^2 E_6(1)+7099312 E_4(2) E_6(1)+9478404 E_4(3) E_6(1)+802584 E_4(6) E_6(1)\hspace{-100cm} && {} && \nonumber\\
&\hspace{1cm}+10000
   E_2(2){}^2 \left(143 E_6(1)-1216 E_6(2)\right)+72429696 E_4(2) E_6(2)+33059232 E_4(3) E_6(2)-49350528
   E_4(6) E_6(2)-230947200 E_4(6) E_6(3)\big)\,,\hspace{-100cm} && {} && \nonumber\\[10pt]
&g^{(5,1)}_0=0\,,&&g^{(5,1)}_2=0\,,&&g^{(5,1)}_4=-18 \left[E_2(1){}^2-E_4(1)\right]\,,\hspace{1cm}
g_6^{(5,1)}=\frac{108}{5} \left[15 E_2(1){}^3-7 E_4(1) E_2(1)-8 E_6(1)\right]\,, \hspace{1cm}&g_8^{(5,1)}=-54 \left[27 E_2(1){}^4+6 E_4(1) E_2(1){}^2-16 E_6(1) E_2(1)-17 E_4(1){}^2\right]\,, \hspace{-30cm} &&{} && \nonumber\\
&g_{10}^{(5,1)}=\frac{36}{25} \left[1215 E_2(1){}^5+1350 E_4(1) E_2(1){}^3-400 E_6(1) E_2(1){}^2-1173 E_4(1){}^2 E_2(1)-992
   E_4(1) E_6(1)\right]\,, \hspace{-100cm} &&{} && \nonumber\\[10pt]
&g^{(4,2)}_0=0\,,&&g^{(4,2)}_2=0\,,&&g^{(4,2)}_4=-24 \left[E_2(1){}^2-2 E_2(2){}^2-E_4(1)+2 E_4(2)\right]\,,\nonumber\\
&g_6^{(4,2)}=\frac{24}{5} \left[55 E_2(1){}^3-21 E_4(1) E_2(1)-80 E_2(2){}^3-48 E_2(2) E_4(2)-34 E_6(1)+128 E_6(2)\right]\,, \hspace{-30cm} &&{} && \nonumber\\
&g_8^{(4,2)}=-\frac{48}{175} \left[1575 E_2(1){}^4+1155 E_4(1) E_2(1){}^2-10 \left(61 E_6(1)+1088 E_6(2)\right) E_2(1)-24
   \left(280 E_4(2) E_2(2){}^2-800 E_6(2) E_2(2)+83 E_4(1){}^2+72 E_4(2){}^2\right)\right]\,, \hspace{-100cm} &&{} && \nonumber\\
&g_{10}^{(4,2)}=\frac{96}{2695} \big[385 \left(63 E_4(1)-64 E_4(2)\right) E_2(1){}^3+17380 E_6(1) E_2(1){}^2-26565 E_4(1){}^2
   E_2(1)+2 \big(98560 E_4(2) E_2(2){}^3-204160 E_6(2) E_2(2){}^2+4224 \left(3 E_4(1){}^2+22
   E_4(2){}^2\right) E_2(2)\hspace{-100cm} && {} &&\nonumber\\
&\hspace{1cm}-9935 E_4(1) E_6(1)+14720 E_4(2) E_6(2)\big)\big]\,, \hspace{-100cm} &&{} && \nonumber\\[10pt]
&g^{(3,3)}_0=0\,,&&g^{(3,3)}_2=6 [E_2(3)- E_2(1)]\,,&&g^{(3,3)}_4=-\frac{4}{5} \left[50 E_2(1){}^2-135 E_2(3){}^2-23 E_4(1)+108 E_4(3)\right]\,,\nonumber\\
&g_6^{(3,3)}=\frac{24}{35} \left[245 E_2(1){}^3-7 \left(14 E_4(1)+117 E_4(3)\right) E_2(1)+6 \left(126 E_2(3) E_4(3)-23
   E_6(1)+9 E_6(3)\right)\right]\,, \hspace{-30cm} &&{} && \nonumber\\
&g_8^{(3,3)}=\frac{4}{175} \left[-11025 E_2(1){}^4-21630 E_4(1) E_2(1){}^2+80 \left(167 E_6(1)+3024 E_6(3)\right)
   E_2(1)+22319 E_4(1){}^2+271836 E_4(3){}^2+34020 E_2(3){}^2 E_4(1)-550800 E_2(3) E_6(3)\right]\,, \hspace{-100cm} &&{} && \nonumber\\
&g_{10}^{(3,3)}=\frac{8}{35} \big[315 E_2(1){}^5+3150 E_4(1) E_2(1){}^3+40 \left(28 E_6(1)+243 E_6(3)\right)
   E_2(1){}^2-\left(5248 E_4(1){}^2+755397 E_4(3){}^2\right) E_2(1)-3 \big(29160 E_6(3) E_2(3){}^2-\left(3721
   E_4(1){}^2+783099 E_4(3){}^2\right) E_2(3) \hspace{-100cm} &&{} && \nonumber\\
&\hspace{1cm}+16 \left(1763 E_4(1)+30042 E_4(3)\right) E_6(3)\big)\big]\,, \hspace{-100cm} &&{} && \nonumber\\[10pt]
&g^{(1,4,1)}_0=0\,,&&g^{(1,4,1)}_2=0\,,&&g^{(1,4,1)}_4=0\,,\hspace{1cm}g_6^{(1,4,1)}=-48 \left[E_2(1){}^3-3 E_4(1) E_2(1)+2 E_6(1)\right]\,, \hspace{-30cm} &&{} && \nonumber\\
&g_8^{(1,4,1)}=192 \left[3 E_2(1){}^4-5 E_4(1) E_2(1){}^2-2 E_6(1) E_2(1)+4 E_4(1){}^2\right]\,, \hspace{1cm}g_{10}^{(1,4,1)}=192 \left[-6 E_2(1){}^5+3 E_4(1) E_2(1){}^3+8 E_6(1) E_2(1){}^2+5 E_4(1){}^2 E_2(1)-10 E_4(1) E_6(1)\right]\,, \hspace{-100cm} &&{} && \nonumber
\end{align}}}}
\caption{Coefficients for $\widehat{F}^{(k_i)}(\tau,m)$ with $\sum k_i=K=6$.}\label{Tab:Index6a}
\end{center}
\end{table}
%%%%
\begin{table}[h!tbp]
\begin{center}
\scalebox{.68}{\rotatebox{90}{\parbox{15cm}{
\begin{align}
&g^{(4,1,1)}_0=0\,,&&g^{(4,1,1)}_2=0\,,&&g^{(4,1,1)}_4=-14 \left[E_2(1){}^2-E_4(1)\right]\,,\hspace{1cm}g_6^{(4,1,1)}=\frac{28}{3} \left[11 E_2(1){}^3-3 E_4(1) E_2(1)-8 E_6(1)\right]\,, \hspace{-20cm} &&{} && \nonumber\\
&g_8^{(4,1,1)}=\frac{2}{3} \left[49 E_2(1){}^4-294 E_4(1) E_2(1){}^2-16 E_6(1) E_2(1)+261 E_4(1){}^2\right]\,, \hspace{1cm}g_{10}^{(4,1,1)}=\frac{4}{3} E_2(1) \left[-343 E_2(1){}^4-126 E_4(1) E_2(1){}^2+208 E_6(1) E_2(1)+261 E_4(1){}^2\right]\,, \hspace{-20cm} &&{} && \nonumber\\[10pt]
&g^{(2,2,2)}_0=0\,,&&g^{(2,2,2)}_2=4 \left[E_2(2)-E_2(1)\right]\,,&&g^{(2,2,2)}_4=-\frac{8}{5} \left[25 E_2(1){}^2-40 E_2(2){}^2-9 E_4(1)+24 E_4(2)\right]\,,\nonumber\\
&g_6^{(2,2,2)}=-\frac{16}{35} \left[175 E_2(1){}^3-84 E_4(1) E_2(1)-560 E_2(2){}^3+1008 E_2(2) E_4(2)+69 E_6(1)-608
   E_6(2)\right]\,, \hspace{-20cm} &&{} && \nonumber\\
&g_8^{(2,2,2)}=\frac{32}{175} \left[525 E_2(1){}^4-945 E_4(1) E_2(1){}^2-80 \left(7 E_6(1)+16 E_6(2)\right) E_2(1)+468
   E_4(1){}^2-64 \left(105 E_4(2) E_2(2){}^2-250 E_6(2) E_2(2)+117 E_4(2){}^2\right)\right]\,, \hspace{-20cm} &&{} && \nonumber\\
&g_{10}^{(2,2,2)}=-\frac{32}{18865} \big[5390 \left(21 E_4(1)+80 E_4(2)\right) E_2(1){}^3-169400 E_6(1) E_2(1){}^2-171633 E_4(1){}^2
   E_2(1)+59136 E_2(2) E_4(2){}^2-3449600 E_2(2){}^3 E_4(2)+227523 E_4(1) E_6(1)) \hspace{-20cm} &&{} && \nonumber\\
&\hspace{1cm}+4040960 E_2(2){}^2
   E_6(2+142560 E_4(1) E_6(2)-1223936 E_4(2) E_6(2)\big]\,, \hspace{-20cm} &&{} && \nonumber\\[10pt]
&g^{(3,1,2)}_0=0\,,&&g^{(3,1,2)}_2=0\,,&&g^{(3,1,2)}_4=0\,,\hspace{1cm}g_6^{(3,1,2)}=0\,, \hspace{1cm}g_8^{(3,1,2)}=60 \left[E_2(1){}^2-E_4(1)\right]{}^2\,, \hspace{-20cm} &&{} && \nonumber\\
&g_{10}^{(3,1,2)}=-8 \left[E_2(1){}^2-E_4(1)\right) \left(25 E_2(1){}^3-9 E_4(1) E_2(1)-16 E_6(1)\right]\,, \hspace{-20cm} &&{} && \nonumber\\[10pt]
&g^{(2,3,1)}_0=0\,,&&g^{(2,3,1)}_2=0\,,&&g^{(2,3,1)}_4=0\,,\hspace{1cm}g_6^{(2,3,1)}=-48 \left[E_2(1){}^3-3 E_4(1) E_2(1)+2 E_6(1)\right]\,, \hspace{-20cm} &&{} && \nonumber\\
&g_8^{(2,3,1)}=12 \left[21 E_2(1){}^4-26 E_4(1) E_2(1){}^2-32 E_6(1) E_2(1)+37 E_4(1){}^2\right]\,, \hspace{-20cm} &&{} && \nonumber\\
&g_{10}^{(2,3,1)}=-24 \left[3 E_2(1){}^5+18 E_4(1) E_2(1){}^3-16 E_6(1) E_2(1){}^2-37 E_4(1){}^2 E_2(1)+32 E_4(1) E_6(1)\right]\,, \hspace{-20cm} &&{} && \nonumber\\[10pt]
&g^{(3,2,1)}_0=0\,,&&g^{(3,2,1)}_2=0\,,&&g^{(3,2,1)}_4=0\,,\hspace{1cm}g_6^{(3,2,1)}=-32 \left[E_2(1){}^3-3 E_4(1) E_2(1)+2 E_6(1)\right]\,, \hspace{-20cm} &&{} && \nonumber\\
&g_8^{(3,2,1)}=128 \left[2 E_2(1){}^4-3 E_4(1) E_2(1){}^2-2 E_6(1) E_2(1)+3 E_4(1){}^2\right]\,, \hspace{1cm}g_{10}^{(3,2,1)}=-128 E_4(1) \left[E_2(1){}^3-3 E_4(1) E_2(1)+2 E_6(1)\right]\,, \hspace{-20cm} &&{} && \nonumber\\[10pt]
&g^{(1,2,2,1)}_0=0\,,&&g^{(1,2,2,1)}_2=0\,,&&g^{(1,2,2,1)}_4=0\,,\hspace{1cm}g_6^{(1,2,2,1)}=-16 \left[E_2(1){}^3-3 E_4(1) E_2(1)+2 E_6(1)\right]\,, \hspace{-20cm} &&{} && \nonumber\\
&g_8^{(1,2,2,1)}=-16 \left[7 E_2(1){}^4-30 E_4(1) E_2(1){}^2+32 E_6(1) E_2(1)-9 E_4(1){}^2\right]\,, \hspace{1cm}g_{10}^{(1,2,2,1)}=32 \left[E_2(1){}^5-16 E_6(1) E_2(1){}^2+27 E_4(1){}^2 E_2(1)-12 E_4(1) E_6(1)\right]\,, \hspace{-20cm} &&{} && \nonumber\\[10pt]
&g^{(1,3,1,1)}_0=0\,,&&g^{(1,3,1,1)}_2=0\,,&&g^{(1,3,1,1)}_4=0\,,\hspace{1cm}g_6^{(1,3,1,1)}=-32 \left[E_2(1){}^3-3 E_4(1) E_2(1)+2 E_6(1)\right]\,, \hspace{-40cm} &&{} && \nonumber\\
&g_8^{(1,3,1,1)}=64 \left[E_2(1){}^4-4 E_6(1) E_2(1)+3 E_4(1){}^2\right]\,, \hspace{1cm}g_{10}^{(1,3,1,1)}=128 E_2(1) \left[2 E_2(1){}^4-3 E_4(1) E_2(1){}^2-2 E_6(1) E_2(1)+3 E_4(1){}^2\right]\,, \hspace{-40cm} &&{} && \nonumber\\[10pt]
&g^{(3,1,1,1)}_0=0\,,&&g^{(3,1,1,1)}_2=0\,,&&g^{(3,1,1,1)}_4=-10 \left[E_2(1){}^2-E_4(1)\right]\,,\hspace{1cm}g_6^{(3,1,1,1)}=-\frac{4}{3} \left[5 E_2(1){}^3-21 E_4(1) E_2(1)+16 E_6(1)\right]\,, \hspace{-40cm} &&{} && \nonumber\\
&g_8^{(3,1,1,1)}=\frac{8}{3} E_2(1) \left[35 E_2(1){}^3-3 E_4(1) E_2(1)-32 E_6(1)\right]\,, \hspace{1cm}g_{10}^{(3,1,1,1)}=\frac{16}{3} E_2(1){}^2 \left[25 E_2(1){}^3-9 E_4(1) E_2(1)-16 E_6(1)\right]\,, \hspace{-40cm} &&{} && \nonumber\\[10pt]
&g^{(2,1,1,2)}_0=0\,,&&g^{(2,1,1,2)}_2=0\,,&&g^{(2,1,1,2)}_4=0\,,\hspace{1cm}g_6^{(2,1,1,2)}=0 \,,\hspace{1cm}g_8^{(2,1,1,2)}=36 \left(E_2(1){}^2-E_4(1)\right){}^2\,, \hspace{-40cm} &&{} && \nonumber\\
&g_{10}^{(2,1,1,2)}=72 E_2(1) \left(E_2(1){}^2-E_4(1)\right){}^2\,, \hspace{-40cm} &&{} && \nonumber
\end{align}}}}
\caption{Coefficients for $\widehat{F}^{(k_i)}(\tau,m)$ with $\sum k_i=K=6$ (continued).}\label{Tab:Index6b}
\end{center}
\end{table}
%%%%
\begin{table}[h!tbp]
\begin{center}
\scalebox{.68}{\rotatebox{90}{\parbox{15cm}{
\begin{align}
&g^{(2,1,2,1)}_0=0\,,&&g^{(2,1,2,1)}_2=0\,,&&g^{(2,1,2,1)}_4=0\,,\hspace{1cm}g_6^{(2,1,2,1)}=0 \,, \hspace{-10cm} &&{} && \nonumber\\
&g_8^{(2,1,2,1)}=0\,, \hspace{-10cm} &&{} && \nonumber\\
&g_{10}^{(2,1,2,1)}=96 \left[E_2(1){}^2-E_4(1)\right) \left(E_2(1){}^3-3 E_4(1) E_2(1)+2 E_6(1)\right]\,, \hspace{-10cm} &&{} && \nonumber\\[10pt]
&g^{(2,2,1,1)}_0=0\,,&&g^{(2,2,1,1)}_2=0\,,&&g^{(2,2,1,1)}_4=6 [E_4(1)-E_2(1){}^2]\,, \hspace{-10cm} &&{} && \nonumber\\
&g_6^{(2,2,1,1)}=-52 E_2(1){}^3+84 E_4(1) E_2(1)-32 E_6(1)\,, \hspace{-10cm} &&{} && \nonumber\\
&g_8^{(2,2,1,1)}=-8 \left[8 E_2(1){}^4-15 E_4(1) E_2(1){}^2+16 E_6(1) E_2(1)-9 E_4(1){}^2\right]\,, \hspace{-10cm} &&{} && \nonumber\\
&g_{10}^{(2,2,1,1)}=16 E_2(1) \left[2 E_2(1){}^4-3 E_4(1) E_2(1){}^2-8 E_6(1) E_2(1)+9 E_4(1){}^2\right]\,, \hspace{-10cm} &&{} && \nonumber\\[10pt]
&g^{(1,1,2,1,1)}_0=0\,,&&g^{(1,1,2,1,1)}_2=0\,,&&g^{(1,1,2,1,1)}_4=0\,, \hspace{-10cm} &&{} && \nonumber\\
&g_6^{(1,1,2,1,1)}=-16 \left[E_2(1){}^3-3 E_4(1) E_2(1)+2 E_6(1)\right]\,, \hspace{-10cm} &&{} && \nonumber\\
&g_8^{(1,1,2,1,1)}=-64 E_2(1) \left[E_2(1){}^3-3 E_4(1) E_2(1)+2 E_6(1)\right]\,,  \hspace{-10cm} &&{} && \nonumber\\
&g_{10}^{(1,1,2,1,1)}=-64 E_2(1){}^2 \left[E_2(1){}^3-3 E_4(1) E_2(1)+2 E_6(1)\right]\,, \hspace{-10cm} &&{} && \nonumber\\[10pt]
&g^{(1,2,1,1,1)}_0=0\,,&&g^{(1,2,1,1,1)}_2=0\,,&&g^{(1,2,1,1,1)}_4=0\,, \hspace{-10cm} &&{} && \nonumber\\
&g_6^{(1,2,1,1,1)}=-16 \left[E_2(1){}^3-3 E_4(1) E_2(1)+2 E_6(1)\right]\,, \hspace{-10cm} &&{} && \nonumber\\
&g_8^{(1,2,1,1,1)}=-64 E_2(1) \left[E_2(1){}^3-3 E_4(1) E_2(1)+2 E_6(1)\right]\,,  \hspace{-10cm} &&{} && \nonumber\\
&g_{10}^{(1,2,1,1,1)}=-64 E_2(1){}^2 \left[E_2(1){}^3-3 E_4(1) E_2(1)+2 E_6(1)\right]\,, \hspace{-10cm} &&{} && \nonumber\\[10pt]
&g^{(2,1,1,1,1)}_0=0\,,&&g^{(2,1,1,1,1)}_2=0\,,&&g^{(2,1,1,1,1)}_4=6 (E_4(1)-E_2(1){}^2)\,, \hspace{-10cm} &&{} && \nonumber\\
&g_6^{(2,1,1,1,1)}=-36 \left[E_2(1){}^3-E_2(1) E_4(1)\right]\,, \hspace{-10cm} &&{} && \nonumber\\
&g_8^{(2,1,1,1,1)}=-72 \left[E_2(1){}^4-E_2(1){}^2 E_4(1)\right]\,, \hspace{-10cm} &&{} && \nonumber\\
&g_{10}^{(2,1,1,1,1)}=-48 \left[E_2(1){}^5-E_2(1){}^3 E_4(1)\right]\,, \hspace{-10cm} &&{} && \nonumber\\[10pt]
&g^{(1,1,1,1,1,1)}_0=1\,,&&g^{(1,1,1,1,1,1)}_2=10 E_2(1)\,,&&g^{(1,1,1,1,1,1)}_4=40 E_2(1){}^2\,, \hspace{-10cm} &&{} && \nonumber\\
&g_6^{(1,1,1,1,1,1)}=80 E_2(1){}^3\,, \hspace{-10cm} &&{} && \nonumber\\
&g_8^{(1,1,1,1,1,1)}=80 E_2(1){}^4\,, \hspace{1cm}g_{10}^{(1,1,1,1,1,1)}=32 E_2(1){}^5\,, \hspace{-10cm} &&{} && \nonumber
\end{align}}}}
\caption{Coefficients for $\widehat{F}^{(k_i)}(\tau,m)$ with $\sum k_i=K=6$ (continued).}\label{Tab:Index6c}
\end{center}
\end{table}

%%%%%%%%%%%%%%%%%%%%%%%%%%
%\FloatBarrier
%%%%%%%%%%%%%%%%%%%%%%%%%%

\newpage

%%%%%%%%%%%%%%%%
\bibliography{references}

\begin{thebibliography}{999}

%\cite{Vafa:1996xn}
\bibitem{Vafa:1996xn}
  C.~Vafa,
  {\sl Evidence for F theory},
  Nucl.\ Phys.\ B {\bf 469} (1996) 403
  [hep-th/9602022].
  %%CITATION = HEP-TH/9602022;%%

%\cite{Heckman:2013pva}
\bibitem{Heckman:2013pva}
  J.~J.~Heckman, D.~R.~Morrison and C.~Vafa, {\it On the Classification of 6D SCFTs and Generalized ADE Orbifolds,}
  JHEP {\bf 1405} (2014) 028
  [arXiv:1312.5746 [hep-th]].
  %%CITATION = ARXIV:1312.5746;%%

%\cite{DelZotto:2014hpa}
\bibitem{DelZotto:2014hpa}
  M.~Del Zotto, J.~J.~Heckman, A.~Tomasiello and C.~Vafa,
  {\it 6d Conformal Matter,}
  JHEP {\bf 1502} (2015) 054
  [arXiv:1407.6359 [hep-th]].
  %%CITATION = ARXIV:1407.6359;%%

%\cite{Heckman:2014qba}
\bibitem{Heckman:2014qba}
  J.~J.~Heckman,
  {\it More on the Matter of 6D SCFTs,}
  arXiv:1408.0006 [hep-th].
  %%CITATION = ARXIV:1408.0006;%%

%\cite{Haghighat:2014vxa}
\bibitem{Haghighat:2014vxa}
  B.~Haghighat, A.~Klemm, G.~Lockhart and C.~Vafa,
  {\it Strings of Minimal 6d SCFTs,}
  arXiv:1412.3152 [hep-th].
  %%CITATION = ARXIV:1412.3152;%%

%\cite{Heckman:2015bfa}
\bibitem{Heckman:2015bfa}
  J.~J.~Heckman, D.~R.~Morrison, T.~Rudelius and C.~Vafa,
  {\it Atomic Classification of 6D SCFTs,}
  arXiv:1502.05405 [hep-th].
  %%CITATION = ARXIV:1502.05405;%%

%\cite{Ganor:1996mu}
\bibitem{strings}
%\bibitem{Ganor:1996mu}
  O.~J.~Ganor and A.~Hanany,
  {\sl Small E(8) instantons and tensionless noncritical strings},
  Nucl.\ Phys.\ B {\bf 474} (1996) 122
  [hep-th/9602120];\\
  %%CITATION = HEP-TH/9602120;%%
%\cite{Seiberg:1996vs}
%\bibitem{Seiberg:1996vs}
  N.~Seiberg and E.~Witten,
  {\sl Comments on string dynamics in six-dimensions},
  Nucl.\ Phys.\ B {\bf 471} (1996) 121
  [hep-th/9603003];\\
  %%CITATION = HEP-TH/9603003;%%
%\cite{Distler:1996zg}
%\bibitem{Distler:1996zg}
  J.~Distler and A.~Hanany,
  {\sl (0,2) Noncritical strings in six-dimensions},
  Nucl.\ Phys.\ B {\bf 490} (1997) 75
  [hep-th/9611104].
  %%CITATION = HEP-TH/9611104;%%

%\cite{Witten:1996qb}
\bibitem{Witten:1996qb}
  E.~Witten,
  {\sl Phase transitions in M theory and F theory},
  Nucl.\ Phys.\ B {\bf 471} (1996) 195
  [hep-th/9603150].
  %%CITATION = HEP-TH/9603150;%%

\bibitem{Haghighat:2013gba}
  B.~Haghighat, A.~Iqbal, C.~Kozcaz, G.~Lockhart and C.~Vafa,
  {\it M-Strings,}
  arXiv:1305.6322 [hep-th].

\bibitem{Haghighat:2013tka}
  B.~Haghighat, C.~Kozcaz, G.~Lockhart and C.~Vafa,
  {\it Orbifolds of M-strings,}
  Phys.\ Rev.\ D {\bf 89}, no. 4, 046003 (2014)
  [arXiv:1310.1185 [hep-th]].

\bibitem{Hohenegger:2013ala} S.~Hohenegger and A.~Iqbal, {\it M-strings, elliptic genera and $\mathcal{N} = 4$ string amplitudes,} Fortsch.\ Phys.\  {\bf 62} (2014) 155 [arXiv:1310.1325 [hep-th]].
%%CITATION = ARXIV:1310.1325;%%

%\cite{Douglas:2010iu}
\bibitem{Douglas:2010iu}
  M.~R.~Douglas,
  {\sl On D=5 super Yang-Mills theory and $(2,0)$ theory},
  JHEP {\bf 1102} (2011) 011
  [arXiv:1012.2880 [hep-th]].
  %%CITATION = ARXIV:1012.2880;%%

%\cite{Tachikawa:2011ch}
\bibitem{Tachikawa:2011ch}
  Y.~Tachikawa,
  {\sl On S-duality of 5d super Yang-Mills on $S^1$},
  JHEP {\bf 1111} (2011) 123
  [arXiv:1110.0531 [hep-th]].
  %%CITATION = ARXIV:1110.0531;%%

\bibitem{Kim:2011mv}
  H.~C.~Kim, S.~Kim, E.~Koh, K.~Lee and S.~Lee, {\it On instantons as Kaluza-Klein modes of M5-branes,}
  JHEP {\bf 1112}, 031 (2011)
  [arXiv:1110.2175 [hep-th]].

\bibitem{Bak:2014xwa}
  D.~Bak and A.~Gustavsson,
  {\it Elliptic genera of monopole strings,}
  arXiv:1403.4297 [hep-th].

\bibitem{Harvey:2014nha}
  J.~A.~Harvey, S.~Lee and S.~Murthy,
  {\it Elliptic genera of ALE and ALF manifolds from gauged linear sigma models,}
  arXiv:1406.6342 [hep-th].

\bibitem{EG}
E.~Witten, {\it Elliptic Genera and Quantum Field Theory,} Commun.\ Math.\ Phys.\  {\bf 109} (1987) 525;\\
A.~N.~Schellekens and N.~P.~Warner, {\it Anomalies, Characters and Strings,} Nucl.\ Phys.\ B {\bf 287} (1987) 317;\\
W.~Lerche, B.~E.~W.~Nilsson, A.~N.~Schellekens and N.~P.~Warner, {\it Anomaly Cancelling Terms From the Elliptic Genus,} Nucl.\ Phys.\ B {\bf 299} (1988) 91.

\bibitem{taub-nut}
S.A. Connell,
{\sl The dynamics of the SU(3) charge (1,1) magnetic monopole}, preprint (1994) {\tt ftp://maths.adelaide.edu.au/pure/mmurray/oneone.tex};\\
%\cite{Gauntlett:1996cw}
%\bibitem{Gauntlett:1996cw}
  J.~P.~Gauntlett and D.~A.~Lowe,
  {\sl Dyons and S duality in N=4 supersymmetric gauge theory},
  Nucl.\ Phys.\ B {\bf 472} (1996) 194
  [hep-th/9601085];\\
  %%CITATION = HEP-TH/9601085;%%
K.~M.~Lee, E.~J.~Weinberg and P.~Yi,
  {\sl The Moduli space of many BPS monopoles for arbitrary gauge groups},
  Phys.\ Rev.\ D {\bf 54} (1996) 1633
  [hep-th/9602167].

\bibitem{Nekrasov:2009rc} N.~A.~Nekrasov and S.~L.~Shatashvili, {\it Quantization of Integrable Systems and Four Dimensional Gauge Theories,}
  arXiv:0908.4052 [hep-th].
  %%CITATION = ARXIV:0908.4052;%%

\bibitem{Mironov:2009uv}
  A.~Mironov and A.~Morozov,
  {\sl Nekrasov Functions and Exact Bohr-Zommerfeld Integrals},
  JHEP {\bf 1004} (2010) 040
  [arXiv:0910.5670 [hep-th]].
  %%CITATION = ARXIV:0910.5670;%%

\bibitem{Kawai:1993jk} T.~Kawai, Y.~Yamada and S.~K.~Yang, {\it Elliptic genera and N=2 superconformal field theory,} Nucl.\ Phys.\ B {\bf 414} (1994) 191 [hep-th/9306096].
%%CITATION = HEP-TH/9306096;%%

\bibitem{Haghighat:2015coa} B.~Haghighat, {\it From strings in 6d to strings in 5d,} arXiv:1502.06645 [hep-th].
  %%CITATION = ARXIV:1502.06645;%%

%\cite{Nekrasov:2002qd}
\bibitem{Nekrasov:2002qd}
  N.~A.~Nekrasov,
  {\it Seiberg-Witten prepotential from instanton counting},
  Adv.\ Theor.\ Math.\ Phys.\  {\bf 7} (2004) 831
  [hep-th/0206161].
  %%CITATION = HEP-TH/0206161;%%


\bibitem{Gopakumar:1998ii}
  R.~Gopakumar and C.~Vafa,
  {\it M theory and topological strings. 1.,}
  hep-th/9809187.
  %%CITATION = HEP-TH/9809187;%%
  %286 citations counted in INSPIRE as of 21 Nov 2014

\bibitem{Gopakumar:1998jq}
  R.~Gopakumar and C.~Vafa,
  {\it M theory and topological strings. 2.,}
  hep-th/9812127.
  %%CITATION = HEP-TH/9812127;%%

\bibitem{Hollowood:2003cv}
  T.~J.~Hollowood, A.~Iqbal and C.~Vafa,
  {\it Matrix models, geometric engineering and elliptic genera,}
  JHEP {\bf 0803}, 069 (2008)
  [hep-th/0310272].


\bibitem{Leung:1997tw}
  N.~C.~Leung and C.~Vafa, {\it Branes and toric geometry,}
  Adv.\ Theor.\ Math.\ Phys.\  {\bf 2}, 91 (1998)
  [hep-th/9711013].

\bibitem{Witten:1993jg} E.~Witten, {\it On the Landau-Ginzburg description of N=2 minimal models,} Int.\ J.\ Mod.\ Phys.\ A {\bf 9} (1994) 4783 [hep-th/9304026].
  %%CITATION = HEP-TH/9304026;%%

\bibitem{Eguchi:1988vra} T.~Eguchi, H.~Ooguri, A.~Taormina and S.~K.~Yang, {\it Superconformal Algebras and String Compactification on Manifolds with SU(N) Holonomy,} Nucl.\ Phys.\ B {\bf 315} (1989) 193.
  %%CITATION = NUPHA,B315,193;%%

\bibitem{Gritsenko:1999nm} V.~Gritsenko, {\it Complex vector bundles and Jacobi forms,} math/9906191.
%%CITATION = MATH/9906191;%%


%\cite{Troost:2010ud}
\bibitem{Troost:2010ud}
  J.~Troost,
  {\sl The non-compact elliptic genus: mock or modular},
  JHEP {\bf 1006} (2010) 104
  [arXiv:1004.3649 [hep-th]].
  %%CITATION = ARXIV:1004.3649;%%


%\cite{Imbimbo:1983dg}
\bibitem{Imbimbo:1983dg}
  C.~Imbimbo and S.~Mukhi,
  {\sl Topological Invariance in Supersymmetric Theories With a Continuous Spectrum},
  Nucl.\ Phys.\ B {\bf 242} (1984) 81.
  %%CITATION = NUPHA,B242,81;%%

%\cite{Witten:1995ex}
\bibitem{Witten:1995ex}
  E.~Witten,
  {\sl String theory dynamics in various dimensions},
  Nucl.\ Phys.\ B {\bf 443} (1995) 85
  [hep-th/9503124].
  %%CITATION = HEP-TH/9503124;%%

\bibitem{Sethi:1997pa}
  S.~Sethi and M.~Stern,  {\sl D-brane bound states redux},
  Commun.\ Math.\ Phys.\  {\bf 194} (1998) 675
  [hep-th/9705046];\\
  %%CITATION = HEP-TH/9705046;%%
 P.~Yi,
  {\sl Witten index and threshold bound states of D-branes},
  Nucl.\ Phys.\ B {\bf 505} (1997) 307
  [hep-th/9704098].

\bibitem{Green:1997tn}
  M.~B.~Green and M.~Gutperle,
  {\sl D Particle bound states and the D instanton measure},
  JHEP {\bf 9801} (1998) 005
  [hep-th/9711107].
  %%CITATION = HEP-TH/9711107;%%

%\cite{Sen:1993zi}
\bibitem{Sen:1993zi}
  A.~Sen, {\it Magnetic monopoles, Bogomolny bound and SL(2,Z) invariance in string theory},
  Mod.\ Phys.\ Lett.\ A {\bf 8} (1993) 2023
  [hep-th/9303057].
  %%CITATION = HEP-TH/9303057;%%


%\cite{Rey:1989xj}
\bibitem{Rey:1989xj}
  S.~J.~Rey,
  {\sl The Confining Phase of Superstrings and Axionic Strings},''
  Phys.\ Rev.\ D {\bf 43} (1991) 526.
  %%CITATION = PHRVA,D43,526;%%

\bibitem{Font:1990gx}
  A.~Font, L.~E.~Ibanez, D.~Lust and F.~Quevedo,
  {\sl Strong - weak coupling duality and nonperturbative effects in string theory},
  Phys.\ Lett.\ B {\bf 249} (1990) 35.
  %%CITATION = PHLTA,B249,35;%%


%\cite{Montonen:1977sn}
\bibitem{Montonen:1977sn}
  C.~Montonen and D.~I.~Olive,
  {\sl Magnetic Monopoles as Gauge Particles?},
  Phys.\ Lett.\ B {\bf 72} (1977) 117.
  %%CITATION = PHLTA,B72,117;%%

\bibitem{Dorey:2000dt}
  N.~Dorey,
  {\sl Instantons, compactification and S-duality in N=4 SUSY Yang-Mills theory. 1.},
  JHEP {\bf 0104} (2001) 008
  [hep-th/0010115];\\
  %%CITATION = HEP-TH/0010115;%%
 N.~Dorey and A.~Parnachev,
  {\sl Instantons, compactification and S duality in N=4 SUSY Yang-Mills theory. 2.},
  JHEP {\bf 0108} (2001) 059
  [hep-th/0011202].
  %%CITATION = HEP-TH/0011202;%%

\bibitem{toappear} S. Hohenegger, A. Iqbal and S.-J. Rey,
to appear (2015).

\bibitem{kaledin} D. Kaledin, {\sl Hyperkaehler structures on total spaces of holomorphic cotangent bundles},
arXiv:alg-geom/9710026. 

%\cite{Vafa:1994tf}
\bibitem{Vafa:1994tf}
  C.~Vafa and E.~Witten, {\it A Strong coupling test of S duality},
  Nucl.\ Phys.\ B {\bf 431} (1994) 3
  [hep-th/9408074].
  %%CITATION = HEP-TH/9408074;%%

\bibitem{Donaldson:1985id}
  S.~K.~Donaldson,
  {\it Nahm's  equations And The Classification Of Monopoles,}
  Commun.\ Math.\ Phys.\  {\bf 96} (1984) 387.
  %%CITATION = CMPHA,96,387;%%

\bibitem{zagierbook}
M. Eichler and D. Zagier, `The Theory of Jacobi Forms', Birkh\"auser (1985).

\bibitem{Lang} S. Lang, {\it Introduction to Modular Forms}, Grundlehren der Mathematischen Wissenschaften {\bf 222}, Springer Verlag, Berlin (1995).

\bibitem{Stein} W. Stein, {\it Modular Forms, a Computational Approach}, Graduate Studies in Mathematics {\bf 79}, American Mathematical Society, Providence, RI (2007).

\bibitem{Gaberdiel:2010ca} M.~R.~Gaberdiel, S.~Hohenegger and R.~Volpato, {\it Mathieu Moonshine in the elliptic genus of K3,} JHEP {\bf 1010} (2010) 062 [arXiv:1008.3778 [hep-th]].
%%CITATION = ARXIV:1008.3778;%%














\end{thebibliography}

\end{document}